\documentclass[pra,twocolumn,showpacs,superscriptaddress]{revtex4}
\usepackage{psfig}
\usepackage{epsfig}
\usepackage{graphics}

\newcommand{\infig}[2]{\begin{center}\hspace{0mm}\mbox{\epsfig{file=#1,width=#2}}\end{center}}

\def\mco{\multicolumn}
\def\lb{\label}
\def\beq{\begin{equation}}
\def\eeq{\end{equation}}
\def\bea{\begin{eqnarray}}
\def\eea{\end{eqnarray}}

\def\Lc {L_{\mbox{\scriptsize{crit}}} }
\def\rsq {1/r^2}
\def\rw {R_w}

\def\mass{M}
\def\height{h}

\begin{document}
Adv. in At. Mol. Opt. Physics \textbf{48}, 263-356 (2002) \\




\title{Microscopic atom optics: from wires to an atom chip}


\author{Ron Folman, Peter Kr\"uger, J\"org Schmiedmayer}

\address{Physikalisches Institut,
Universit\"at Heidelberg, D-69120 Heidelberg, Germany}

\author{Johannes Denschlag}
\address{Institut f\"ur Experimentalphysik, Universit\"at Innsbruck, A-6020 Innsbruck,
Austria}

\author{Carsten Henkel}
\address{Institut f\"ur Physik, Universit\"at Potsdam, D-14469 Potsdam, Germany}

\begin{abstract}
We give a comprehensive overview of the development of micro
traps, from the first experiments on guiding atoms using current
carrying wires in the early 1990's to the creation of a BEC on an
atom chip.
\end{abstract}

\maketitle


\tableofcontents

\section{Introduction}

Scientific and technological progress in the last decades has
proven that miniaturization and integration are an important step
towards the robust application of fundamental physics. Be it
electronics and semiconductor physics in integrated circuits, or
optics in micro optical devices and sensors. The experimental
effort described in this work aims at the goal to achieve the same
for matter wave optics.

Matter wave optics beautifully illustrates quantum behavior.
Realizations using neutral atoms are attractive because of the
well established techniques to coherently manipulate internal and
external degrees of freedom, and their weak coupling to the
environment.  Miniaturizing electric and magnetic potentials is
essential to building very versatile traps and guides for atoms at
a scale ($<1\mu$m) which will enable controlled quantum
manipulation and entanglement.  Integration with other quantum
optics, micro optics and photonics techniques will allow for the
robust creation, manipulation and measurement of atomic quantum
states in these micro traps. In our vision we see a monolithic
integrated matter wave device which will allow us to establish a
new experimental toolbox and enable new insights into fundamental
quantum physics for example in issues such as decoherence,
entanglement and non linearity, low-dimensional mesoscopic
systems, and degenerate quantum gases (Bosons and Fermions) beyond
mean field. A successful implementation may lead to wide spread
applications from highly sensitive sensors (time and acceleration)
to quantum information technology.

The goal of this review is to sum up our $10$ year long exciting
journey into the miniaturization and integration of matter wave
optics resulting in devices mounted on surfaces, so called {\em
atom chips}. It brought together the best of two worlds, the vast
knowledge of quantum optics and matter wave optics and the mature
techniques of micro fabrication.

The first experiments started in the early 1990s with the guiding
of atoms with free standing wires and investigating the trapping
potentials in simple geometries.  This lead later to the micro
fabrication of atom optical elements down to $1\mu$m size on atom
chips. Very recently the simple creation of a Bose-Einstein
condensation in miniaturized surface traps was achieved, and the
first attempts to integrate light optics on the atom chip are in
progress. Even though there are many open questions, we firmly
believe that we are only at the beginning of a new era of robust
quantum manipulation of atomic systems, with many applications.

The review is organized as follows.  We begin in
section~\ref{s:basics} by describing microscopic atom optical
elements using current carrying and charged structures, acting as
sources for electric and magnetic fields which interact with the
atom. In the following sections we describe first the experiments
with free standing structures - the so called {\em atom wire}
(section~\ref{III}), investigating the basic principles of
microscopic atom optics, and then the miniaturization on the {\em
atom chip} (section~\ref{s:chip-ex}). In section~\ref{s:loss} we
discuss one of the central open questions: what happens with cold
atoms close to a warm surface, how fast will they heat up, and how
fast will they lose their coherence?  The role of technical noise,
the fundamental noise limits and the influence of atom--atom and
atom--surface interactions, are discussed. We then conclude with
an outlook of what we believe the future directions to be, and
what can be hoped for (section~\ref{s:outlook}).

The scientific progress regarding manipulation of atoms close to
surfaces was enormous within the last decade.  Besides the {\em
atom-wire} and {\em atom chip} described here, it ranges a whole
spectrum: from reflection experiments on atom mirrors, to studying
Van der Waals interactions and quantum reflection; from the use of
micro magnets to trap atoms, to evanescent light field traps. Many
of these have been reviewed recently, and will not be included
here.  We will almost exclusively concentrate on manipulation of
atoms with static microscopic electric and magnetic fields created
by charged and/or current carrying (microscopic) structures. For
related experiments and proposals, which are not discussed in this
review, we refer the reader to the excellent reviews referenced
throughout the text e.g. \citep{Dow96-1,Gri00-95,Hin99-R119}.

\section{Designing microscopic atom optics}
\label{s:basics} \label{II}

Neutral atoms can be manipulated by means of their interaction
with magnetic, electric, and optical fields. In this review the
emphasis is put on the magnetic and the electric interaction. The
designing of traps and guides using charged and current carrying
structures and the combination of different types of interaction
to form devices for guided matter wave optics are discussed. It is
shown how miniaturization of the structures leads to great
versatility where a variety of potentials can be tailored at will.
We start with some general statements and focus then on the
concepts that are important for surface mounted structures and
address issues of miniaturization and its technological
implications.

\subsection{Magnetic interaction}
\label{II-A}

A particle with total spin~${\bf F}$ and magnetic moment
$\mbox{\boldmath$\mu$} = g_F \mu_B {\bf F}$ experiences the
potential
\begin{equation}
V_{\rm mag}= -\mbox{\boldmath$\mu$} \cdot {\bf B}=- g_F\mu_B m_F B
\label{eq:magpot}
\end{equation}
where $\mu_B$ is the Bohr magneton, $g_F$ the Land\'{e}-factor of
the atomic hyperfine state, and $m_F$ the magnetic quantum number.
In general, the vector coupling $\mbox{\boldmath$\mu$} \cdot {\bf
B}$ results in a complicated motion of the atom. However, if the
Larmor precession ($\omega_{L}=\mu_B B/\hbar$) of the magnetic
moment is much faster than the apparent change of direction of the
magnetic field in the rest frame of the moving atom, an adiabatic
approximation can be applied. The magnetic moment then follows the
direction of the field adiabatically, $m_F$ is a constant of
motion, and the atom is moving in a potential proportional to the
modulus of the magnetic field $B=|{\bf B}|$.

Depending on the orientation of \mbox{\boldmath$\mu$} relative to
the direction of a static magnetic field, we distinguish two
cases:

(1) If the magnetic moment is pointing in the same direction as
the magnetic field ($V_{\rm mag}<0$), an atom is drawn towards
increasing fields, therefore it is in a {\em strong field seeking}
state. This state is the lowest energy state of the system. Minima
of the potential energy are found at {\em maxima} of the field.
Maxima of the magnetic field in free space are, however, forbidden
by the Earnshaw theorem\footnote{The Earnshaw theorem can be
generalized to any combination of electric, magnetic and
gravitational field \citep{Win84-181,Ket92-403}.}. This means that
for trapping atoms in the strong field seeking state, a source of
the magnetic field, such as a current carrying material object or
an electron beam, has to be located inside the trapping region.

(2) If the magnetic moment of an atom is pointing in the direction
opposite to the magnetic field ($V_{\rm mag}>0$), the atom is
repelled from regions with high magnetic fields; it is then in the
metastable {\em weak field seeking} state. In this case, {\em
minima} of the modulus of the field correspond to potential
minima. Because a minimum of the modulus of the magnetic field in
free space is not forbidden by the Earnshaw theorem, traps of this
type are most common for neutral atom trapping. Losses from the
traps are a potential problem (see section~\ref{s:loss}),
especially when non adiabatic transitions to the energetically
lower high field seeking states become likely in regions of low or
even vanishing fields.

\subsubsection{Kepler guide}
\label{II-A-1}

A possible realization of a trap for an atom in the {\em strong
field seeking} state is a current carrying wire with the atom
orbiting around it
\citep{Vla61-740,Sch92-284,Sch95-169,Sch95-R13,Sch96-R2525,Sch96-693,Den98,Den99-2014}.
The interaction potential is given by:\footnote{This and all other
expressions for magnetic and electric fields in this section are
given in the limit of an infinitely thin wire, unless stated
otherwise.}
\begin{equation}
  V_{\rm mag} = -\mbox{\boldmath$\mu$} \cdot {\bf B} =
  -\left(\frac{\mu_{0}}{2 \pi }\right) I_w \frac{1}{r}  \bf{e}_{\varphi} \cdot \mbox{\boldmath$\mu$} ,
  \label{MagField}
\end{equation}
where $I_w$ is the current through the wire, $\bf{e}_{\varphi}$ is
the azimuthal unit vector in cylindrical coordinates, and $\mu_0=4
\pi$~mmG/A is the vacuum permeability.  This potential has the
$1/r$ form of a Coulomb potential, but the coupling
$\mbox{\boldmath$\mu$} \cdot {\bf B}$ is vectorial. Using the
adiabatic approximation, $V_{\rm mag}$ corresponds to a
2-dimensional scalar ($1/r$) potential, in which atoms move in
Kepler orbits.\footnote{From corrections to the adiabatic
approximation to the next order, we obtain an effective
Hamiltonian for the orbital motion of the atom where the
Coulomb-like binding potential is corrected by a small repulsive
$1/r^2$ interaction
\citep{Sha88,Aha92-3593,Ste92-1022,Lit93-924,Sch96-R2525,Sch96-693}.
As a result, the adiabatic orbits are Kepler-like, and show an
additional precession around the wire. A very similar potential
can be realized for small polar molecules with a permanent dipole
moment interacting with the electric field of a charged wire
\citep{Sek96-407}.}

\begin{figure}[t]
        \infig {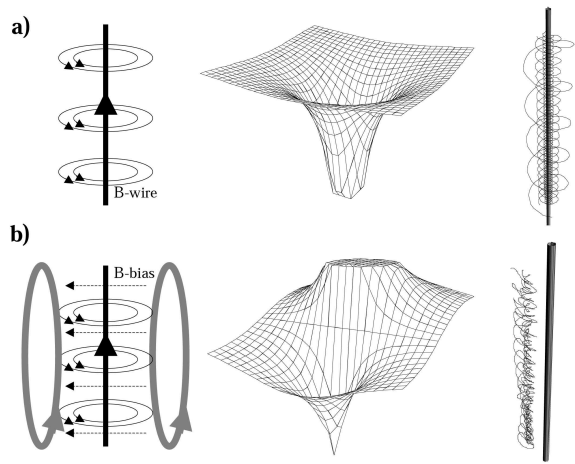} {0.9 \columnwidth}
        \caption{Guiding neutral atoms using a
        current carrying wire. ({\em a}) Guiding the atoms in their {\em
        strong field seeking} state as they circle around the wire.
        ({\em b}) Atoms in the {\em
        weak field seeking} state can be held in a
        2-dimensional magnetic quadrupole field which is created by adding
        a constant bias field to the wire field. Typical trajectories
        of atoms are shown on the
        right hand side of the figure.}
        \label{f:cw_guides}
\end{figure}

In the quantum regime, the system looks like a 2-dimensional
hydrogen atom in a (nearly circular) Rydberg state.  The wire
resembles the {\em ``nucleus"} and the atom takes the place of the
{\em ``electron"}. Considerable theoretical work has been
published on the quantum mechanical treatment of this system
showing a hydrogen-like energy spectrum
\citep{Pro77-1075,Blu89-236,Blu91-22,Vor91-29,Hau95-3138,Bur96-3225,Ber96-1653,Sch96-R2525,Sch96-693}.

The magnetic field, the potential, and typical classical
trajectories are presented in Fig.~\ref{f:cw_guides}a.

\subsubsection{Side guide}
\label{II-A-2}

Originally, \cite{Fri33-610} presented the idea that a current
carrying wire ($I_{w}$) and a homogeneous bias field ($B_{b}$) may
produce a well defined 2-dimensional field minimum, in the form of
a quadrupole field (Fig.~\ref{f:cw_guides}b). The bias field
cancels the circular magnetic field of the wire along a line
parallel to the wire at a distance
\begin{equation}
 r_0 =\left(\frac{\mu_{0}}{2\pi}\right)\frac{I_{w}}{B_{b}}.\label{eq:trap_height_in_chap.2}
\end{equation}
Around this line the modulus of the magnetic field increases in
all directions and forms a tube with a magnetic field minimum at
its center. Atoms in the {\em weak field seeking} state can be
trapped in this 2-dimensional quadrupole field and be guided along
the side of the wire, i.e. in a {\em side guide}. At the center of
the trap the magnetic field gradient is
\begin{equation}
 \left.\frac{dB}{dr}\right|_{r_{0}}= \left( \frac{2\pi}{\mu_{0}} \right) \frac{B_{b}^2}{I_{w}}=\frac{B_b}{r_0}.
\end{equation}
If the bias field is orthogonal to the wire, the two fields cancel
exactly, and trapped atoms can be lost due to Majorana transitions
between trapped and untrapped spin states (see
section~\ref{s:loss2}). This problem can be circumvented by adding
a small B-field component $B_{ip}$ along the wire direction which
lifts the energetic degeneracy between the trapped and untrapped
states. This potential is conventionally called a Ioffe-Pritchard
trap \citep{Got62-1045,Pri83-1336,Bag87-2194}. At the same time,
the potential form of the guide near the minimum changes from
linear to harmonic. The guide is then characterized by the
curvature in the transverse directions
\begin{equation}
 \left.\frac{d^{2}B}{dr^2}\right|_{r_{0}}=
\left( \frac{2\pi}{\mu_{0}} \right)^2
\frac{B_{b}^4}{B_{ip}I_{w}^2}=\frac{B_b^2}{r_0^2 B_{ip}}.
\end{equation}
In the harmonic oscillator approximation, the trap frequency is
given by
\begin{equation}
\frac{\omega}{2\pi}= \frac{1}{2\pi}\sqrt{\frac{\mu_B g_F
m_F}{M}\left(\frac{d^2B}{dr^2}\right)}\propto\frac{B_b}{r_0}\sqrt{\frac{1}{M
B_{ip}}} \label{eq:freq}\label{trap_omega_chap.2}
\end{equation}
where $M$ is the mass of the atom.

\begin{figure}[t]
 \infig{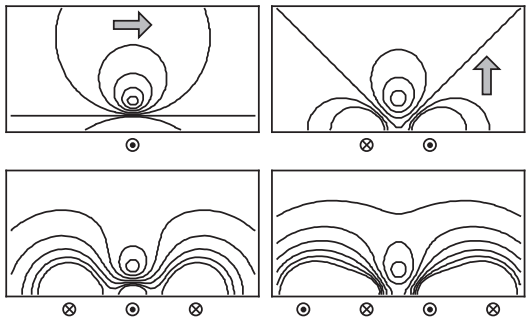}{0.8\columnwidth}
 \caption{
 The {\em upper left} picture shows the potential for a side guide
 generated by one wire and an external bias field perpendicular to
 the wire direction. The external bias field can be replaced by two
 extra wires ({\em lower left}). The
 {\em upper right} picture shows the field configuration for a two-wire
 guide with an external bias field perpendicular to the plane
 containing the wires. This external bias field may also be
 replaced by surface mounted wires ({\em lower right}).}
 \label{fig:Guides}
\end{figure}

When mounting the wire onto a surface, the bias field has to have
a component parallel to the surface in order to achieve a side
guide above the surface. The bias field can be formed by two
additional wires on each side of the guiding wire. The direction
of the current flow in these wires has to be opposite to the
current in the guiding wire (Fig.~\ref{fig:Guides}). This is
especially interesting because the wires can be mounted on the
same chip, and a self sufficient guide can be obtained.

\subsubsection{Two wire guides}
\label{subs:2wires} \label{II-A-3}

\paragraph{Counter-propagating currents}

A different way to create a guide is by using two parallel wires
with counter-propagating currents with a bias field which has a
component $B_{b}$ orthogonal to the plane containing the two wires
(Fig.~\ref{fig:Guides})\citep{Thy99-361}.

The important advantage of this configuration is that the two
wires and therefore the atom guide can be bent in an arbitrary way
in the plane perpendicular to the bias field, whereas in the
single wire guide the direction is restricted to angles close to
the line perpendicular to the bias field. If there is an
additional bias field $B_{ip}$ applied along the wires, a
Ioffe-Pritchard guide is obtained. Again, two added wires can
replace the external bias fields (see section~\ref{s:chip-ex} for
an example of an experimental implementation).

\begin{figure}
  \infig{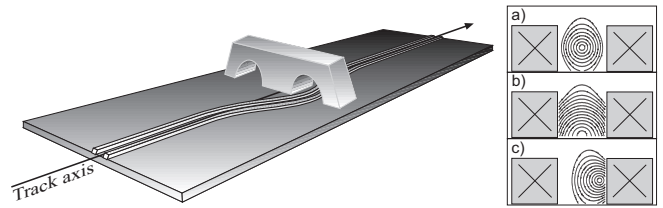}{\columnwidth}
  \caption{Atoms are guided in a two wire guide that is self
  sufficient without external bias fields. insets ({\em a}), ({\em b}), and
  ({\em c}) show the magnetic field contour lines for no bias,
  horizontal bias, and vertical bias fields, respectively.
  Courtesy E.~Cornell.}
  \lb{f:TrapSplit}
\end{figure}

The field generated by the wires compensates the bias field
$B_{b}$ at a distance:
\begin{equation}
 r_{0}=\frac{d}{2}\sqrt{\left(\frac{2\mu_{0}}{\pi}\right)\frac{I_{w}}{d B_{b}}-1},
\end{equation}
where $d$ is the distance between the two wires. When
$B_{b}>2\mu_{0}I_{w}/\pi d$ the field from the wires is not
capable of compensating the bias field. Two side guides are then
obtained, one along each wire in the plane of the wires.

In the case $B_{b}<2\mu_{0}I_{w}/\pi d$, the gradient in the
confining directions is given by
\begin{equation}
 \left.\frac{dB}{dr}\right|_{r_{0}}=
 \left(\frac{4 \pi}{\mu_0}\right)\frac{B_b^2}{I_w}\frac{r_0}{d}.
\end{equation}
If there is a field component $B_{ip}$ along the wire, the
position of the guide is unchanged. However, the shape of the
potential near its minimum is harmonic, the curvature in the
radial direction is given by:
\begin{equation}
 \left.\frac{d^2 B}{dr^2}\right|_{r_{0}}=\left(\frac{4\pi}{\mu_{0}}\right)^2
 \frac{B_{b}^4}{B_{ip}I_{w}^2}\frac{r_0^2}{d^2}.
\end{equation}
In the special case of $r_0=d/2$, the gradient and, for the case
of a non vanishing $B_{ip}$, the curvature of the potential at the
minimum position, are exactly equal to the corresponding
magnitudes for the single wire guide.

\paragraph{Co-propagating currents}

The magnetic fields formed by two parallel wires carrying
co-propagating currents vanishes along the central line between
the wires and increases and changes direction like a 2-dimensional
quadrupole. The wires form a guide as shown in
Fig.~\ref{f:TrapSplit} allowing atoms to be guided around curves
\citep{Mue99-5194}. It is even possible to hold atoms in a storage
ring formed by two closed wire \citep{Sau01-270401-1}
(section~\ref{III-B-7}). When aiming at miniaturized, surface
mounted structures, the fact that the potential minimum is located
between the wires rather than above them, has to be considered.

\begin{figure}
  \infig{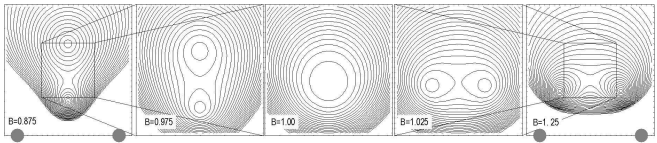}{\columnwidth}
  \caption{Potential for a two wire guide formed by copropagating
  currents. The plots show from left to right the equipotential
  lines for increasing bias fields. As the field is raised, two
  (quadrupole) minima approach each other in the vertical
  direction and merge at the characteristic bias field denoted as
  $B=1$ into a harmonic (hexapole) minimum. At higher bias fields
  this minimum splits into a double (quadrupole) well again, this
  time the splitting occurs in the horizontal direction.}
  \lb{f:TrapBiasSplit}
\end{figure}

When a bias field parallel to the plane of the wires is added, the
potential minimum moves away from the wire plane and a second
quadrupole minimum is formed at a distance far above the wire
plane where the two wires appear as a single wire carrying twice
the current (see side guide in section~\ref{II-A-2}). Depending on
the distance $d$ between the wires with respect to the
characteristic distance
\begin{equation}
d_{\rm split}=\left(\frac{\mu_0}{2 \pi}\right)
\frac{I_w}{B_b}\label{eq:dsplit}
\end{equation}
one observes three different cases (Fig.~\ref{f:TrapBiasSplit}):
({\em i}) If $d/2<d_{\rm split}$, two minima are created one above
the other on the axis between the wires. In the limit of $d$ going
to zero, the barrier potential between the two minima goes to
infinity and the minimum closer to the wire plane falls onto it;
({\em ii}) if $d/2=d_{\rm split}$, the two minima fuse into one,
forming a harmonic guide; ({\em iii}) if $d/2>d_{\rm split}$ two
minima are created one above each wire. Splitting and
recombination can be achieved by simply increasing and lowering
the bias field \citep{Den98,Zok00-93,Hin01-1462}.

Finally we mention a proposal by \cite{Ric98-481} where a tube
consisting of two identical, interwound solenoids carrying equal
but opposite currents can be used as a low field seeker guide. The
magnetic field is almost zero throughout the center of the tube,
but it increases exponentially as one approaches the walls formed
by the current carrying wires. Hence, cold low-field-seeking atoms
passing through the tube should be reflected by the high magnetic
fields near the walls, which form a magnetic mirror.

Examples of typical guiding parameters for the alkali atoms
lithium and rubidium trapped in single and two wire guides are
given in table~\ref{table:QW}. Trap frequencies in the order of
1~MHz or above can be achieved with moderate currents and bias
fields. The guided atoms are then located a few $\mu$m above the
surface.

\subsubsection{Simple traps}
\label{II-A-4}

An easy way to build traps is to start from the guides discussed
above, and close the trapping potential by `endcaps'. This can be
accomplished by taking advantage of the fact that the magnetic
field is a vector field, and the interaction potential is scalar
(eq.~\ref{eq:magpot}). By changing the angle between the wire and
the bias field, one can change the minimum of the potential and
close the trap. Simple geometries are either a straight guide and
an inhomogeneous bias field, or a homogeneous bias field in
combination with a bent wire.

\begin{figure}[tbp]
    \infig {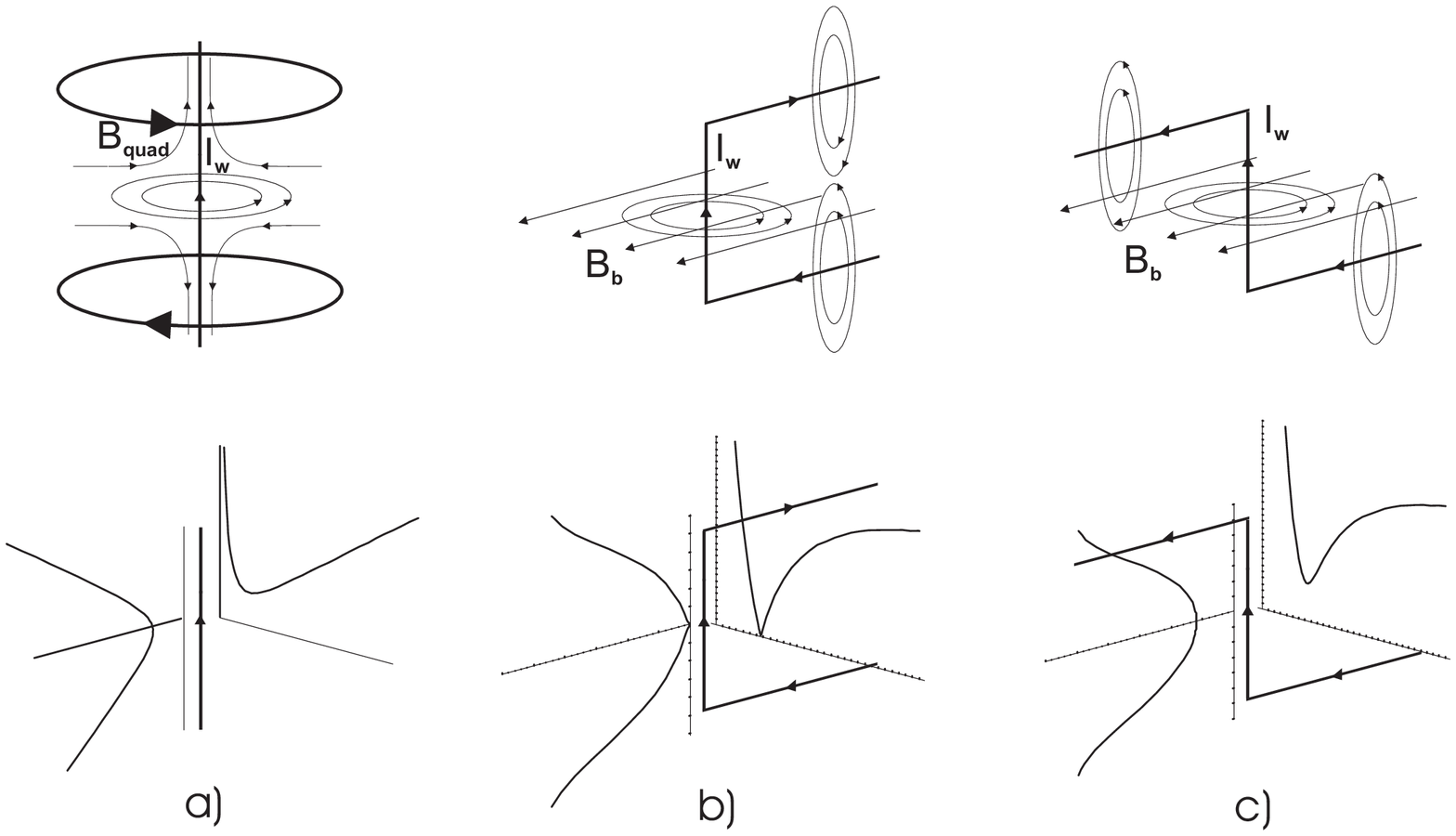} {\columnwidth}
    \caption{Creating wire traps:  The upper row shows the
    geometry of various trapping wires, the currents and the bias fields.
    The lower column shows the corresponding radial and axial trapping potential.
    ({\em a}) A straight wire on the axis of a quadrupole bias
    field creates a ring shaped 3-dimensional non-zero trap
    minimum.
    ({\em b}) A ``U''-shaped wire creates a field configuration similar to
    a 3-dimensional quadrupole field with a zero in the trapping center.
    ({\em c}) For a ``Z''-shaped wire a Ioffe-Pritchard type of trap is
    obtained.}
    \label{fig:zscheme}
\end{figure}

\begin{table}[tbp]
        \caption{Typical potential parameters for wire guides,
        based on tested atom chip components:
        ({\em Top}) {\em Side guide} created by a thin current
        carrying wire mounted on a surface with an added bias field
        parallel to the surface but orthogonal to the wire.
        ({\em Bottom}) {\em Two wire guide} created by two
        thin current carrying wires mounted on a surface with an
        added bias field orthogonal to the plane of the wires. In
        this example the two wires are $10 \mu$m apart. The
        parameters are given for the two different atoms lithium
        and rubidium, both assumed to be in the (internal) ground state
        with the strongest confinement ($F=2,m_F=2$). For both types of
        guide small bias field components $B_{ip}$ pointing along the guide were
        added in order to get a harmonic bottom of the potential and to enhance
        the trap life time that is limited by Majorana spin flip transitions
        (see eq.~\ref{eq:Majorana-rate} in section~\ref{s:loss}). It was
        confirmed in a separate calculation that the trap ground state
        is always small enough to fully lie in the harmonic region
        of the Ioffe-Pritchard potential. See also Fig.~\ref{fig:Guides}.
        \label{table:QW} }
        \begin{center}
        \begin{tabular}{ccccccccc}
        \mco{9}{c}{side guide}\\
        \hline
        wire    &\mco{2}{c}{bias fields} &\mco{3}{c}{potential}        &\mco{3}{c}{ground state } \\
        current & $B_{b}$ & $B_{ip}$     & depth & dist.  & grad.   & energy & size & life time \\
         mA   & [G]     & [G]          & [mK]  & [$\mu$m] & [kG/cm] & [kHz]  & [nm] & [ms] \\
        \hline
          \mco{9}{c}{Li}\\
        \hline
         1000& 80  & 2  & 5.4 & 25  & 32   & 100  & 120  & $>1000$ \\
         500 & 200 & 10 & 13  & 5   & 400  & 570  & 50   & $>1000$ \\
         200 & 400 & 30 & 27  & 1   & 4000 & 3300 & 21   & 7 \\
        \hline
          \mco{9}{c}{Rb}\\
        \hline
         1000& 80  & 1  & 5.4 & 25  & 32   & 41   &  53  & $>1000$\\
         500 & 200 & 4  & 13  & 5   & 400  & 250  &  21  & $>1000$\\
         200 & 400 & 20 & 27  & 1   & 4000 & 1100 & 10   & $>1000$\\
         1000& 2000& 50 & 130 & 1   & 20000& 3600 & 6    & $>1000$\\
         \hline
        \hline
        \mco{9}{c}{two wire guide (counter-propagating currents)}\\
        \hline
           \mco{9}{c}{Li}\\
        \hline
          1000& 80  & 2 & 5.4 & 25 & 32  & 100 & 120 & $>1000$\\
          500 & 200 & 10& 13  & 5  & 400 & 570 & 50  & $>1000$\\
          100 & 130 & 10& 8.7 & 1.5& 870 & 1200& 34  & 5\\
        \hline
          \mco{9}{c}{Rb}\\
        \hline
          1000& 80  & 1 & 5.4 & 25 & 32  & 41  & 53  & $>1000$\\
          500 & 200 & 4 & 13  & 5  & 400 & 250 & 21  & $>1000$\\
          100 & 130 & 5 & 8.7 & 1.5& 870 & 490 & 15  & 185 \\
        \end{tabular}
        \end{center}
\end{table}

\begin{figure}[t]
    \infig {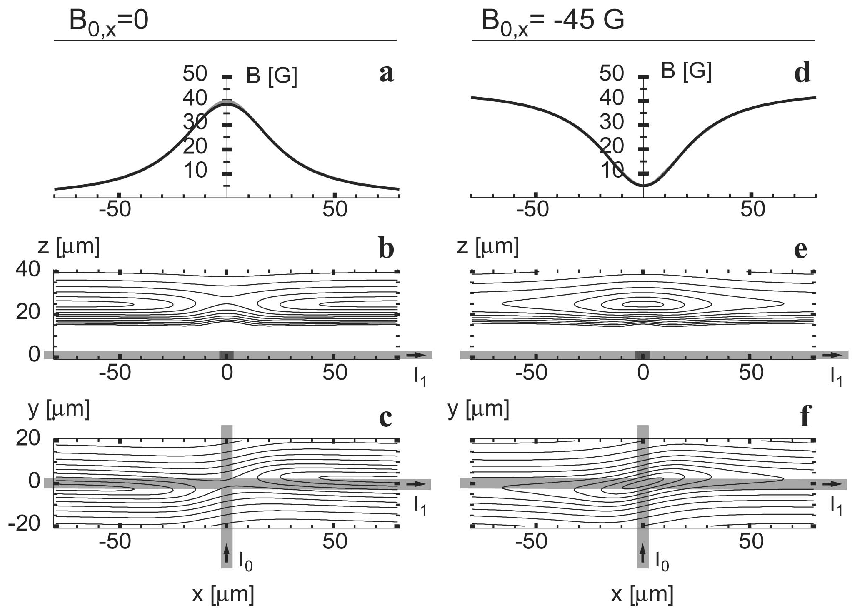} {0.8\columnwidth}
    \infig {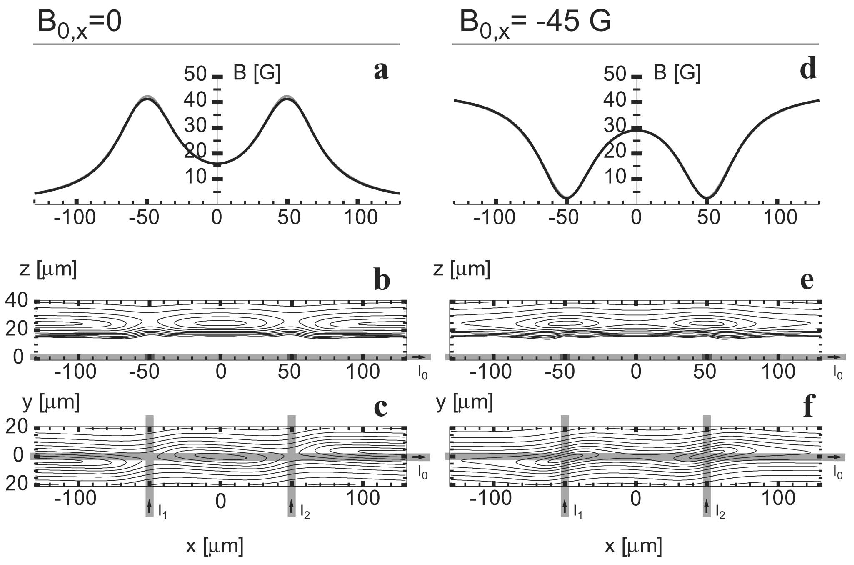} {0.8\columnwidth}
    \caption{Two geometries of crossed wire traps: Different
    cuts through the potential are displayed without and with a longitudinal bias field
    component in the left and right column, respectively. The 1-dimensional
    plots show the potential along the direction of the side
    guide, in the contour plots the wire configuration is illustrated by
    light gray bars. Courtesy J.~Reichel.}
    \label{fig:crosswire}
\end{figure}

\paragraph{Straight guide and an inhomogeneous bias field}

Examples for traps formed by superposing an inhomogeneous bias
field and the field of a straight wire use quadrupole fields
because the complete change of direction in addition to the
inhomogenity is needed to close the trap. An interesting fact is
that a current carrying wire on the symmetry axis of a quadrupole
field can be used to `plug' the zero of the field. In this
configuration a ring shaped trap is formed
(Fig.~\ref{fig:zscheme}a) that has been demonstrated
experimentally \citep{Den98,Den99-291}. In the Munich (now
T\"ubingen) group of C.~Zimmermann a modified version of this type
of trap with the wire displaced from the quadrupole axis
\citep{For98-5310,For00-701} was used to create a Bose-Einstein
condensate on an atom chip \citep{Ott01-230401}.

\paragraph{Bent wire traps:  the U- and Z-trap}

3-dimensional magnetic traps can be created by bending the current
carrying wire of the side guide
\citep{Cas99-XXX,Rei99-3398,Haa01-043405}. The magnetic field from
the bent leads creates endcaps for the wire guide, confining the
atoms along the central part of the wire. The size of the trap
along this axis is then given by the distance between the endcaps.
Here we describe two different geometries:

({\em 1}) Bending the wire into a ``U"-shape
(Fig.~\ref{fig:zscheme}b) creates a magnetic field that in
combination with a homogeneous bias field forms a 3-dimensional
quadrupole trap\footnote{The minimum of the U-trap is displaced
from the central point of the bar, in a direction opposite to the
bent wire leads. A more symmetric quadrupole can be created by
using 3 wires in an H configuration. There the side guide is
closed by the two parallel wires crossing the central wire
orthogonally. The trap is then in between the two wires, along the
side guide wire.}. The geometry of the bent leads results in a
field configuration where a rotation of the bias field displaces
the trap minimum but the field always vanishes completely at this
position.

({\em 2}) A magnetic field zero can be avoided by bending the wire
ends to form a ``Z'' (Fig.~\ref{fig:zscheme}c). Here, one can find
directions of the external bias field where there are no zeros in
the trapping potential, for example when the bias field is
parallel to the leads. This configuration creates a
Ioffe-Pritchard type trap.

The potentials for the U- and the Z-trap scale similarly as for
the side guide, but the finite length of the central bar and the
directions of the leads have to be accounted for. Simple scaling
laws only hold as long as the distance of the trap from the
central wire is small compared to the length of the central bar
\citep{Cas99-XXX,Rei99-3398,Haa01-043405}. Finally, one should
note that bending both Z leads once more, resulting in 3 parallel
wires, again supplies the needed bias field, creating a self
sufficient Z-trap.

\paragraph{Crossed wires}

Another way to achieve confinement in the direction parallel to
the wire in a side guide is to run a current $I_1<I_w$ through a
second wire that crosses the original wire at a right angle. $I_1$
creates a magnetic field ${\bf B_1}$ with a longitudinal component
which is maximal at the position of the side guide that is closest
to the additional wire. Adding a longitudinal component to the
bias field, i.e. rotating $B_b$, results in an attractive
potential confining the atoms in all three dimensions. As a side
effect position and shape of the potential minimum are altered by
the vertical component of ${\bf B_1}$. Fig.~\ref{fig:crosswire}
illustrates this type of trap configuration. Experiments of the
Munich group have proven this concept to be feasible (see
section~\ref{IV-C-1} and Fig.~\ref{fig:crossing}) and it was
suggested to use the two wire cross as a basic module for
implementing complex trapping and guiding geometries
\citep{Rei01-81}.

\subsubsection{Weinstein-Libbrecht traps}
\label{II-A-5}

\begin{figure}[tbp]
    \infig {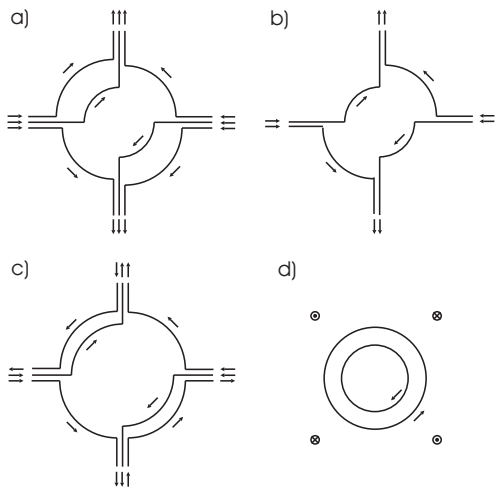} {0.75 \columnwidth}
    \caption{Four planar (and pseudoplanar) Ioffe
    trap configurations, as described in the text. Coutesy J.~Weinstein/K.~Libbrecht}
    \label{fig:libbrecht}
\end{figure}

Even more elaborate designs for traps than the ones described
previously can be envisioned. As an example we refer the reader to
ideas of \cite{Wei95-4004}. They describe planar current
geometries for constructing microscopic magnetic traps (multipole
traps, Ioffe traps and dynamical traps). We focus here on the
Ioffe trap proposals. Fig.~\ref{fig:libbrecht} shows four possible
geometries: (a) three concentric half loops; (b) two half loops
with an external bias field; (c) one half loop, one full loop and
a bias field; (d) two full loops with a bias field and external
Ioffe bars. The first of these, which we refer to as Ioffe (a), is
essentially a planar analog of the nonplanar Ioffe trap with two
loops and four Ioffe bars. The Ioffe (b) configuration replaces
one of the loops with a bias field. The Ioffe (c) configuration is
similar to Ioffe (b) but provides a steeper trapping potential on
axis and weaker trapping in the perpendicular directions; this
makes an overall deeper trap with greater energy-level splitting
for given current and size. The Ioffe (d) is a hybrid
configuration, which uses external (macroscopic) Ioffe bars to
produce the 2-dimensional quadrupole field, while deriving the
on-axis trapping fields from two loops and a bias field. This is a
reasonable configuration since macroscopic coils can generate
quadrupole fields that are nearly as large as those from
microscopic coils. Typical energy splittings in the range of 1MHz
are achievable using experimentally realistic parameters
\citep{Drn98-2906}.

\subsubsection{Arrays of traps}
\label{II-A-6}

The various tools for guiding and trapping discussed above can be
combined to form arrays of magnetic microtraps on atom chips.
Especially suitable for this purpose is the technique of the
crossed wires which requires, however, a multilayer fabrication of
the wires on the surface. Arrays of traps and their applications,
especially in quantum information processing, are discussed in the
outlook section~\ref{s:outlook}.

\begin{figure}[t]
   \infig{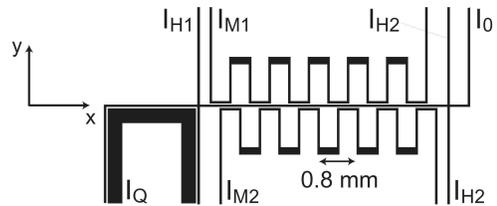}{0.8\columnwidth }
   \caption{Magnetic `conveyor belt': The wires are configured in
   a way that allows to transport atoms from one trap to another along a
   side guide. Together with a homogeneous time independent bias field,
   the currents $I_Q$, $I_{H1}$, and $I_{H2}$ are used
   for the confining fields of the source and collecting traps,
   $I_0$ is the current through the side guide wire. The currents
   $I_{M1}$ and $I_{M2}$ alternate sinosoidally with a phase
   difference of $\pi/2$ and provide the moving potential.
   Courtesy J.~Reichel.}
   \label{fig:motor}
\end{figure}

\subsubsection{Moving potentials}
\label{subs:movpot} \label{II-A-7}

Introducing time dependent potentials facilitates arbitrary
movement of atoms from one location to another. There are
different proposals for possible implementations of such `motors'
or `conveyor belts', one of which has already been demonstrated
experimentally by \cite{Hae01-608}: Using solely magnetic fields
it is based on an approximation of the crossed wire configuration.
Atoms trapped in a side guide potential are confined in the
longitudinal direction by two auxiliary meandering wires
(Fig.~\ref{fig:motor}). By running an alternating current through
both auxiliary wires with a relative phase difference of $\pi/2$,
the potential minimum moves along the guide from one side to the
other in a controllable way. In section~\ref{s:chip-ex} we present
experimental results of the above scheme.

\subsubsection{Beam splitters}
\label{II-A-8}

\begin{figure}[t]
     \includegraphics[angle=-90, width=\columnwidth]{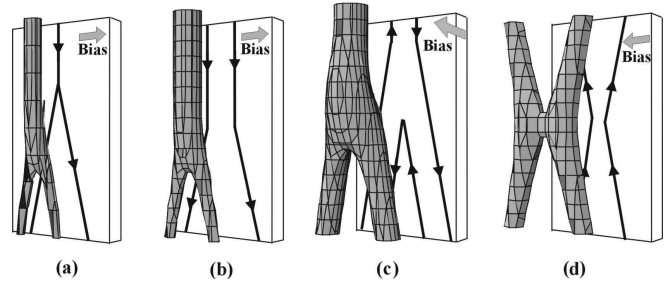}
     \caption{Different wire geometries for a beam splitting
     potential: The plots show the wires arrangement on the surface of an atom chip,
     the
     direction of current flow, and where the additional bias
     field is pointing to. Each picture also shows a typical
     equipotential surface to illustrate the shape of the resulting
     potential.
     ({\em a}) A simple Y-beam splitter consisting of a single wire that is split
     into two: The output side guides are tighter and
     closer to the surface than the input guide. Note that a second minimum
     closer to the chip surface occurs in the region between the wire
     splitting and the actual split point of the potential;
     ({\em b}) a two wire guide split into two single wire guides does not
     exhibit this `loss channel'. ({\em c}) Here, the output guides have the same characteristics as
     the input guide minimizing the backscattered amplitude.
     The vertical orientation of the bias field ensures exact symmetry of the two output
     guides. ({\em d}) In an X-shaped wire pattern the splitting occurs because of tunneling
     between two side guides in the region of close approach of the
     two wires.}
     \label{fig:bs3}
\end{figure}

By combining two of the above described guides, it is possible to
design potentials where at some point two different paths are
available for the atom. This can be realized using different
configurations (examples are shown in Fig.~\ref{fig:bs3}) some of
which have already been demonstrated experimentally (see
section~\ref{s:chip-ex}).

\paragraph{Y-beam splitters}
A side guide potential can be split by forking an incoming wire
into two outgoing wires in a Y-shape (Fig.~\ref{fig:bs3}a).
Similar potentials have been used in photon and electron
interferometers\footnote{The Y-configuration has been studied in
quantum electronics by \cite{Pal92-237,Wes99-2564}.}
\citep{Buk98-871}. A Y-shaped beam splitter has one input guide
for the atoms, that is the central wire of the Y, and two output
guides corresponding to the right and left wires. Depending on how
the current $I_w$ in the input wire is sent through the Y, atoms
can be directed to the output arms of the Y with any desired
ratio. This simple configuration has been investigated by
\cite{Cas00-5483} (see sections \ref{III-B-3} and \ref{IV-C-3} for
experimental realizations). Its disadvantages are: 1) In a single
wire Y-beam splitter the two outgoing guides are tighter and
closer to the surface than the incoming guide. The changed trap
frequency and the angle between incoming and outgoing wires lead
to a change of field strength at the guide minimum and can cause
backscattering from the splitting point. 2) In the Ioffe-Pritchard
configuration (i.e. with an added longitudinal bias field), the
splitting is not fully symmetric due to different angles of the
outgoing guides relative to the bias field. 3) A fourth guide
leads from the splitting point to the wire plane, i.e. the surface
of the chip.

The backscattering and the inaccessible fourth guide of the Y-beam
splitter may be, at least partially, overcome using different beam
splitter designs like the ones shown in Fig.~\ref{fig:bs3}b,c. The
configuration shown in Fig.~\ref{fig:bs3}b has two wires which run
parallel up to a given point and then separate. If the bias field
is chosen so that the height of the incoming guide is equal to the
half distance $d/2$ of the wires ($d/2=d_{\rm split}$ as defined
in eq.~\ref{eq:dsplit} in section~\ref{II-A-3}), the height of the
potential minimum above the chip surface is maintained throughout
the device (in the limit of a small opening angle) and no fourth
port appears in the splitting region. The remaining problem of the
possible reflections from the potential in the splitting region
can be overcome by the design presented in Fig.~\ref{fig:bs3}c.
Here, a wave guide is realized with two parallel wires with
currents in opposite directions and a bias field perpendicular to
the plane of the wires. This type of design creates a truly
symmetric beam splitter where input and output guides have fully
identical characteristics.

\paragraph{X-beam splitters}
A different possible beam splitter geometry relies on the
tunneling effect: Two separate wires are arranged to form an X,
where both wires are bent at the position of the crossing in such
a way that they do not touch (see Fig.~\ref{fig:bs3}d). An added
horizontal bias field forms two side guides that are separated by
a barrier that can be adjusted to be low enough to raise the
tunneling probability considerably at the point of closest
approach. If the half distance between the wires becomes as small
as $d_{\rm split}$, the barrier vanishes completely, resulting in
a configuration that is equivalent to the combination of two
Y-beam splitters \citep{Mue00-1382}. The choice of the parameters
in the wire geometry, the wire current, and the bias field govern
the tunneling probability and thereby the splitting ratio in this
type of beam splitter. The relative phase shift between the two
split partial waves in a tunneling beam splitter allows to combine
two beam splitters to form a Mach-Zehnder interferometer. Another
advantage of the X-beam splitter is that the potential shape in
the inputs and outputs stays virtually the same all over the
splitting region as opposed to the Y-beam splitter. A detailed
analysis of the tunneling X-beam splitter can be found in
\cite{And99-3841}.

For an ideal symmetric Y-beam splitter, coherent splitting for all
transverse modes should be achieved due to the definite parity of
the system \citep{Cas00-5483}. This was confirmed with numerical
2-dimensional wave packet propagation for the lowest 35 modes. The
50/50 splitting independent of the transverse mode is an important
advantage over four way beam splitter designs relying on tunneling
such as the X-beam splitter described above. For the X-beam
splitter, the splitting ratios for incoming wave packets are very
different for different transverse modes, since the tunneling
probability depends strongly on the energy of the particle. Even
for a single mode, the splitting amplitudes, determined by the
barrier width and height, are extremely sensitive to experimental
noise.

\subsubsection{Interferometers}
\label{II-A-9}

Following the above ideas of position dependent multiple
potentials and time dependent potentials which are able to split
minima in two and recombine them, several proposals for chip based
atom interferometers have been put forward
\citep{Hin01-1462,Hae01-063607,And01-100401}.

\paragraph{Interferometers in the spatial domain}

\begin{figure}
    \infig{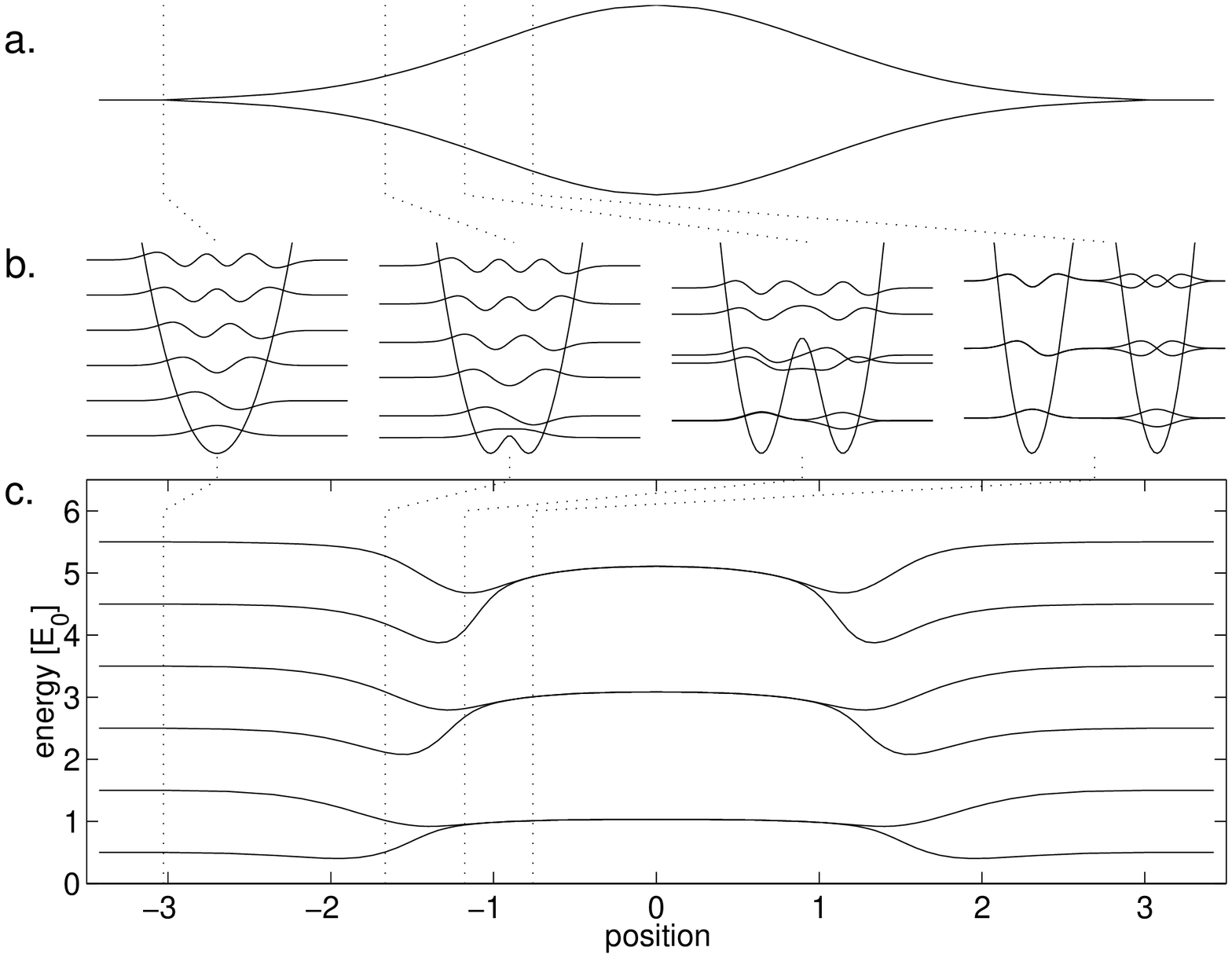}{\columnwidth}
    \infig{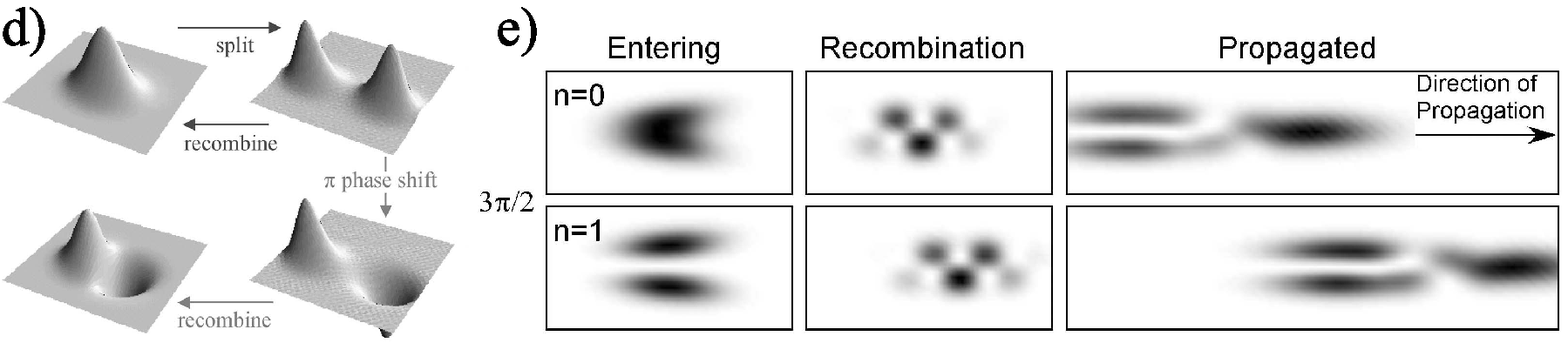}{\columnwidth}
    \caption{
    Basic properties of the guided matter wave interferometer:
    ({\em a})  Two Y-beam splitters are joined together to form the
    interferometer.
    ({\em b}) Transverse eigenfunctions of the guiding potentials in
    various places along the first beam splitter.
    When the two outgoing guides are separated far
    enough, i.e. no tunnelling between left and right occurs,
    the symmetric and the antisymmetric states become degenerate.
    ({\em c}) Energy eigenvalues for the lowest transverse
    modes as they evolve along the interferometer.  One clearly sees that
    pairs of transverse eigenstates form disjunct interferometers.
    ({\em d}) The wavefunction of a cold atom cloud starts out in the vibrational ground state of a
    guide or trap. The wavefunction splits when the guide divides, leaving
    a part of the wavefunction in each arm of the interferometer.
    If the phases of the two parts evolve identically on each
    side, then the original ground state is recovered when the two
    parts of the wavefunction are recombined. But if a phase
    difference of $\pi$ accumulates between the two parts (for
    example due to a different gravitational field acting on
    each),
    then recombination generates the first excited state of the
    guide with a node in the center. Courtesy E.~Hinds.
    ({\em e}) Basic properties of a wave packet propagating
    through a guided matter wave interferometer for $|0 \rangle$
    and $|1 \rangle$ incoming transverse modes calculated by solving the time-dependent Schr\"odinger equation
    in two dimensions (x,z,t) for realistic guiding potentials, where z is the longitudinal propagation axis.
    The probability density of the wave function just before entering,
    right after exiting the interferometer, and
    after a rephasing time $t$ are shown for a phase shift of
    $3\pi/2$.
    One clearly sees the separation of the two outgoing packets
    due to the energy conservation in the guide, e.g. for $n=0$ the first
    excited outgoing state is slower than the ground state.}
    \label{fig:schema}
\end{figure}

To build an interferometer for guided atoms \citep{And01-100401}
two Y-beam splitters can be joined back to back
(Fig.~\ref{fig:schema}a). The first acts as splitter and the
second as recombiner. The eigenenergies of the lowest transverse
modes along such an interferometer in 2-dimensional
geometry\footnote{In 2-dimensional confinement the out of plane
transverse dimension is either subject to a much stronger
confinement or can be separated out. For experimental realization
see \citep{Gau98-5298,Spr00-053604,Hin01-1462,pfa}} are depicted
in Fig.~\ref{fig:schema}c. From the transverse mode structure one
can see that there are many disjunct interferometers in Fock
space. Each of them has two transverse input modes ($|2n \rangle$
and $|2n+1 \rangle$, $n$ being the energy quantum number of the
harmonic oscillator) and two output modes. In between the two
Y-beam splitters, the waves propagate in a superposition of $|n
\rangle_{l}$ and $|n \rangle_{r}$ in the left and right arm,
respectively. With adiabaticity fulfilled, the disjunct
interferometers are identical.

Considering any one of these interferometers, an incoming
transverse state evolves after the interferometer into a
superposition of the same and the neighboring transverse outgoing
state (Fig.~\ref{fig:schema}c), depending on the phase difference
acquired between $|n \rangle_{l}$ and $|n \rangle_{r}$ during the
spatial separation of the wave function\footnote{The relative
phase shift $\Delta\phi$ between the two spatial arms of the
interferometer can be introduced by a path length difference or by
adjusting the potentials to be slightly different in the two arms.
In general, $\Delta\phi$ is a function of the longitudinal
momentum $k$.}. While the propagation remains unchanged if the
emerging transverse state is the same as the incoming state, a
transverse excitation or deexcitation translates into an altered
longitudinal propagation velocity ($\Delta v\simeq\pm\omega/k$
where $\hbar k$ is the momentum of a wave packet moving through
the interferometer and $\omega/2\pi$ is the transverse trapping
frequency), since transverse oscillation energy is transferred to
longitudinal kinetic energy, and vice versa.

As presented in Fig,~\ref{fig:schema}e, integrating over the
transverse coordinate results in a longitudinal interference
pattern observable as an atomic density modulation. As all
interferometers are identical, an incoherent sum over the
interference patterns of all interferometers does not smear out
the visibility of the fringes.

\paragraph{Interferometers in the time domain}

Two different proposals are based on time dependent potentials
\citep{Hin01-1462,Hae01-063607}. These proposals differ from the
interferometer in the spatial domain in several ways: 1) The
adiabaticity of the process may be controlled to a better extent
due to easier variation of the splitting and recombination time.
2) The interferometers are based on a population of only the
ground state. 3) The interference signal amounts to different
transverse state populations in the recombined single minimum
trap, whereas the above proposal anticipates a spatial
interference pattern which may be easier to detect.

The first proposal \citep{Hin01-1462} is based on a two parallel
wire configuration with co-propagating currents (see
section~\ref{subs:2wires}). Changing the bias field in this
configuration as a function of time produces the cases (i), (ii),
and (iii) discussed in section~\ref{subs:2wires} depending on the
strength of the bias field as compared to the critical bias field
$B_c=\frac{\mu_0}{\pi}\left(\frac{I_w}{d}\right)$. Starting with
$B_b<B_c$ and an atom cloud in the ground state of the upper
minimum, a coherent splitting of the corresponding wave function
is achieved when $B_b$ is raised to be larger than $B_c$. As shown
in Fig.~\ref{fig:schema}d, the symmetry of the wave function now
depends on the relative phase shift introduced between its two
spatially separated parts. Thus, when the bias field is lowered
again to $B_b=B_c$, a superposition of the symmetric and the
antisymmetric state forms in the recombined guide.

If the spatial resolution of the detection system is not
sufficient to distinguish between the two output states, the
following scheme is proposed:
The node plane of the excited state is rotated by $90^{\circ}$ by
turning an additional axial bias field while the guides are
combined. If after such an operation the bias field is lowered,
the ground state goes to the upper guide whereas the excited state
is found in the lower guide.

The second proposal \citep{Hae01-063607} utilizes the crossed wire
concept introduced in section~\ref{II-A-4}. Here, in contrast to
the interferometer described above, the splitting of the atomic
wavefunction occurs in one dimension whereas the confinement in
the other two dimensions is the constant strong confinement of a
side guide. Longitudinally, the atom is trapped by two currents
running through wires crossing the side guide wire. The resulting
Ioffe-Pritchard potential well is split into a double well and
then recombined by a third crossing wire carrying a time dependent
current flowing in the opposite direction.

Starting with a wavefunction in the ground state of the combined
potential, a relative phase shift introduced between the two parts
of the potential after splitting leads to a wavefunction in a
(phase shift dependent) superposition of the ground and first
excited states upon recombination. A state selective detection
then displays a phase shift dependent interference pattern. A
detailed analysis of realistic experimental parameters has shown
that in this scheme non-adiabatic excitations to higher levels can
be sufficiently suppressed. The position and size of the
wavefunction are unchanged during the whole process. Therefore,
the interferometer is particularly well suited to test local
potential variations.

\subsubsection{Permanent magnets}
\label{II-A-10}

Although beyond the scope of this paper, we mention configurations
with permanent magnets
\citep{Sid96-713,Mes97-191,Sab99-468,Hin99-R119,Dav99-408}. Though
less versatile in the sense of not enabling the ramping up and
down of fields, permanent magnets might reward us with advantages
such as less noise, strong fields, and large scale periodical
structures. As described in section~\ref{s:loss}, technical noise
in the currents which induce the magnetic fields, may have severe
consequences in the form of heating and decoherence. In the
framework of extremely low decoherence, such as that demanded by
quantum computation proposals, permanent magnets might be a better
choice.

\begin{figure}[tb]
    \infig{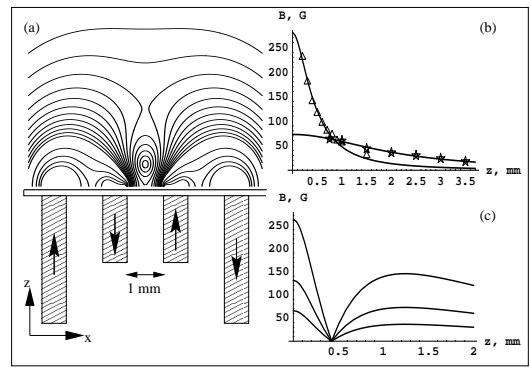}{0.8 \columnwidth}
    \caption{({\em a}) Two pairs of differently sized magnetic sheets ({\em bottom}) are magnetized
    using current carrying wires wound around them. By choice of
    the direction of current flow in these wires, the direction of
    magnetization is defined, the arrows show a possible
    configuration for which the equipontential lines are plotted
    ({\em top}). ({\em b})~The field produced by the sheet pairs measured in the
    symmetry plane. (c)~Scaling of the field due to the combined inner and
    outer pair of sheets in the plane of symmetry. Courtesy M.~Prentiss.} \label{fig:Prentiss}
\end{figure}

An interesting tool is a magnetic atom mirror formed by
alternating magnetic dipoles \citep{Opa92-396}, creating an
exponentially growing field strength as the mirror is approached.
This situation can be achieved by running alternating currents in
an array of many parallel wires or by writing alternating magnetic
domains into a magnetic medium such as a hard disk or a video
tape. This has been demonstrated by \cite{Sab99-468} and may
achieve a periodicity in the order of 100nm. Current carrying
structures have the disadvantage of large heat dissipation,
especially when the structure size is in the submicron region.

Another possibility is based on a combination of current carrying
wires and magnetic materials and was experimentally demonstrated
at Harvard in the group of M.~Prentiss: Two pairs of ferromagnetic
foils that were magnetized by current carrying wires wound around
them were used for magnetic and magnetooptic trapping
\citep{Ven01}. The setup and the potential achieved is illustrated
in Fig.~\ref{fig:Prentiss}. The advantages of such a hybrid scheme
over a purely current carrying structure are larger capturing
volumes of the traps, less heat dissipation, and enhanced trap
depths and gradients because the magnetic field of the wires is
greatly amplified by the magnetic material. The magnets can still
be switched by means of time dependent currents through the wires.

\subsection{Electric interaction}
\label{secElecInteract} \label{II-B}

The interaction between a neutral atom and an electric field is
determined by the electric polarizability $\alpha$ of the atom. In
general, $\alpha$ is a tensor.  For the simple atoms we consider,
i.e. atoms with only one unpaired electron in an s-state, the
electric polarizability is a scalar and the interaction can be
written as
\begin{equation}
        V_{\rm pol}(r) = -\frac{1}{2} \alpha E^2 (r).
        \label{vpol}
\end{equation}

\subsubsection{Interaction between a neutral atom and a charged wire}
\lb{s:ChFiber} \label{II-B-1}

We now consider the interaction of a neutral polarizable atom with
a charged wire (line charge $q$) inside a cylindrical ground plate
\citep{Hau92-6468,Sch95-169,Den97-405,Den98-737}.

\begin{figure}
        \infig {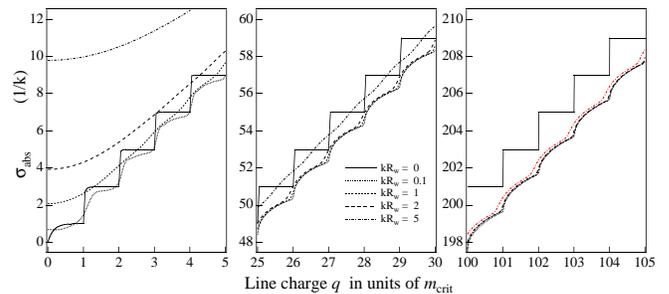} {\columnwidth}
        \caption{Theoretical absorption cross section for
        a charged wire.  The calculations are made for
        several different relative thicknesses ($k \rw$) of the
        wire, the charge is given in units of the angular momentum
        $m_{\rm crit}=\Lc/\hbar$.}
        \label{f:abs_cross}
\end{figure}

The interaction potential (in cylindrical coordinates) given by
\begin{equation}
        V_{\rm pol}(r) =
         - \left( \frac{1}{4 \pi \epsilon_0} \right)^2 \frac{2 \alpha \, q^2}{r^2}
        \label{Vpol2}
\end{equation}
is {\em attractive}. It has exactly the same radial form ($\rsq$)
as the centrifugal potential barrier ($V_L = L^{2}_{z}/2M r^2$)
created by an angular momentum $L_{z}$. $V_L$ is {\em repulsive}.
The total Hamiltonian for the radial motion is
\begin{eqnarray}
        H &=& \frac{p_{r}^2}{2 \, M}
         + \frac{L_{z}^2}{2 \, M \, r^2} -  \left( \frac{1}{4 \pi \epsilon _0} \right)^2
        \frac{2 \alpha \, q^2}{r^2} \\
           &=& \frac{p_{r}^2}{2M} + \frac{ L_{z}^{2}-\Lc^{2}}{2 M r^2},
        \label{e:clHam}
\end{eqnarray}
where $\Lc=\sqrt{M \, \alpha} \, |q| / 2 \pi \epsilon _0 $ is the
critical angular momentum characteristic for the strength of the
electric interaction.  There are no stable orbits for the atom
around the wire. Depending on whether $L_{z}$ is greater or
smaller than $\Lc$, the atom either falls into the center and hits
the wire ($|L_{z}| < \Lc$) or escapes from the wire towards
infinity ($|L_{z}| > \Lc$). In the quantum regime, only partial
waves with $\hbar l< \Lc$ ($l$ is the quantum number of the
angular momentum $L_{z}$) fall towards the singularity and thus
the absorption cross section of the wire should be quantized
(Fig.~\ref{f:abs_cross}).

To build stable traps and guides one has to compensate the
strongly attractive singular potential of the charged wire.  This
can be done either by adding a repulsive potential, for example
from an atom mirror or an evanescent wave (see
section~\ref{II-C-1}), or by oscillating electric fields (see
section~\ref{II-B-2}).

\subsubsection{Stabilizing the motion with an oscillating electric
charge: the Kapitza wire} \lb{s:KapizzaFiber} \label{II-B-2}

The motion in the attractive electric potential can be stabilized
by oscillating the charges.  The mechanism is similar to the RF
Paul trap \citep{Pau90-531} where an oscillatory part of the
electric fields creates a 3-dimensional confinement for ions. An
elementary theoretical discussion of the motion in a sinusoidally
varying potential shows that Newton's equations of motion can then
be integrated approximately, yielding a solution that consists of
a fast oscillatory component superimposed on a slow motion that is
governed by an effective potential \citep{Lan76}.

An example of a 2-dimensional atom trap based on a charged wire
with oscillating charge was proposed by \cite{Hau92-6468}. By
sinusoidally varying the charge on a wire, it is possible to add
an effective repulsive $1/r^6$ potential which stabilizes the
motion of the atoms around the wire. Sizeable electrical currents
appear when the charge of a real wire (with capacitance) is
rapidly varied. Magnetic fields are produced which interact with
the magnetic moment of an atom. This leads to additional
potentials which have not been taken into account in the original
calculations.

Another AC-electrical trap with several charged wires was proposed
by \cite{Shi92-L1721}. Their setup is reminiscent of a quadrupole
mass filter and consists of 4 to 6 charged electrodes that are
grouped around the trapping center. The atom experiences an
oscillatory micromotion synchronous with the oscillation of the
quadrupole potential, which leads to an overall trapping force.

\subsubsection{Guiding atoms with a charged optical fiber}
\lb{s:ChFiberSub} \label{II-B-3}

\begin{figure}
        \infig {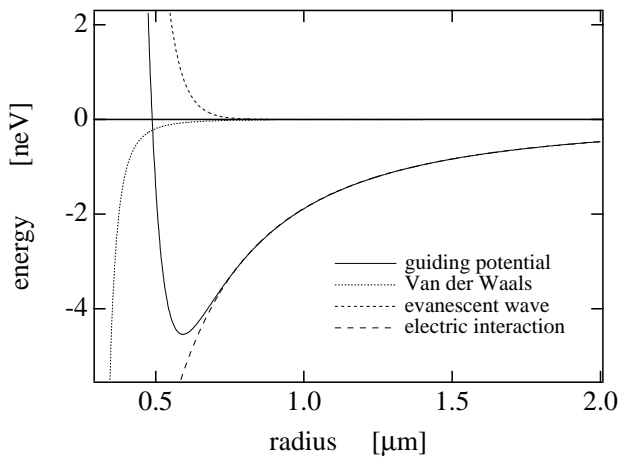} {\columnwidth}
        \caption{Typical radial potential for a neutral lithium
        atom trapped around a
        charged (5V) optical quartz fiber (diameter 0.5$\mu$m) with
        1mW light and a detuning of  $\Delta/\Gamma = 3
        \times 10^5$.  The attractive potential ($\rsq$) is created by the
        interaction of the induced dipole moment with the electric field of the
        charged fiber.  The repulsion is due to the evanescent wave
        from blue detuned light propagating in the fiber.
        Close to the wire surface  the Van der
        Waals interaction becomes important. }
        \protect\label{f:q_fiber}
\end{figure}

Stable orbits for the motion of an atom around a line charge are
obtained if the atom is prevented from hitting the wire by a
strong repulsive potential near the surface of the wire.  Such a
strong repulsion can be obtained by the exponential light shift
potential of an evanescent wave that is blue detuned from an
atomic resonance. This can be realized by replacing the wire with
a charged {\em optical fiber} with the cladding removed and the
blue detuned light propagating in the fiber \citep{Bat94-13}.  The
fiber itself has to be conducting or coated with a thin ($\ll
\lambda$) conducting layer to allow uniform charging. For the
simple case of a TE$_{01}$ mode propagating in the fiber, the
light shift potential is independent of the polar angle and the
combined guiding potential is given by: \beq \lb{e:guidPot}
        V_{\rm guid}(r)= A K_0^2(B r) -  \left( \frac{1}{4 \pi \epsilon _0} \right)^2  \frac{2 \alpha q^2}{r^2},
\eeq where $A$ and $B$ are constants that depend on specifics of
the optical fiber as well as on light power, wavelength and atomic
properties \citep{Bat94-13}.  $K_{0}$ is the modified Bessel
function of the second kind. Fig.~\ref{f:q_fiber} shows a typical
example of such a potential. Cold atoms are bound in radial
direction by the effective potential but free along the
z-direction, the direction of the charged optical fiber.

\subsection{Traps and guides formed by combining the interactions}
\label{II-C}

\subsubsection{Charged wire on a mirror}
\label{II-C-1}

\begin{figure}
        \infig {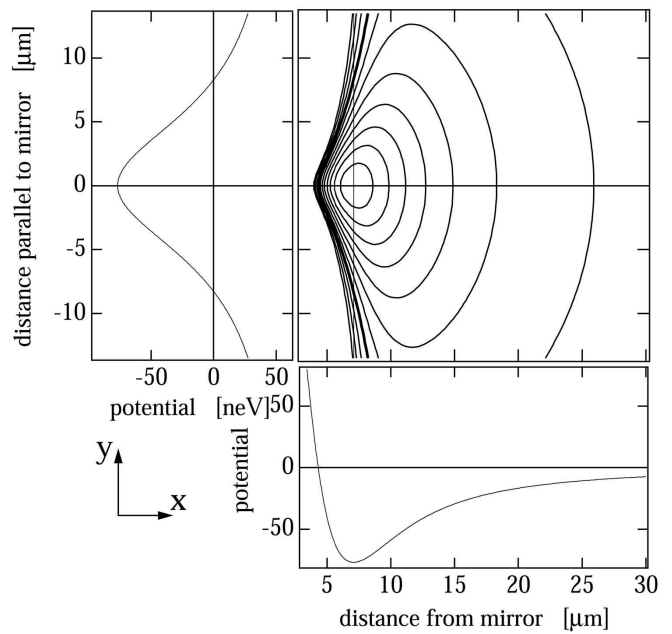} {\columnwidth}
        \caption{Typical potential for a neutral atom guide.  The attractive
        potential ($\rsq$) is created by the interaction of the induced dipole
        moment in the electric field of the charged wire mounted directly on the
        surface of an atomic mirror.  The action of the atomic mirror (evanescent wave or
        magnetic mirror) prevents the atom from reaching the surface and creates a
        potential tube close to the surface illustrated by the contour graph.
        The two adjacent plots give the potential in a direction orthogonal
        to the charged wire and orthogonal to the
        mirror surface. Distances are given from the
        location of the charged wire and the surface of the atom mirror.
        }
        \label{f:qw_potential}
\end{figure}

As we have seen previously, a static charged wire alone cannot
form the basis for stable trapping. Cylindrical solutions such as
the charged light fiber have the disadvantage that they cannot be
mounted on a surface. An alternative solution would be to mount a
charged wire onto the surface of an atom mirror. The combination
of the attractive $\rsq$ potential with the repulsive potential of
the atom mirror\footnote{There are two main types of atom mirrors:
The first type utilizes evanescent waves (e.g. above the surface
of a reflecting prism) of blue detuned light which repels the
atoms. Here the potential takes the form $V_m(z)=V_0
\exp(-\kappa_m z)$ where $\kappa_m$ is of the order of the light
wave number and $z$ the distance from the mirror
\citep{Coo82-258}. The second type is based on a surface with
alternating magnetic fields. Here, $\lambda$ is the periodicity of
the alternating magnetic field. The approaching atom experiences
an exponentially increasing field, and consequently the weak field
seekers are repelled
\citep{Opa92-396,Roa95-629,Sid96-713,Hin99-R119}} $V_m(z)$ gives:
\begin{equation}\lb{e:qw_pot}
        V_{\rm guid}({\bf r})=V_m(z)-\left( \frac{1}{4 \pi \epsilon_0} \right)^2 \frac{2 \alpha \, q^2}{r^2}
\end{equation}
where $z$ is the height above the mirror and $r$ the distance from
the wire. This creates a potential tube for the atoms as shown in
Fig.~\ref{f:qw_potential} which can be viewed as a wave guide for
neutral atoms.

Typical parameters for guides formed by a magnetic mirror and a
charged wire are given by \cite{Sch98-57}. They can be very
similar to the magnetic guides discussed in section~\ref{II-A}.
Using typical mirror parameters \citep{Roa95-629,Sid96-713}, one
can easily achieve deep and narrow guides with transverse level
spacings in the kHz range for both light (Li) and heavy (Rb)
atoms.

In a similar fashion microscopic traps can be created by mounting
a charged tip (point) at or close beneath the atom mirror surface.
A point charge on the surface of an atom mirror creates an
attractive $1/r^4$ interaction potential:
\begin{equation}
        V_{\rm pol}(r) =-\left(\frac{1}{8 \pi \epsilon_0}\right)
        \frac{ 2 \alpha q^2 }{ r^4 }.
        \label{e:vpol_dot}
\end{equation}
where q is the tip charge. Together with the atomic mirror it
forms a microscopic cell for the atoms. It can be viewed as the
atom optical analog to a quantum dot \citep{Sch98-57,Sek01-197}.

This approach of combining a charged structure with an atom mirror
is compatible with well developed nanofabrication techniques. This
opens up a wide variety of possibilities ranging from curved and
split guides to interferometers or even complex networks.

\subsubsection{Combined electric-magnetic state dependent traps}
\label{II-C-2}

\begin{figure}[t]
    \infig {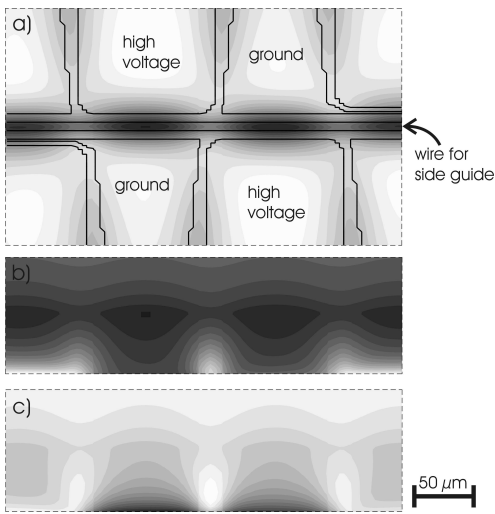} {0.75 \columnwidth}
    \caption{State dependent potential: ({\em a}) top view of an actual
    chip design; the wire in the center is used as a side guide
    wire, the additional electrodes create a spatially oscillating
    electric field providing confinement also along the wire. The
    contour plot shows a typical potential configuration for $^7$Li atoms
    in the $|F=2,m_{f}=2\rangle$ magnetic substate using experimentally
    accessable parameters. Dark areas correspond to attractive
    potentials, the trap minima are located 50$\mu$m above the surface.
    The side views show that only one state ($|F=2,m_{f}=2\rangle$) is trapped
    in the combined potential ({\em b}), while the other ($|F=1,m_{f}=-1\rangle$) is
    not because the weaker magnetic barrier to the surface is
    compensated by the attractive electric potential ({\em c}). The parameters used in a simulation
    with the electromagnetic field solver MAFIA were
    $I_w=$500mA, $B_b=$20G for the side guide and a voltage of
    600V on the electrodes.}
    \label{fig:StateDepTrap}
\end{figure}

The magnetic guides and traps (section~\ref{II-A}) can be modified
by combining them with the electric interaction, thereby creating
tailored potentials depending on internal ({\em e.g.} spin)
states. For example, supplementary electrodes located between
independent magnetic traps can be used to lower the magnetic
barrier between them by the attractive electric potential the
electrodes create. Since the magnetic barrier height depends on
the magnetic substate of the atom, whereas the electric potential
does not, this allows state selective operation. This is
especially interesting since it can lead to implementing quantum
information processing with neutral atoms in microscopic trapping
potentials where the logical states are identified with atomic
internal levels (see section~\ref{s:outlook}).

A simple example, showing such a controllable state dependence, is
a magnetic wire guide approached by a set of electrodes
(Fig.~\ref{fig:StateDepTrap}a). Applying a high voltage to the
electrodes introduces an electrostatic potential which provides
confinement along the direction parallel to the magnetic side
guide, and also shifts the trapping minimum towards the surface,
possibly breaking the magnetic potential barrier in the direction
perpendicular to the surface itself. The charge can be adjusted in
a way that depending on the strength of the magnetic barrier
created by the wire current, the atoms either impact onto the
surface or are trapped above it. Since the strength of the
magnetic barrier depends on the magnetic substate of the atom or,
more precisely, depends linearly on the quantum number $m_F$, this
can be exploited to form a state selective magnetic trap
(Fig.~\ref{fig:StateDepTrap}b,c).

\subsubsection{The electric motor}
\label{II-C-3}

In general, electric fields are always present in magnetic wire
traps since an electric potential difference is needed to drive a
current through a wire with finite resistance. For large wires,
the voltages in question are low and if the distances of the atoms
from the wire are large enough, the attractive electric
interaction can be neglected. However, for micron sized wires, one
finds that if the current carrying wire becomes long, at some
point the voltage is strong enough to create a significant driving
force for the atoms or even to destroy the traps.

\begin{figure}
    \infig {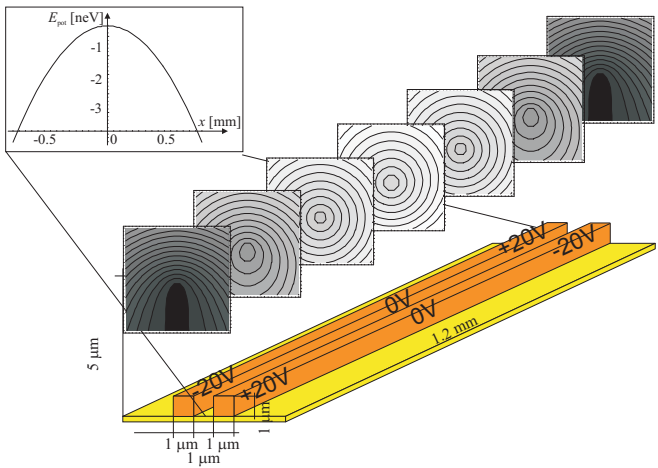} {\columnwidth}
    \caption{Two wire guide configuration with currents of 1A
    running in opposite directions with a vertical bias field of 150G.
    The combined magnetic and electric potential is shown in contour
    plots perpendicular to the wire and along the wire at the minimum
    height (inset). The parabolic potential shape offers the
    possibility to drive the atoms ($^{87}$Rb) along the wire.
    In the example, the voltages applied to the wires are chosen
    to be 0V with respect to ground in the wire center.}
    \label{fig:emotor}
\end{figure}

On the other hand, one can actually exploit this effect and turn
it into an `electric motor' by using the electric potential
gradient inside the magnetic minimum to accelerate and decelerate
the atoms at will. Fig.~\ref{fig:emotor} illustrates the mechanism
used for the motor for the example of a two wire guide with a
vertical bias field. The wires carry counter-propagating currents,
and the polarization interaction is zero in the middle of the
guide (see inset) where both wires have the same voltage. By
adding a homogeneous electric potential relative to ground, the
zero electric field point may be moved at will to achieve any
acceleration or deceleration rate. A constant acceleration is
obtained when the zero electric field point is maintained at a
constant distance from the position of the atoms.

\subsection{Miniaturization and technological considerations}
\label{sec:tech} \label{II-D}

To achieve very robust and highly controlled atom manipulation one
would like to localize atoms in steep traps or guides which can be
fabricated with high precision.  The large technological advances
in precise nanofabrication, with the achievable size limit on
chips smaller than 100nm, makes the adaptation of these processes
for mounting the wires onto surfaces very attractive.

\subsubsection{Miniaturization} \label{II-D-1}

The main motivations behind miniaturization and surface
fabrication are:

\begin{itemize}

\item{Large trap level spacings help to suppress heating rates. To
achieve the necessary {\em large trapping gradients} and {\em
curvatures} with reasonable power consumption, miniaturization is
unavoidable (see section~\ref{II-D-3}).}

\item{The tailoring resolution of the potentials used for atom
manipulation is given by the resolution of the fabrication of the
structures used. It is, for example, important for the realization
of atom--atom entanglement by controlled collisions as suggested
by \cite{Cal00-022304}(see section~\ref{s:outlook}) to reduce the
distances between individual trapping sites to the micron regime.
This would be virtually impossible with (large) free standing
structures.}

\item{Nanofabrication is a mature field which allows one to place
wires on a surface with great accuracy ($<100$nm). Surface mounted
structures are very {\em robust} and the substrate serves as a
{\em heat sink} allowing larger current densities (see
section~\ref{II-D-3}). In addition, nanofabrication allows
parallelism in production of manipulating elements ({\em
scalability}).}

\item{Nanofabrication also allows us to contemplate the {\em
integration} of other techniques on the chip (see
section~\ref{s:outlook} for details).}

\end{itemize}

\subsubsection{Finite size effects} \label{II-D-2}

\begin{figure}
\infig{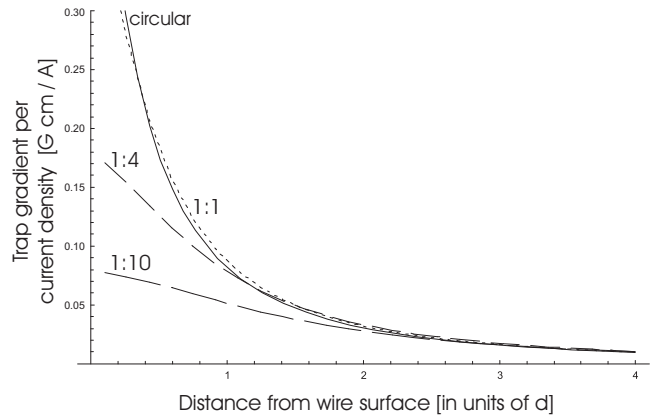}{\columnwidth}
 \caption{Deviations from the field of an
 infinitely thin wire become important as the surface of a
 physical wire is approached. The plot shows the trap gradient for
 a side guide (see section~\ref{II-A-2}) when differently shaped wires
are
 used. The solid line corresponds to a circular cross section as a
 reference since the field outside the wire equals that of an
 infinitely thin wire at the wire center. A wire with a square cross section (dotted line) shows
very small deviations, while broad and thin wires (dashed lines)
deviate more and more as the thickness/width ratio decreases.
Here, all wires are chosen to  have the same cross section $d^2$.
Therefore, the width of  the rectangular wires are $2d$ and
$\sqrt{10} d=3.2d$ for the ratios 1:4 and 1:10, respectively.}
\label{fig:finitesize}
\end{figure}

The formulae presented in section \ref{II-A} to \ref{II-C} are
exact only for infinitely small wire cross sections. In the case
of a physical wire with a finite cross section, they are a good
approximation only as long as the height above the wire is greater
than the width of the wire. For experiments requiring a trap
height smaller than the width of the wire, finite size effects
have to be taken into account. In Fig.~\ref{fig:finitesize}, we
present examples of numerical calculations showing how the trap
gradient is limited by finite size wires. One clearly sees that at
trap heights in the order of the width of the wire the resulting
gradient starts to deviate from the expected value. The effect is
small for wires with a square cross section, while it becomes
considerably more important when rectangular wires with high
ratios of width to thickness are used.

\subsubsection{Van der Waals interaction}
\label{II-C-4}

The Van der Waals interaction becomes important at distances of
the order of a few 100nm from the surface. The interaction can be
strong enough to significantly alter the magnetic and electric
trapping potentials \citep{Gri00-95}. Traps much closer than 100nm
from the surface will be very hard to achieve since the Van der
Waals potential attracts the atoms to the surface and increases
with $1/d^3$ (in the non-retarded regime where the distance $d$ is
smaller than the optical wavelength).

\subsubsection{Current densities} \label{II-D-3}

A limiting factor in creating steep traps and guides is the
maximally tolerable current density of a current carrying
structure. Considering a side guide potential created by a wire
with finite width $d$ and a constant thickness, the highest
possible gradient is achieved at a distance from the wire
comparable with $d$. The bias field needed for such a trap is
given by the ratio of the maximum current that can be pushed
through the wire and $d$; therefore the bias field is proportional
to the maximum current density $j$. This leads to the conclusion
that the highest possible gradient is given by $j/d$ which favors
smaller wires. If a square wire cross section $d^2$ is assumed,
the maximum gradient is proportional to $j$. Even in this case,
smaller $d$ will allow for larger gradients because $j$ has been
observed to increase with smaller wire cross sections. The drive
for smaller width is stopped at a distance of about 100nm where
surface decoherence effects (see section~\ref{s:loss}) and Van der
Waals forces may be too strong to endure. Hence, it is probably of
no use to make the wires even thinner.

\subsubsection{Multi-layer chips}

Last, one should also note that as more complex operations are
demanded from the atom chip (see section~\ref{s:outlook}), it will
have to move on from a 2-dimensional structure into a
3-dimensional structure in which not only current and charge
carrying wires are embedded, but also light elements and wave
guides. These highly complex devices will force upon the
fabrication a whole range of material and geometrical constraints.

\section{Experiments with free standing structures}
\label{III}

The basic principles of microscopic atom optics have been
demonstrated using free standing structures: current carrying and
charged wires. The interaction potentials are in general shallow,
typically only a few mK deep. Hence experiments use cold atoms
from a MOT or a well collimated atom beam (even the moderate
collimation of 1mm over 1m results in a typical transverse
temperature of $<1$mK).

Free standing wire structures can be installed close to a standard
six beam MOT without significantly disturbing its operation (as
long as the wire is thin enough), and offer large optical access
which has advantages when probing the dynamics of the atoms and
their spatial distribution within the wire potentials.  They have
the disadvantages that they are not very sturdy, they deform
easily due to external forces, and they cannot be cooled
efficiently to dissipate energy from ohmic heating. This limits
the achievable confinement and the potential complexity of wire
networks. Nevertheless there are some special potentials which can
only be realized with free standing wires.

\subsection{Magnetic interaction}
\label{III-B}

As discussed in section~\ref{II-A}  there are two possibilities
for magnetically trapping a particle with a magnetic moment: traps
for {\em strong} field seekers and traps for {\em weak} field
seekers. In the following we describe experiments with magnetic
microtraps which are based on small, free standing wires or other
magnetic structures. Typical wire sizes range from 10$\mu$m to a
few of mm and the wires carry electrical currents of up to 20A.
All experiments but the first example start with a conventional
MOT of alkali atoms (lithium or rubidium) which is initially
situated a few mm away from the magnetic field producing
structures. This distance prevents the atoms in the MOT from
coming into contact with the structure surface where they would be
absorbed. It also provides the necessary optical access for the
MOT laser beams.

To load the magnetic wire traps and guides, the MOT laser light is
simply switched off and the magnetic trap fields are turned on.
The loading rate into the miniature magnetic traps has been
enhanced in some experiments a) by optically pumping the
unpolarized MOT atoms to the right trapping state
\citep{Key00-1371}; b) by first loading the MOT atoms in a size
matched magnetic trap which is then further adiabatically
compressed
\citep{Vul96-349,Vul98-1634,Key00-1371,For98-5310,Haa01-043405};
c) by moving the MOT closer to the trapping region shortly before
the light is turned off which can be done with an additional
magnetic bias field \citep{Den99-2014}. In this way the efficiency
of transferring the atoms into the miniature magnetic traps
reached between 1 and 40 \%. In general the spatial distribution
of the trapped atoms was imaged with a CCD camera  by shining a
resonant laser beam onto the atoms and detecting its absorption or
the atomic fluorescence.

\subsubsection{Magnetic strong field seeking traps: the Kepler guide}
\label{III-B-1}

A magnetic strong field seeker trap for cold neutral atoms was
experimentally demonstrated in two experiments: in 1991 by guiding
an effusive beam of thermal sodium atoms (mean velocity $\sim$
600m/s) along a $1$m long current carrying wire
\citep{Sch92-284,Sch95-169,Sch95-R13} and in 1998 with cold
lithium atoms loaded from a MOT \citep{Den99-2014}.

\begin{figure}[t]
    \infig {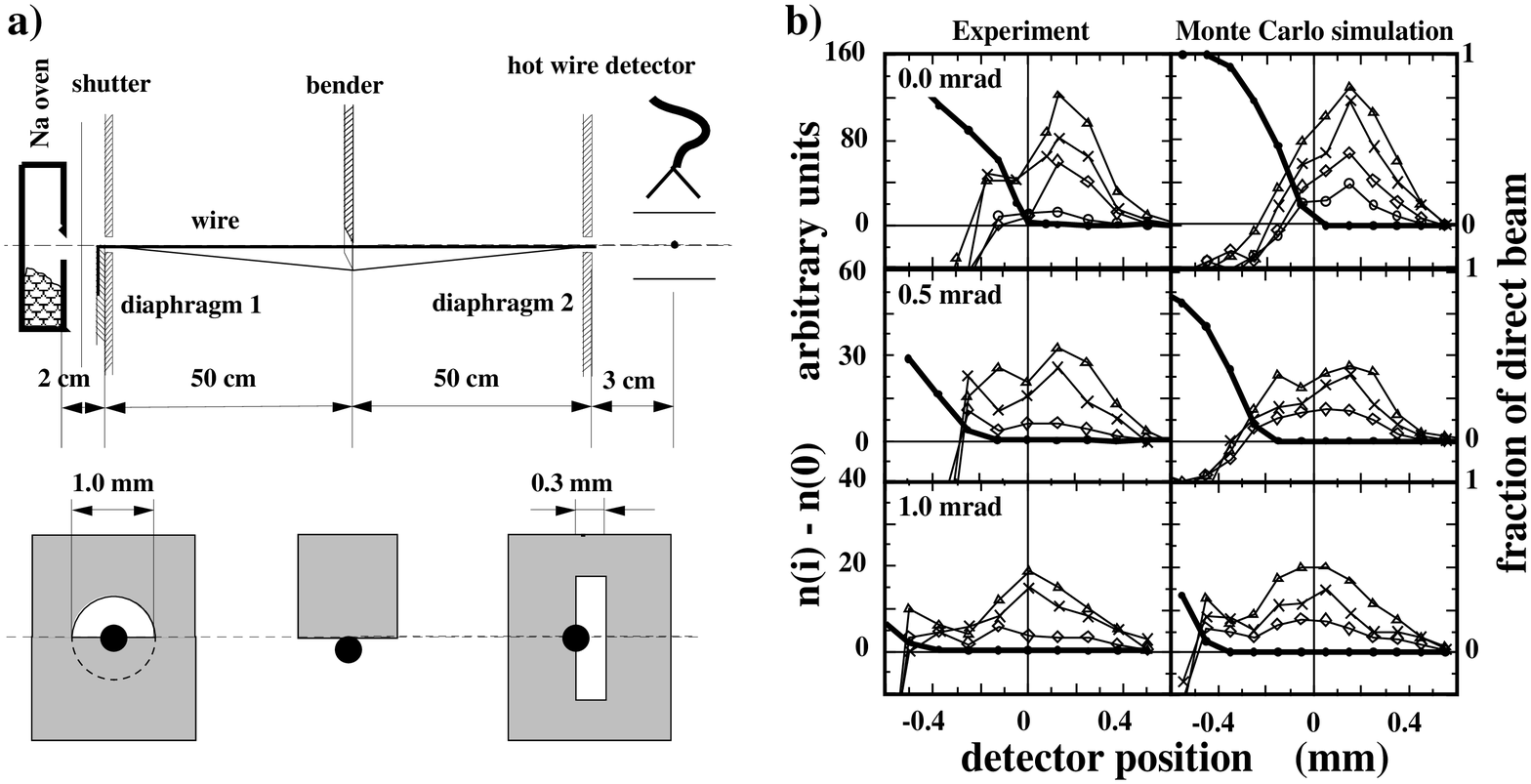} {\columnwidth}
    \caption{({\em a}) Experimental setup: The insets below show in detail the
    relative geometric arrangement between the apertures, the movable beam
    shutter used to bend the wire, and how the wire is mounted.  ({\em b})
    Guiding of Na atoms along the 1m long, 150$\mu$m in diameter, tungsten
    wire (at detector position 0 indicated by the vertical line).  Experimental
    count rates, $n(I)-n(0)$ (left), and Monte Carlo simulations (right), are
    shown for a 0.0, 0.50 and 1.00mrad bend in the wire.  The different
    symbols represent 0.5A (o), 1.0A ($\diamond$), 1.5A ($\times$) and 2.0A
    ($\triangle$) current through the wire.  The thick line shows the fraction
    of atoms of the direct beam that get to the detector when no current is on
    (use right y-axis of graph).  Its form corresponds to the shadow of the
    bender that is cast onto the detector.}
    \label{beamschema}
\end{figure}

The experimental setup for the beam experiment is given in Fig.
\ref{beamschema}. The atom beam is emitted from  a 1mm-diameter
nozzle in a 100$^{\circ}$C  oven and is collimated to 3mrad.
Introducing a small bend in the wire ($\approx1$mrad), one can
guide some of the atoms along the wire around the beam stop. The
atomic flux was measured with a hot wire detector. The guiding
wire was 150$\mu$m thick and carried 2A of electrical current.

In the second experiment, lithium atoms were cooled in a MOT
(1.6mm diameter FWHM) to about 200$\mu$K (which corresponds to a
velocity of about 0.5 m/s). By shifting the MOT onto a 50$\mu$m
thick wire and releasing the atoms from the MOT, about 10$\% $ of
the unpolarized atomic gas could magnetically be trapped in orbits
of about 1mm diameter around the wire that carried about 1A of
current. Monte Carlo calculations indicate that by optically
pumping the atoms and optimizing the trap size and current through
the wire, it should be possible to guide over 40\% of the atoms
from a thermal cloud with the Kepler guide.  The loading
efficiency is limited to this amount, because atoms in highly
eccentric orbits hit the wire and are lost.

\begin{figure}
    \infig{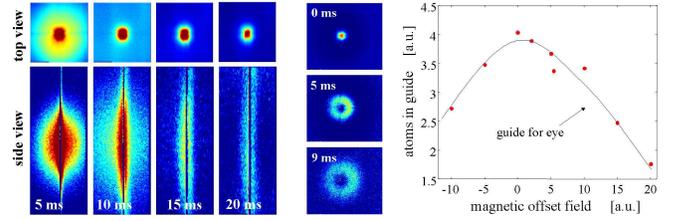}{\columnwidth}
    \caption{
    ({\em left}) Guiding of atoms along a current carrying wire in their
     {\em strong field seeking} state (Kepler guide). Pictures of the atomic clouds
     are shown, taken in axial and transverse direction with respect to the wire.
     For times shorter than 15ms the expanding cloud of untrapped atoms is also visible.
     The location of the wire is indicated by a line (dot).
     The pictures show a 2cm long section of the wire that is illuminated
     by the laser beams.
     ({\em center})
     Atomic distribution after free expansion of 0 to 9ms for atoms
     that have been guided in Kepler orbits around the wire.
     The expanded cloud is doughnut-shaped
     due to the orbital motion of the atoms around the wire.
    ({\em right})
     Experimentally measured stability of the Kepler guide
     as a function of the magnitude of bias fields. The
     signal is proportional to the trapped atom number
     in the guide after an interaction time of 20ms.
    }\label{f:WireGuideExp}
\end{figure}

The bound atoms are guided along the wire corresponding to their
initial velocity component in this direction.  Consequently a
cylindrical atomic cloud forms that expands along the wire. After
40ms of guiding, the atoms typically had propagated over a 2cm
distance along the wire (see Fig. \ref{f:WireGuideExp} (left)).
For long guiding times  the bound atoms leave the field of view,
and the fluorescence signal of the atoms decreases. The top view
images of Fig. \ref{f:WireGuideExp} (left) show a round atom cloud
that is centered on the wire suggesting that atoms circle around
it.

By studying the ballistic expansion of the bound atoms after
switching off the guiding potentials, the momentum distribution of
the guided atoms can be extracted.
Fig.~\ref{f:WireGuideExp}(center) shows a picture sequence
demonstrating how the atomic cloud expands as a function of time.
Starting from a well localized cylindrical cloud of guided atoms
at t = 0 the spatial atomic distribution transforms into a
doughnut-like shape.  This shows that there are no zero-velocity
atoms in the Kepler guide. In order to be trapped in stable orbits
{\em around} the wire the atoms need sufficient angular momentum
and therefore velocity. Atoms with too little angular momentum hit
the wire and are lost.

Guiding in the Kepler guide is very sensitive to the presence of
uncompensated bias fields.  Such additional magnetic bias fields,
even if homogeneous, destroy the rotational symmetry of the Kepler
potential and angular momentum is not conserved anymore. Over the
course of time, the Kepler orbits become increasingly eccentric
and thus finally hit the current carrying wire leading to loss,
which was confirmed by Monte Carlo calculations. The right side of
Fig.~\ref{f:WireGuideExp} shows a qualitative experiment
investigating the dependence of the magnetic trap stability on the
magnetic bias field. The remaining atom number in the Kepler guide
was measured after 20ms of interaction time. It clearly decreases
with increasing bias field strength: the larger the bias field,
the faster the atoms get lost \citep{Den98}.  In a case of a weak
disturbance the orbits can be stabilized by an additional $1/r^2$
potential which leads to a precession of the orbits.

\subsubsection{Magnetic weak field seeking traps and guides}
\label{III-B-2}

The development of miniature weak field seeker traps, as discussed
in Section \ref{II-A-2} and \ref{II-A-3}, lays the foundations for
microscopic atom optics. Here and in the following sections we
restrict our discussion explicitly to experiments with free
standing structures. Surface mounted guides and traps are
discussed in Section \ref{s:chip-ex}.

\begin{figure}
    \infig{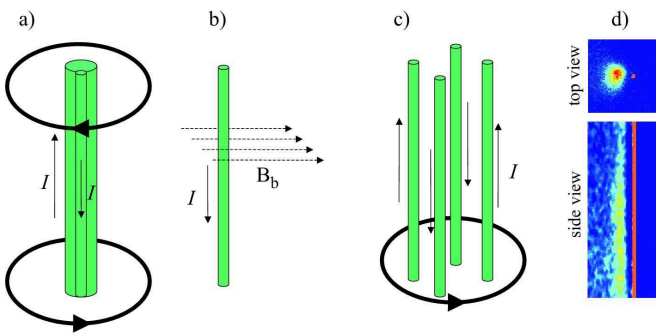}{\columnwidth}
    \caption{
   Three realizations of magnetic quadrupole traps with straight wires.
    ({\em a}) Trap realized
   by \protect\cite{For98-5310}
   with a thin wire (50$\mu$m) glued onto a thick wire (1mm).
   The current through both wires flows in opposite directions.
   ({\em b}) A homogeneous bias field is combined with a  single straight wire
   \protect\citep{Den98,Den99-2014}.
   ({\em c}) Four wires with alternating current direction produce a quadrupole field
   minimum in the center. In the experiment the four
   wires were embedded in a silica fiber \protect\citep{Key00-1371}.
   ({\em d}) Images of atoms in guide {\em b}.}
    \label{f:schema3}
\end{figure}

In the following experiments the circular symmetric magnetic field
of a straight current carrying wire is combined with a magnetic
bias field as described in Section \ref{II-A-2}. The two fields
cancel each other along a line that is parallel to the wire
creating a magnetic field minimum (side guide). In the simplest
case, the bias field can be created by an additional wire
(Fig.~\ref{f:schema3}a) \citep{For98-5310} or by an homogeneous
external field (Fig.~\ref{f:schema3}b) \citep{Den98,Den99-2014}.
Four wires also create a 2-dimensional quadrupole field
(Fig.~\ref{f:schema3}c) \citep{Key00-1371}.

The experiments of the group of C.~Zimmermann \citep{For98-5310}
used additional endcap (`pinch') coils (see Fig. \ref{f:schema3}a)
to confine the atoms also in the direction along the wire. They
succeeded in adiabatically transferring and compressing the
magnetic trap - reaching a relatively high transfer efficiency of
14\% from the MOT into a microtrap without losing in phase space
density. In experiments in Innsbruck \citep{Den98,Den99-2014} and
Sussex \citep{Key00-1371} (Figs. \ref{f:schema3}b and c,
respectively) cold atoms released from a MOT were guided along the
wires a distance of one to two centimeters (Fig.
\ref{f:schema3}d). In addition the vertical Sussex experiment used
one bottom pinch coil to confine the falling atoms from exiting
the guide. The atoms bounced back and were imaged at the top exit.

By choosing appropriate bias field strengths and wire currents, a
wide range of traps with different gradients have been realized
and the interesting scaling properties (see section~\ref{II-A-4})
were studied. With a fixed trap depth (given by the magnitude of
the bias field $B_{b}$) the trap size and its distance from the
wire can be controlled by the current in the wire. The trap gets
smaller and steeper (gradient $\propto B_{b}^{2}/I$) for
decreasing the current in the wire, which was confirmed
experimentally \citep{Den98,Den99-2014}. For example, a trap with
a gradient over 1000G/cm can be achieved with a moderate current
of $0.5$A and an offset field of $10$G. The trap is then be
located $100\mu$m away from the wire center.

A different interesting low field seeker trap has been
experimentally realized by placing a current carrying wire right
through the minimum of a magnetic quadrupole field
\citep{Den98,Den99-291}. If the wire is aligned along the
direction of the symmetry axis of the quadrupole field a ring
shaped potential is obtained with a non-vanishing minimum field
strength (see section~\ref{II-A-4} and Fig.~\ref{fig:zscheme}d).

\subsubsection{Beam splitters}
\label{III-B-3}

\begin{figure}
 \infig{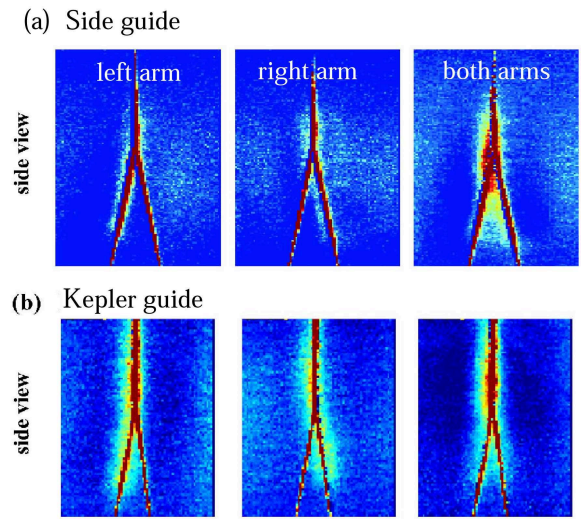}{0.9\columnwidth}
   \caption{Atomic beam switch for guided atoms
    using a ``Y"-shaped current carrying wire.  By controlling the
    current through the arms, one can send cold lithium atoms along
    either arm or split the beam in two. The images here show the
    switch operated in the Kepler guide mode and the ``weak field seeker" mode.
   }
    \label{atomfork}
\end{figure}

Although free-standing wire experiments are certainly limited in
their architectural complexity because of mechanical stability,
some variations of the straight wire geometry have been explored.
By combining two free standing wires one can form  a ``Y" or fork,
which can be used as an atomic switch (see Fig. \ref{atomfork})
\citep{Cas98Glas,Den99-291}. Choosing an arm of the fork through
which an electrical current is conducted, the atomic flow can be
switched from one arm to the other.  If current is sent through
both arms the atom beam is split in two.

\subsubsection{Free-standing bent wire traps}
\label{III-B-4}

\begin{figure}
    \infig{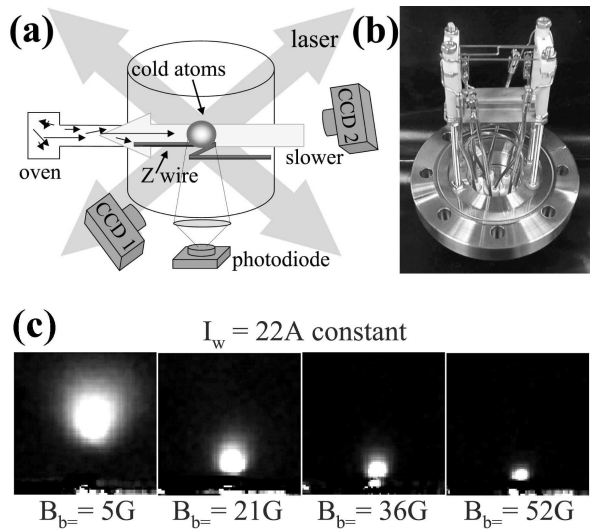}{0.9\columnwidth}
    \caption{
    ({\em a}) A schematic description of the experiment is
    shown. Camera 1 is looking along the central bar of the magnetic
    trap and camera 2 along the leads. In addition to the two laser beams
    shown in the figure, there is the third MOT beam parallel to the
    central bar.
    ({\em b}) The Z-wire held by two Macor blocks is mounted on a flange.
    ({\em c})
    The cloud of trapped atoms monitored by camera 1.
    By changing the bias field B$_b$ from 5 to 52G,
    the trap size and position change. Also the trap frequency
    increases from 30 to 1600 Hz.
    The experiment confirms the predicted scaling laws concerning
    trap distance, frequency and bias field.}
    \label{haasescale}
\end{figure}

Experiments with free standing wires that are bent in shape of a
``U'' or ``Z'' have been reported by
\cite{Den99-291,Haa00,Haa01-043405}. Bending the wire has the
effect of putting potential endcaps on the wire guide, which turns
it into a 3-dimensional low field seeker trap. As explained in
detail in section~\ref{s:basics}, in the ``U" case a quadrupole
trap is formed and for the Z-configuration the trap is of the
Ioffe-Pritchard type. Such a Z-wire trap can achieve trapping
parameters similar to the ones currently used in conventional BEC
production, here, however, with moderate currents of a few
amperes. In their experiment, Haase {\it et al.} used a 1mm thick
copper wire, with the central bar being about 6mm long. The wire
can carry 25A without any sign of heating. Fig. \ref{haasescale}c
shows the scaling properties of the Z-trap. The atomic cloud can
be compressed by raising the bias field or by lowering the wire
current.

\subsubsection{The tip trap}
\label{III-B-5}

\begin{figure}
   \infig{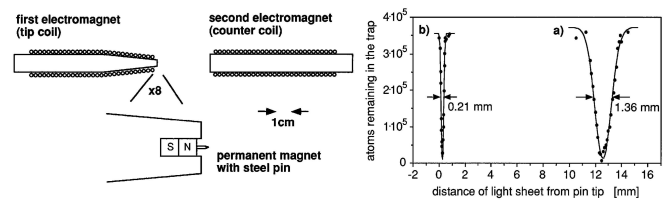}{\columnwidth}
   \caption{
   ({\em left}) Setup of the tip trap by Vuletic {\it et al}. A sharp steel pin
   is magnetized by a permanent magnet and exposed to a variable
   magnetic field that is generated by two electromagnets.
    ({\em right}) Observed shape of the atomic cloud (a) in the
    shallow field after loading from a magneto-optical trap and
    (b) after compression in the steep potential of the  tip trap
    at a current in the tip coil of 1.2A.
   Courtesy V. Vuletic. }
    \label{f:vuletictip}
\end{figure}

\cite{Vul96-349,Vul98-1634} have demonstrated a miniature magnetic
quadrupole trap (the tip trap) by mounting small coils on a
combination of permanent magnets and ferromagnetic pole pieces
(see Fig. \ref{f:vuletictip}). In this way they exploited the fact
that for a given magnetic field $B_{o}$ the maximum possible field
gradient scales like  $B_{o}/R$ where R is the geometric size of
the smallest relevant element. The central element of the tip trap
is a 0.65mm steel pin of which one tip is sharpened to a radius of
curvature of  10$\mu$m. Thus with $R =$ $10 \mu$m and $B_{o} =
1000$G the magnetic field gradient exceeded $10^{5}$G/cm. Working
with lithium atoms, this gradient implies a ground state size of
the atomic wavefunction smaller than the wavelength of the optical
transition at 671nm. The microtrap was loaded by adiabatic
transport and compression: The atoms of the lithium MOT are
transferred to a volume matched, but still relatively shallow
magnetic potential after turning off the MOT light. By
adiabatically changing the currents through the miniature coils
the magnetic trap  compresses its size by a factor of 6.5  within
100ms. A total of 3\% of the MOT atoms could be transferred to the
microtrap at moderate currents of 3A through the tip trap coils.

\subsubsection{Scattering experiments with a current carrying wire}
\label{III-B-6}

\begin{figure}
    \infig{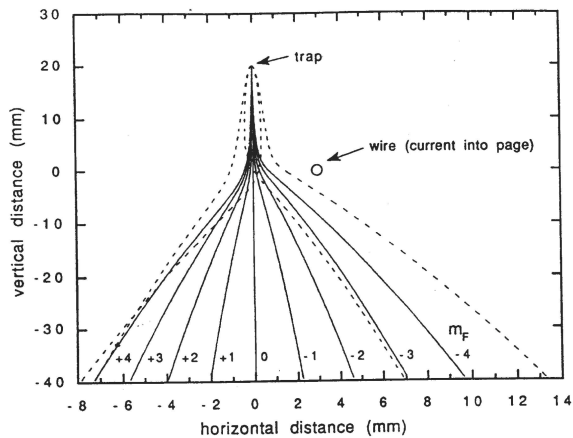}{0.9\columnwidth}
    \caption{Computer simulation of trajectories of cesium atoms
    deflected by the magnetic field from a wire carrying 20A. The
    solid lines indicate the trajectories for atoms in the nine
    possible magnetic substates, assuming zero initial velocity.
    The broken lines are for atoms in the $m_F = \pm F$ substates
    with initial transverse velocities of $\pm 1$cm/s.
    Courtesy P. Hannaford.
   }
    \label{waynetrajectories}
\end{figure}

In 1995 the Melbourne group \citep{Row95,Row96-55,Row96-577}
performed an experiment where a beam of laser cooled cesium atoms,
after being released from a MOT, is scattered off a current
carrying wire. As the atoms pass through the static inhomogeneous
magnetic field of the wire they are deflected by a force $\nabla
(\mu B)$ dependent on the magnetic substate of the atom (see Fig.
\ref{waynetrajectories}). With currents of up to 45A through the
wire, the positions of the atoms in the individual magnetic
substates are resolved and deflection angles as large as
25$^\circ$ are observed. State preparation of the atoms using
optical pumping increases the number of atoms deflected through
essentially the same angle.

\subsubsection{A storage ring for neutral atoms}
\label{III-B-7}

\begin{figure}
    \infig{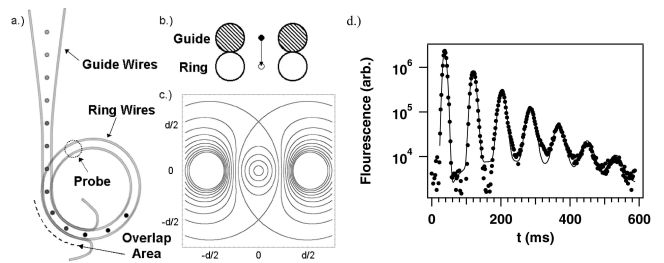}{\columnwidth}
    \caption{ ({\em a}) A schematic of the storage ring.
    ({\em b}) A cross section of the overlap region. The trap minimum is
    shifted from between the guide wires to the ring wires by
    adjusting the current. ({\em c}) A contour plot of a two wire
    potential. The contours are drawn every 0.5mK for the wire distance
    $d = 0.84$mm and $I = 8$A. ({\em d}) Successive revolutions in the
    storage ring.  The points represent experimental data, the curve
    is a theoretical model.
    Courtesy M. Chapman.
   }
    \label{storagering}
\end{figure}

Very recently \cite{Sau01-270401-1} have demonstrated a storage
ring for neutral atoms using a two wire guide
(section~\ref{II-A-3}). A pair of wires (separation $\approx
840\mu$m) which forms a ring of 2cm diameter, produces a
2-dimensional quadrupole magnetic field (see Fig.
\ref{storagering}). The wires carry currents of 8A in the same
direction which produces a field minimum between the two wires
with a field gradient of 1800G/cm and a trap depth of 2.5 mK for
the $F=1, m_F = -1$ ground state of $^{87}$Rb (weak field seeker).
The ring has a diameter of 2cm and is loaded from a MOT via a
similar second two wire waveguide. The MOT is turned off and the
second waveguide is ramped up in 5ms. Approximately $10^6$ laser
cooled rubidium atoms (longitudinal temperature 3$\mu$K) fall 4cm
under gravity along the guide after which they enter the storage
ring with a velocity of about 1m/s. To transfer the atoms to the
ring, the current in the guide is ramped off while simultaneously
increasing the current in the ring. Using fluorescence imaging the
position and the number of the atom cloud can be probed. Up to
seven revolutions of the atoms in the ring have been observed.

\subsection{Charged wire experiments}
\label{III-A}

Two types of experiments have used the $1/r^2$ potential
(eq.~\ref{Vpol2}) of a charged wire. One investigated the effect
of a charged wire in atom interferometry. The other investigated
atomic motion in the singularity of the $1/r^2$ potential.
Here, laser cooled atoms fall into the attractive singularity and
are lost as they hit the charged wire.

\subsubsection{A charged wire and interferometry}
\label{III-A-1}

\begin{figure}
  \infig{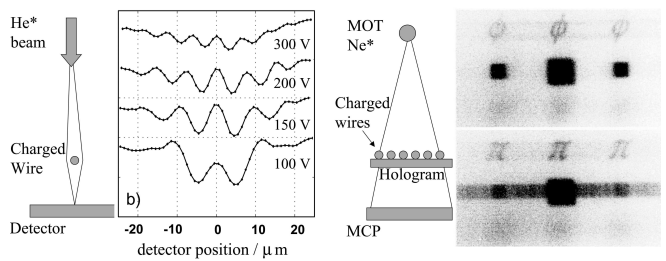}{\columnwidth}
  \caption{ The experimental set-ups and data for interferometry
  and holography experiments with charged wires.
  ({\em left}) Charged wire interferometer for
     metastable helium. Different voltages
     applied to the electrodes: The data sets are plotted with a
     vertical offset. The dotted horizontal lines indicate the zero
     level for the respective measurements. Courtesy J. Mlynek.
  ({\em right}) Selective atom holography: switching between
    atomic images ``$\phi$" and ``$\pi$". For the upper figure the
    wire array is uncharged, whereas for the lower figure it is
    electrically charged. The squares in the lower part of each
    figure are nondiffracted atom patterns. Courtesy F. Shimizu. }
  \label{chargeandinterferometers}
\end{figure}

\cite{Shi92-L436} used a straight charged wire to shift (deflect)
the interference patterns of a matter wave interferometer in a
Young's double slit configuration. In a recent experiment of the
same group \citep{Fuj00-4027} (Fig.~\ref{chargeandinterferometers}
({\em right})), this work is expanded by combining a binary matter
wave hologram with an array of straight charged wires. By changing
the electric potential applied to the electrodes on the hologram
the holographic image patterns can be shifted or erased, and it is
even possible to switch between two arbitrary holographic image
patterns\footnote{Similarly it was suggested by \cite{Eks92-369}
that charged patches on a grating can be used to modify the
diffraction properties.}. These experiments were performed using
laser cooled metastable neon in the 1s$_3$ state. After releasing
them from the MOT, the atoms fell under gravity onto a double slit
or a binary hologram. A few centimeters further down the atoms
formed an interference pattern which was detected by a
multi-channel plate (MCP).

The binary hologram pattern held an array of 513 regularly spaced
parallel wires of platinum on its surface. Each electrode was
either grounded or connected to a terminal. The width and the
spacing of each wire was 0.5$\mu$m and the holes for the binary
hologram in between the wires were 0.5$\mu$m $\times$ 0.5$\mu$m in
size. The electric field $E$ generated between two wires shifted
the energy of the neon atom by $-\alpha E^2/2$.  When two adjacent
electrodes had the same potential, the atoms in the gap were
unaffected. If they had different potentials, the atoms
accumulated an additional phase while passing through the hole.

In an experiment in Konstanz, \cite{Now98-5792} sent a collimated
thermal beam of metastable helium atoms onto a charged wire
(tungsten, 4 $\mu$m diameter) where it was diffracted (see Fig.
\ref{chargeandinterferometers} ({\em left})). 1.3m further
downstream they observed an interferometric fringe pattern which
depended on the wire charge and on the de Broglie wavelength. The
data agreed well with the theoretical predictions for scattering
polarizable particles off a $1/r^{2}$ potential.

\subsubsection{A charged wire in gas of cold atoms: studying a singular potential}
\label{III-A-2}

\begin{figure}
    \infig{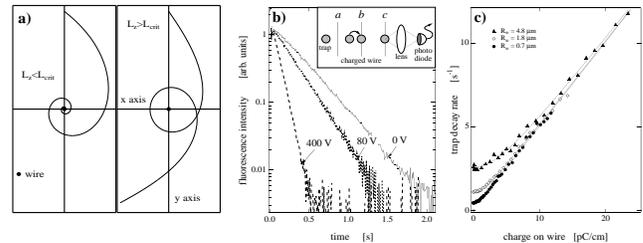}{\columnwidth}
    \caption{
     ({\em a})
    Two classical trajectories: An atom falls into the
    $1/r^2$ singularity of a electrically charged wire if the atomic
    angular momentum $L_z < L_{\rm crit}$.
    If $L_z > L_{\rm crit} $ it
    scatters and escapes from the singularity.
    ({\em b}) When moved onto the wire the atom trap decays
    exponentially, as can be seen by monitoring the atomic
    fluorescence signal. Charging the wire (100V $\leftrightarrow$ 6.4pC/cm)
    creates an attractive $1/r^2$ potential and enhances the decay
    rate. Inset: experimental steps. Loading of the trap, shifting it
    onto the wire and observing its decay. ({\em c}) Dependence of the
    trap decay rates on the wire charge for different wire thickness.
    The decay rate for uncharged wires is proportional to their actual
    diameters. For increasing charges the absorption rate becomes a
    linear function of the charge, a characteristic of the $1/r^2$
    singularity. The slope is independent of the wire diameter. }
    \label{decayrate}
\end{figure}

The motion in a $1/r^{2}$ singularity can be studied by placing a
cloud of cold atoms in the  potential of a charged wire. In this
experiment  the number of cold lithium atoms of a MOT is monitored
while the atoms move in the $1/r^{2}$ potential of the wire
\citep{Den98-737}. At extremely low light levels the MOT acts as a
box holding a gas of atoms. Atoms falling into the attractive
$1/r^{2}$ singularity are lost as they hit the wire. This loss
mechanism leads to an exponential decay of the trapped atom number
(see Fig. \ref{decayrate}b).

The corresponding loss rate is characteristic for the $1/r^{2}$
singularity and its strength. Atoms with angular momentum $L_{z}<
L_{\rm crit}$ (see Eq. (\ref{e:clHam}) in section
\ref{secElecInteract}) fall into the singularity. The loss rate is
a linear function of $q$ because $ L_{\rm crit}$ is proportional
to the line charge $q$ and the atoms are uniformly distributed
over angular momentum states (see Fig. \ref{decayrate}c). This is
actually true only for high charges, since for lower $q$, the
finite thickness of the charged wire becomes apparent. The MOT
decay rate for an uncharged wire is proportional to its actual
diameter. The diameters of the wires in the experiments ranged
between 0.7$\mu$m and 5$\mu$m. A detailed analysis of the
absorption data reveals that Van der Waals forces also contribute
to the atomic absorption rate \citep{Den98}. This effect was found
to be important  for thin wires with diameters of less than
1$\mu$m. Hence this system should allow for detailed future
studies of Van der Waals interaction and retardation in nontrivial
boundary conditions.

The $1/r^2$ potential would be especially interesting to study in
the quantum regime where the de Broglie wavelength of the atoms is
much larger than the diameter of the charged wire; the
quantization of angular momentum then begins to play a role
\citep{Den97-405}. This can be used for example in order to build
an angular momentum filter for atoms \citep{Sch95-169}.


\section{Surface mounted structures: The atom chip}
\label{s:chip-ex} \label{IV}

Free standing structures, as those described in the previous
section, are extremely delicate, and one arrives quickly at their
structural limit, when miniaturizing traps and guides. Wires
mounted on a surface are more robust, can be made much smaller and
heat is dissipated more easily which allows significantly more
current density to be sent through the wires. This together with
strong bias fields allows for tighter confinement of atoms in the
traps. Consequently, ground state sizes $< 10$nm become feasible.
Existing accurate nanofabrication technology provides rich and
well established production procedures, not only for conducting
structures, but also for micromagnets. Optical elements such as
microoptics, photonic crystals and microcavities can also be
included to arrive at a highly integrated device. The small ground
state size of such microtraps implies that we know the exact
location of the atom relative to other structures on the surface
to the precision of the fabrication process (typically $< 100$nm),
allowing extremely close sites to be addressed individually for
manipulation and measurement.

We have named nanofabricated surfaces for cold atom manipulation
`atom chips' in reminder of the similarity of these atom optical
circuits to electronic integrated circuits. In designing atom
chips one attempts to bring together the best of two worlds:  the
well developed techniques of quantum manipulation of atoms, and
the mature world of nanofabrication in electronics and optics, to
build complex experiments utilizing the above techniques.

In the following we describe the {atom chip} and its present
experimental status. Future goals will be addressed in section VI.

\subsection{Fabrication}
\label{Fabrication} \label{IV-A}

There are many different techniques of atom manipulation which can
be integrated into an atom chip. Present atom chip experiments
follow a simple scheme based on wires that carry currents or
charges. These allow to miniaturize the free standing devices
discussed in section~\ref{III}. We will focus here on these simple
integrated structures, leaving issues of further integration to
the outlook in section VI.

To build an atom chip one has to solve the following problems:
first of all, the microstructures have to withstand high current
densities and high electric fields. This requires structures with
low electrical resistance and good heat conductivity.  The
material of choice for the wires is gold, though other materials
such as copper are also used. For the substrate one wants good
heat conductivity with high electrical insulation withstanding
large electric fields (created at sharp ($r\approx 1\mu$m) corners
even by small voltages), and ease of fabrication. Typical
materials are silicon, gallium arsenide, aluminum nitride,
aluminum oxide and sapphire (${\rm Al}_2 {\rm O}_3$), though glass
has also been used.

Another requirement lies in the fact that cold atoms have to be
collected and then transferred towards the small traps on the
chip. If one wants to avoid transferring the atoms from a distant
MOT, the chip has to be either transparent or reflecting, to allow
lasers to address the atoms from all directions near the surface.
Nevertheless, experiments exist in which the atoms have been
brought from a distance to a chip \citep{Ott01-230401,Gus01}.

Presently, atom chips are built mainly using two technologies:
thin film hybrid technology, or plain nanofabrication which is the
first step of the two stage hybrid technology.

\subsubsection{Thin film hybrid technology}

In this approach one starts from an insulating substrate (e.g.
sapphire) and patterns, using lithographic techniques, a layout of
the desired structure onto a thin ($< 100$nm) metallic layer.
In the second stage, the wires are grown by electroplating: Metal
ions from a solution are deposited onto the exposed metallic
layer, which is now charged. With this process one obtains wires
with quite large cross sections (typical structure widths are $3$
to $100\mu$m) that support high currents. However, miniaturization
will be limited to a few micron wire width. Furthermore, surface
roughness is quite large, which makes such surfaces less suitable
for the reflection MOT and atom detection. These drawbacks and the
expected shadows from large etchings between wires, have been
dealt with successfully by covering the chip with an insulating
layer and then with a metallic reflection layer e.g., with the
Munich chip as shown in Fig.~\ref{f:ChipFabMunich})
\citep{Rei01-81}. This, however, carries the price of not enabling
atoms to be closer than some $20\mu$m from the wires themselves. A
technical advantage of electroplating is that it wastes less gold
because one avoids evaporation of large amounts of gold, which
mostly cover the evaporation chamber and not the chip.

Atom chips fabricated using using this technique have been used
sucessfully by the groups in Harvard (M. Prentiss), Munich (J.
Reichel and T. W. H\"ansch), JILA, Boulder (D. Anderson and E.
Cornell) and T\"ubingen (C. Zimmermann).

\begin{figure}[tbp]
    \infig{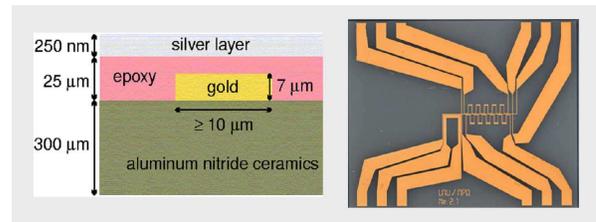}{0.9\columnwidth}
    \caption[]{Electroplating: ({\em left}) Cross section of the Munich
    group chip.  The metallic layer on top of the wires gives the chip
    enhanced surface quality in order to form a mirror MOT.
    ({\em right}) The layout of the chip.  The magnetic
    `conveyer belt' explained in section \ref{s:basics} is visible.
    The wires are connected to the chip pads from the outside by means of wire
    bonding.  Recently, this chip was used to achieve Bose--Einstein
    condensation. Courtesy J. Reichel.}
    \lb{f:ChipFabMunich}
\end{figure}

\subsubsection{Nanofabrication}

\begin{figure}[tb]
    \infig{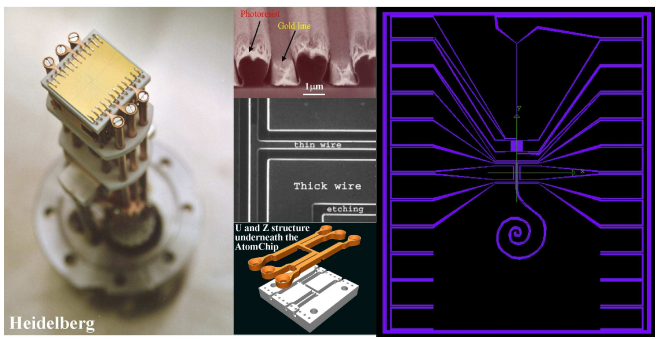}{\columnwidth}
    \caption{Nanofabricated atom chips (Heidelberg):
    ({\em left}) A mounted chip, ready to be put into the vacuum
    chamber.  The mechanical clamp contacts to the pads are visible. The mounting also includes cooling
    in order to remove heat produced by the currents.
    ({\em center, from top to bottom}) Details of fabrication and assembly:
    (i) A chip in the middle of the fabrication process,
    after some gold has been evaporated and before the photoresist
    has been removed.  The visible wires have a cross section of
    $1\times1\,\mu{\rm m}^2$.
    (ii) An electron microscope view of the surface.  A
    `T' junction of a $10\mu$m wide wire is visible as well as the
    $10\mu$m etchings which define it.
    (iii) Typical design of the U-- and Z--shaped wires placed underneath
    the chip to help in the initial loading process.  The wires
    can support $>50$A of current in DC operation without
    degrading a $p < 10^{-11}$mbar vacuum.
    ({\em right})
    A typical design of an atom chip. On both
    sides contact pads are visible. The center of the chip is used for the loading
    the atoms, which are then released into the physics areas: on top,
    a magnetic guide with
    arrays of electric leads, on the bottom, a spiral formed by
    two parallel wires enables atom guiding in all directions on the chip.
    }
    \label{f:ChipFabWIS}
\end{figure}

Atom chip structures can also be fabricated into an evaporated
conductive layer with state--of--the--art processes used for
electronic chips. To the best of our knowledge, this approach is
only used by the Heidelberg (formerly Innsbruck) group. In these
atom chips a $1-2.5\mu$m gold layer is evaporated onto a $0.6$mm
thick semiconductor substrate (GaAs or Si). As GaAs or Si tend to
leak current, especially in the presence of light, a thin
isolating layer of ${\rm SiO}_2$ is put between the substrate and
the gold layer. The chip wires are defined by $2-10\mu$m wide
etchings from which the conductive gold has been removed. This
leaves the chip as a gold mirror that can be used to reflect MOT
laser beams (the $10\mu$m etchings impede the MOT operation only
in a slight way). The mirror surface quality is very high,
achieving an extremely low amount of scattered light. The chips
were produced at the microfabrication centers of the University of
Vienna and of the Weizmann Institute of Science (Rehovot), see
Fig.~\ref{f:ChipFabWIS}.

Atom chips fabricated with this method have the advantage that the
structure size is only limited by nanofabrication ($<100$nm). The
drawback is that the conductive layer cannot be too thick. This is
due mainly to restrictions on the available thickness of the
photoresist used in the process.   The thin wires support only
smaller currents, and therefore only smaller traps closer to the
surface can be built. This disadvantage can be corrected by adding
larger wires below the chip surface, as presented in
Fig.~\ref{f:ChipFabWIS}.

At this stage it is hard to judge what is the best fabrication
process. There are still many open questions. For example, is
there a sizable difference in the specific resistance between
evaporated gold and electroplated gold? For a direct comparison,
one would have to unify all other parameters such as substrates
and intermediate layers. Another question concerns the final
fabrication resolution one wishes to realize. Assuming one aspires
to achieve the smallest possible trap height above the chip
surface for sake of low power consumption and high potential
tailoring resolution, the limit will be at a height below which
surface induced decoherence becomes too strong (see
sec.~\ref{s:loss}). This height, together with the finite size
effects described in section~\ref{s:basics}, will determine the
fabrication resolution needed. Smoothness of resolution will also
be required as fluctuating wire widths will change the current
density and therefore the trap frequency in a way that may hinder
the transport of BEC due to potential hills. Finally, as
multi--layer chips using more elaborate 3-dimensional designs are
introduced e.g. for wire crossings and more complicated structures
including photonic elements, it may be that conductor layers
thicker than a few microns will have to be abandoned. In order to
fully exploit the potential of the atom chip in the future, the
technology used will have to be such that all elements could be
made with a suitable process into a monolithic device.

Finally, we note that although usual current densities used in the
experiments range between $10^6-10^7$A/cm$^2$ (higher with smaller
cross sections and depending on pulse time, work cycle and heat
conductivity of substrate), densities of up to $10^8$A/cm$^2$ have
been reported for cooled substrates \citep{Drn98-2906}. Gold wires
have been found to be the best, achieving superior performance
even when compared to superconductors.

\subsection{Loading the chip}
\label{Loading} \label{IV-B}

In general there are two different approaches to loading cold
atoms into the chip traps:

(i) Collect and cool the atoms at a different location and
transport the cold ensemble to the surface traps.  This may be
achieved using direct injection from a cold atomic beam coming
from a low--velocity--intense--source (LVIS)
\citep{Mue99-5194,Mue00-1382,Mue01-041602} or a released MOT
whereby the atoms are pulled by gravity \citep{Dek00-1124}.
Transferring the atoms with magnetic traps has also been achieved
\citep{Ott01-230401}. Experiments transferring a Bose--Einstein
condensate (BEC) using optical tweezers also exist \citep{Gus01}.

(ii) Cool and trap atoms close to the surface in a surface MOT,
and transfer the atoms from there to the microtraps on the chip
\citep{Rei99-3398,Fol00-4749}.  For this method to be implemented,
a MOT has to be formed close to the surface, and consequently the
atom chip has either to be transparent or reflecting.

In the following, we describe experiments performed at Heidelberg
(resp.\ Innsbruck), Sussex and Munich using the second approach.
Further on, several experiments using the first approach will also
be discussed (e.g. Figs. \ref{fig:vertical} and \ref{jila}).

\subsubsection{Mirror MOT}

\begin{figure}[tb]
    \infig{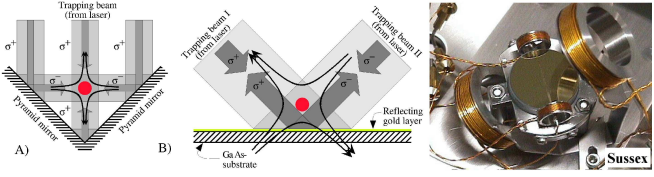}{\columnwidth}
    \caption{({\em left})
    A `pyramid MOT' is obtained when one single laser beam is
    retro--reflected by a four-sided pyramid in the center of a
    magnetic quadrupole.  The reflections ensure the correct
    helicities of the laser beams when the quadrupole field (field
    lines) has the same symmetry as the pyramid.
    ({\em center}) The mirror MOT is generated from the pyramid by leaving
    out 3 of the 4 reflecting walls.  Two MOT beams (I and II)
    impinge from opposite directions on the reflecting surface of the
    atom chip.  The correct MOT configuration is ensured together with
    the magnetic quadrupole field rotated $45^\circ$ to the atom chip
    surface as illustrated by the field lines.  The magnetic field can
    be obtained either by a set of external quadrupole coils, or by a
    U-shaped wire on the chip.
    ({\em right}) The Sussex mirror MOT chip setup with the external quadrupole
    coils on the mounting, inside the vacuum. Two parallel wires embedded in a fiber, are positioned on
    the surface of the mirror, forming a two wire guide and a time dependent interferometer
    (see section \ref{s:basics}). Two small `pinch' coils visible
    at the edges of the mirror provide longitudinal confinement.
    Courtesy E. Hinds.}
    \label{fig:pyramid}
\end{figure}

The first problem to solve is how to obtain a MOT configuration
close to a surface. This problem has an easy solution if we recall
that a circularly polarized light beam changes helicity upon
reflection from a mirror. To the best of our knowledge, this idea
was first put into practice with a pyramid of mirrors and one beam
\citep{Lee96-1177}, as presented in Fig.~\ref{fig:pyramid}a.
Almost in parallel, a single planar surface with four beams
impinging at $45^\circ$ degrees onto the surface was used, thus
realizing an eight beam MOT \citep{Pfa77,Sch98,Gau98-5298,Sch01}
\footnote{It is interesting to note that in another version of
this experiment, an evanescent field just above the extremely thin
metal surface, formed by light beams impinging on the back of the
surface, was used as an atom mirror. This allowed to produce a MOT
with reasonable surface induced losses even at the extreme
proximity of $100$nm from the surface.}. In the surface MOT most
common today one generates the MOT beam configuration from four
beams by reflecting only two beams off the chip surface (see
Fig.~\ref{fig:pyramid}b) \citep{Rei99-3398,Fol00-4749}.
The magnetic quadrupole field for the MOT can be obtained either
by a set of external quadrupole coils, or by superimposing a
homogeneous bias field with the field generated from a U--shaped
wire on or below the chip (`U--MOT').  External quadrupole coils
generate the correct magnetic field configuration if one of the
reflected light beams is in the coil axis.  If the U--MOT is used,
the reflected light beams must lie in the symmetry plane of the U.
Trapping in the U--MOT has the advantage that the MOT is well
aligned with respect to the chip and its microtraps. If the mirror
MOT is sufficiently far from the surface (a few times the MOT
radius), its loading rate and final atom number are very similar
to a regular free space MOT under the same conditions (laser
power, vacuum, supply of cold atoms, etc.).  In agreement with
earlier observations using wires, the shadows (diffraction
patterns) from the $10\mu$m etchings in the gold surface of the
nanofabricated atom chip do not disturb the MOT significantly
\citep{Den99-2014,Den98}.

Such atom chip mirror MOTs, have been loaded from an atomic beam
in Innsbruck/Heidelberg \citep{Fol00-4749}, from the background
vapor in Munich \citep{Rei99-3398} and in a double MOT system in
Innsbruck/Heidelberg ($>10^8$ atoms at lifetime $>100$s), using
either external coils or the U--wire for the quadrupole field. In
addition, at Sussex and Harvard surface MOTs were realized using
permanent and semi-permanent (magnetizable cores) magnetic
structures.

\begin{figure}[tb]
    \infig{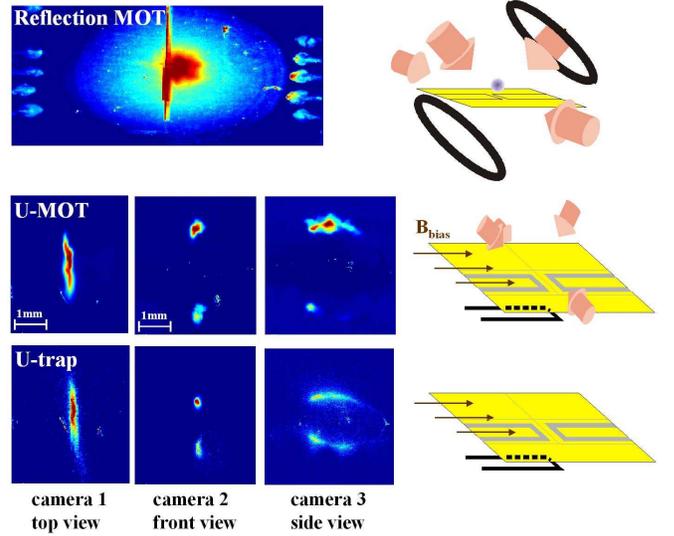}{\columnwidth}
    \caption{Pictures showing the loading of cold atoms close to the
    surface of an atom chip.  ({\em top left}) Picture of
    the mirror MOT, taken from above. The cloud is visible at the
    center while the electrical contacts can be seen at the edges.
    ({\em top right}) Schematic of the MOT beams and quadrupole coils.
    ({\em center row}) Atoms trapped in the U--MOT created by a current
    in the large U--shaped wire underneath the chip and a homogeneous
    bias field.
    ({\em bottom row}) Atoms in a magnetic trap generated by the
    U-wire field. The columns show from left to right the
    top, front and side (direction of bias field) views respectively,
    the far right column
    shows the schematics of the wire configuration.
    Current carrying wires are highlighted in black.  The front and
    side views show two images: the upper is the actual atom cloud and
    the lower is the reflection on the gold surface of the chip.  The
    distance between both images is an indication of the distance of
    the atoms from the chip surface. The
    pictures of the magnetically trapped atomic cloud are obtained by
    fluorescence imaging using a short laser pulse (typically
    $<1$ms).}
    \label{fig:MOT2Trap}
\end{figure}

As an example we describe the Innsbruck/Heidelberg lithium setup.
Fig.~\ref{fig:MOT2Trap} shows a top view of the mirror MOT sitting
above the chip with some of its electric connections. For the
transfer into the U--MOT, the large external quadrupole coils are
switched off while the current in the U--shaped wire underneath
the chip is switched on (up to $25$A), together with an external
bias field ($8$G). This forms a nearly identical, but spatially
smaller quadrupole field as compared to the fields of the large
coils. By changing the bias field, the U--MOT can be compressed
and shifted close to the chip surface (typically $1-2$mm). The
laser power and detuning are changed to further cool the atoms,
giving a sample with a temperature of about $200\mu$K.

\subsubsection{Transferring atoms to the chip surface}

After the U--MOT phase, atoms are cooled using optical molasses,
optically pumped and transferred into a matched magnetic trap,
typically produced by a thick Z-shaped wire plus bias field. From
there atoms are transferred closer and closer to the chip and
loaded sequentially into smaller and smaller traps.  In general,
it is favorable to lower the trap towards the surface by
increasing the magnetic bias field.  This way the trap depth
increases and less atoms are lost due to adiabatic heating during
compression. Unfortunately this is not feasible all the way:
Finite size effects limit small traps to thin wires, at the price
of not being able to push high currents.

\begin{figure}[tb]
    \infig{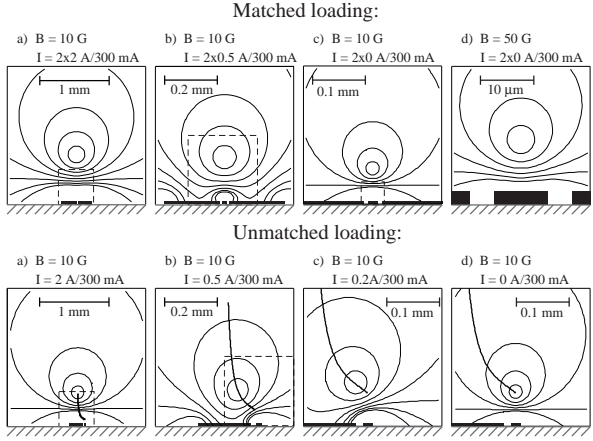}{0.9\columnwidth}
     \caption{Principle of compressing and loading wire guides.
     The position of the surface mounted wires and
     equipotential lines of the trapping potential are shown. The
     {\em top row} shows the transfer from {two} large $200\mu$m
     wires to one small $10\mu$m wire.  In ({\em a})--({\em c}) the
     current in the small wire is constant at $300$mA and the bias
     field is constant at $10$G. The current in the two large wires is
     decreased from $2$A in each wire to zero.  This transfers the
     atoms to the small wire.  ({\em d}) By increasing the bias field
     the trap can be compressed further.  The {\em bottom row} shows
     the transfer from {one} large $200\mu$m wire to one small
     $10\mu$m wire.  In ({\em a})--({\em d}) the current in the
     large wire drops from $2$A to zero.  The thick line shows how
     the trap center moves during transfer. A much weaker confinement
     during transfer is obtained in this configuration. }
     \label{fig:TransferToSmall}
\end{figure}

The basic transfer principle from a large wire to a small wire is
to switch on first the current for the smaller trap, and then to
ramp down the current in the bigger trap maintained by a thicker
wire (Fig.~\ref{fig:TransferToSmall}). Further compression is
achieved by using smaller and smaller currents. Care has to be
taken, that the transfer is adiabatic, especially with respect to
the motion of the potential minimum. By an appropriate change of
the bias field, the compression of the atoms in the shrinking trap
can be performed very smoothly. Transferring into more complicated
potential configurations one has to avoid the opening of escape
routes for the trapped atoms.

For an adiabatic transfer of relatively hot atoms, the main loss
is due to heating: when compressing by lowering the current,
high--lying levels may eventually spill over the potential
barrier. Significant loss occurs if the trap depth is much smaller
than $10$ times the temperature of the atomic ensemble. Other loss
mechanisms are described in section \ref{s:loss}. For thermal
clouds, typical achieved transfer efficiencies from the MOT to the
magnetic chip trap are as high as $60\%$.

\begin{figure}[tbp]
    \infig{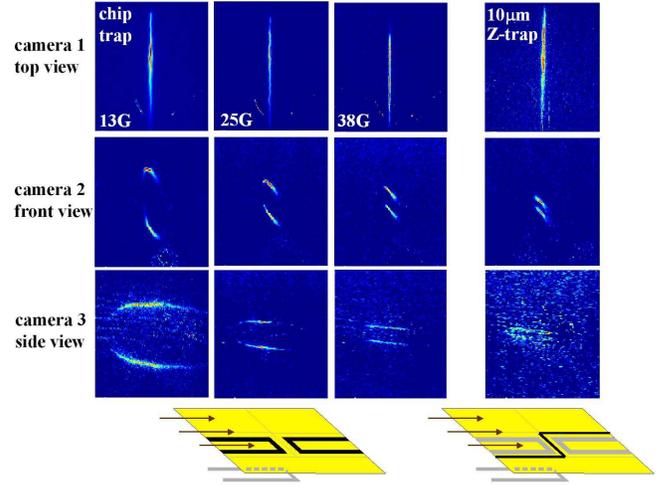}{\columnwidth}
    \caption{Compressing a cloud of cold atoms on an {atom chip}: the {\em top
    row} shows the view from the top, the {\em center row} the front
    view, and the {\em bottom row} the side view.  The first three
    columns show atoms trapped on the chip with the two U--shaped
    wires.  The compression of the trap is accomplished by increasing
    the bias field.  The last row displays images from a Z-trap
    created by $300$mA current through the $10\mu$m wire in the
    center of the chip.  The pictures of the magnetically trapped
    atomic cloud are obtained by fluorescence imaging using a
    short ($<1$ms) molasses laser pulse. }
    \label{fig:doubleUtrap}
 \end{figure}

As a detailed example we describe the loading of the first of the
Innsbruck experiments \citep{Fol00-4749}.  After accumulating
atoms in a mirror MOT and transferring them to the U--MOT, the
laser beams are switched off and the quadrupole field generated by
the U--shaped wire below the chip surface serves as a magnetic
trap for low--field seekers (Fig.~\ref{fig:MOT2Trap}: U--trap).
The magnetic trap is lowered further towards the surface of the
chip by increasing the bias field. Atoms are now close enough to
be trapped by the chip fields. Next, a current of $2$A is sent
through two $200\mu$m U--shaped wires on the chip, and the current
in the U--shaped wire located underneath the chip is ramped down
to zero. This procedure brings the atoms closer to the chip,
compresses the trap considerably, and transfers the atoms to a
magnetic trap formed by the currents on the chip surface.  This
trap is further compressed and lowered towards the surface
(typically $< 100\mu$m) by increasing the bias field
(Fig.~\ref{fig:doubleUtrap}). From there the atoms are transferred
to a microtrap created by a $10\mu$m Z--shaped wire.

In the lowest height and most compressed trap achieved to date, a
$1 \times 1\mu$m$^2$ Z--shaped wire is used with a current of
$100$mA (Heidelberg). With a bias field of $30$G the atoms are
trapped at a height of about $7\mu$m above the surface and at an
angular oscillation frequency $\omega \approx 2\pi \times 200$kHz
(magnetic field gradients of $50$kG/cm) for several tens of ms
(see eq.~\ref{trap_omega_chap.2} for typical trap frequencies). At
such a small trap height several problems come into play: First,
with the present $2$mm distance between the bends of the Z--shaped
wire, the Ioffe--Pritchard configuration is lost and one is left
with a single wire quadrupole field where atoms can suffer
Majorana flips. Two easy remedies would involve smaller Z lengths
or a slight tilt of the bias field direction. Second, the trap is
so tight that the number of atoms that survive the transfer and
compression is small. This limitation should not be applicable in
the case of a BEC as there are no high-lying states where atoms
run the risk of spilling over the finite trap barrier. A third
problem has to do with the observation of the atoms: even with
negligible stray light from surface scattering or blurring by
atomic motion, it is found that direct observation of extremely
tight traps close to the surface ($<20\mu$m) is very hard. The
signal suppression is probably due to large Zeeman shifts in the
cloud, which together with optical pumping processes dramatically
reduce the scattered light. In such a case, one can observe the
atoms after trapping by `pulling' them up, away from the surface
into a less compressed trap. This may be done simply by increasing
the wire current or decreasing the bias field.

\subsubsection{Observing atoms on the chip}

A simple way to observe the trapped atoms is by fluorescence
imaging.  For this, one illuminates the cloud with near--resonant
molasses laser beams for a short time (typically much less than
$1$ms).  The scattered light is imaged by CCD cameras as shown in
Fig.~\ref{fig:MOT2Trap} and Fig.~\ref{fig:doubleUtrap}. One should
use short enough exposure times to avoid blurring of the image due
to atomic motion.  One also has to select the camera positions
wisely to avoid stray light caused by scattering off the etchings
in the atom chip surface. Furthermore, it is important that the
metal surface itself shows minimal light scattering.  Here, the
excellent surface quality of evaporation on semiconductor surfaces
is essential.

A different possibility is to use absorption or phase contrast
imaging.  If the probe beam is directed parallel to the chip
surface, the surface quality is not as critical, and one does not
have to take care of diffraction peaks from the etchings.  Such
absorption imaging is used by the Munich and T\"ubingen groups.
With an excellent surface mirror quality, one could also implement
absorption imaging with laser beams reflected from the chip
surface.  More sophisticated methods such as phase contrast
imaging will be important for more complicated atom optical
devices on atom chips, where non-destructive observation very
close to the chip surface becomes essential. For an overview of
these methods, we refer the reader to the many BEC review papers
(see for example \cite{Ket99}).

Finally, future light optical elements incorporated on the chip,
such as microspheres or cavities, will allow for much better
detection sensitivity, possibly at the single atom level (see
section \ref{VI-det}). Such work has been started in several of
the labs.

\subsection{Atom chip experiments}
\label{AtChExp} \label{IV-C}

Since the first attempts two years ago, the atom chip has now
become a `tool box in development' in numerous labs around the
world. To the best of our knowledge these include (in alphabetical
order) the groups at Boulder/JILA (D. Anderson and E. Cornell),
CalTech (H. Mabuchi), Harvard (M. Prentiss), Heidelberg (J.
Schmiedmayer), MIT (W. Ketterle), Munich (J. Reichel and T. W.
H\"ansch), Orsay (C. Westbrook and A. Aspect), Sussex (E. Hinds),
and T\"ubingen (C. Zimmermann). We will unfortunately not be able
to present in detail all the extensive work done, nor we will be
able to touch upon other surface related projects such as the atom
mirror.

\subsubsection{Traps} \label{IV-C-1}

The simplest traps (i.e 3-dimensional confinement) are usually
based on a straight wire guide with some form of longitudinal
confinement, which is produced either by external coils or by
wires on the chip (section \ref{II-A-4}). Additional wires for
on--board bias fields may also be added. More sophisticated
designs have been suggested by \cite{Wei95-4004} (see also section
\ref{II-A-5}).

As an example, we start with the simple microtraps realized in
Innsbruck/Heidelberg with lithium \citep{Fol00-4749} and Munich
with rubidium atoms \citep{Rei99-3398,Rei01-81}. Here, the traps
are based on wires of $1$ to $30\mu$m width with which
surface--trap distances below $10\mu$m were achieved. The wires
used are either U--, Z--, or H--shaped.

In these experiments, the compression of traps and guides was also
investigated \citep{Fol00-4749,Rei01-81}. This is done by ramping
up the bias magnetic field.  In this process one typically
achieves gradients of $>25$kG/cm.  With lithium atoms, trap
parameters with a transverse ground state size below $100$nm and
angular frequencies of $2\pi \times 200$kHz were achieved
\citep{Fol00-4749}, thus reaching the parameter regime required by
quantum computation proposals \citep{Cal00-022304,Bri00-415}.

In addition, an on--board bias field for the thin wire trap was
also created by sending currents through two U--shaped wires in
the opposite direction with respect to the thin wire current.
These create a magnetic field parallel to the chip surface,
substituting the external bias field.  Hence, trapping of atoms on
a self contained chip was demonstrated \citep{Fol00-4749}.

\begin{figure}[tbp]
    \infig{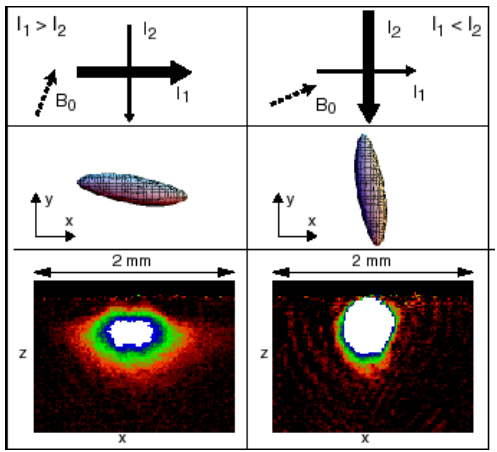}{0.75\columnwidth}
     \caption{Ioffe--Pritchard trap created by two intersecting wires.
     The {\em left column} corresponds to $I_{1}>I_{2}$ and
     $|B_{0,y}|>|B_{0,x}|$, in the {\em right column} both relations
     are reversed.  ({\em top row}) Conductor pattern; the thickness of
     the arrows corresponds to the magnitude of the current.  Dashed
     arrows indicate the bias field direction.  ({\em middle row})
     Calculated contours of the magnetic field modulus $|B(x,y)|$
     indicating how the long trap axis turns.  The left potential
     continuously transforms into the right one when the parameters
     are changed smoothly.  ({\em bottom row}) Absorption images
     corresponding to the two situations. Courtesy J. Reichel.}
     \label{fig:crossing}
\end{figure}

An example of a different configuration was realized in Munich
with rubidium atoms. In this experiment, three-dimensional
trapping was achieved by crossing two straight wires and choosing
an appropriate bias field direction, as discussed in section
\ref{II-A-4} \citep{Rei01-81} (Fig.~\ref{fig:crossing}). Here the
additional wire actually provides the endcaps that were previously
provided in the Z-- and U--shaped traps by the same wire. This
type of trap might be useful for the realization of arrays of
nearby traps. In T\"ubingen and Sussex longitudinal confinement
has been achieved by additional coils.

\begin{figure}[tb]
    \infig{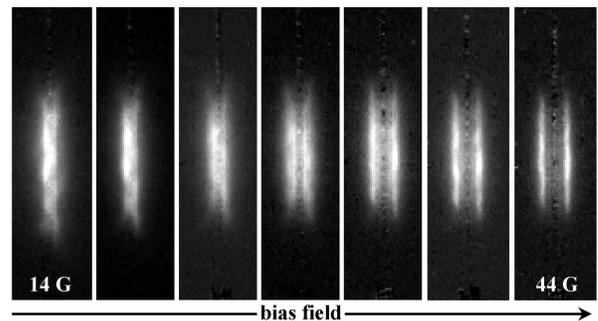}{0.9\columnwidth}
     \caption{A top view of a thermal $200\mu$K cloud of lithium atoms in
     a double well potential $40\mu$m above the chip surface. The
     minima are separated by $350\mu$m. The imaging flash light pulse is
     $100\mu$s long.
     The splitting may be done as slow as needed in order to achieve adiabaticity.}
     \label{fig:doublewell}
\end{figure}

Finally, the splitting of a single trap into two has been
demonstrated in Heidelberg, Munich and Sussex. Such a time
dependent potential is presented in Fig.~\ref{fig:doublewell} and
as explained in section \ref{s:basics}, may form the basis of an
interferometer. It is also the first step in creating multi well
traps or arrays of traps.

More sophisticated designs have been suggested by
\cite{Wei95-4004} (see section \ref{II-A-5}) and fabricated (e.g.
Harvard, \citep{Drn98-2906}).

\subsubsection{Guiding and Transport}
\label{GuideTransport}

\begin{figure}[tb]
    \infig{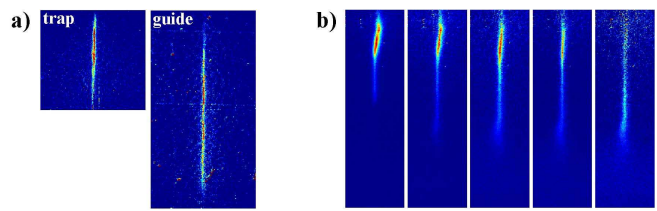}{\columnwidth}
    \caption[]{({\em a}) Cold atoms in a microtrap (left) and released
    all at once into a linear guide (right).
    ({\em b}) Continuous loading of an atom guide by leaking atoms
    from a reservoir created by a U--trap into the guide by ramping
    down the current in the U. Propagation is due to thermal velocity.
   These pictures from are taken at time spacings of $1$ms.}
    \label{fig:ChipGuide}
\end{figure}

To achieve mesoscopic atom optics on a chip, it is essential to
have reliable means of transporting atoms.  One such device is an
atomic guide using a single wire with a bias field. Such an
experiment is shown in Fig.~\ref{fig:ChipGuide}a.  The Z--trap is
transformed into an L--shaped guide by re-routing the current from
one of the Z leads. The atoms expand along the guide due to their
thermal velocity \citep{Fol00-4749}. Similarly, it was
demonstrated that one can directly load the guide from a larger
magnetic trap on the chip and skip the small surface trap.

It is also possible to achieve a continuously loaded magnetic
guide using a leaky microtrap (see Fig.~\ref{fig:ChipGuide}b).
This is achieved by lowering the barrier between the trap and the
guide, the barrier being simply the trap end cap, whose height may
be controlled by changing the current in the microtrap
\citep{Bru00-2789}.

\begin{figure}
    \infig{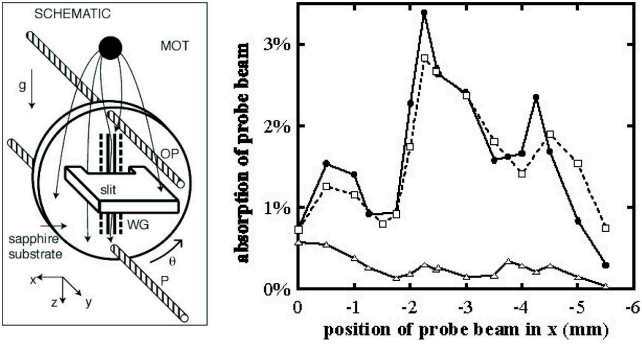}{\columnwidth}
    \caption[]{Vertical bias field: In this Harvard experiment,
    two wire vertical guides were realized, enabling the guiding
    of atoms in a variety of directions. ({\em left}) The setup. ({\em
    right})
    Absorption of probe beam versus position along x at a height
    of $3.5$mm below the output of the guide. The left and right
    peaks are attributed to unaffected atoms and atoms deflected
    by the outside of the guide potential, respectively. The open
    triangles are the data while the guide is turned off. Courtesy
    M. Prentiss.
    }
    \label{fig:vertical}
\end{figure}

However, there are limitations to such a simple guide. Using a
homogeneous external bias field, such a guide has to be straight
(linear), since the bias field must be perpendicular to the wire
as discussed in section \ref{II-A-2}. This considerably limits the
potential use of the whole chip surface. A possible solution is to
create the bias field using on--chip wires (3-wire configuration
shown in Fig.~\ref{fig:Guides}) or the two--wire guide
configuration discussed in section \ref{II-A-3}, in which the
currents are counter propagating and the bias field is
perpendicular to the chip surface.  A first experiment was
conducted by M. Prentiss' group in Harvard \citep{Dek00-1124}.
Here, cesium atoms were dropped from a MOT onto a vertically
positioned chip, on which a two--wire guide managed to deflect the
atoms from their free fall (see Fig. \ref{fig:vertical}).
Furthermore, a four-wire guide was realized whereby the two extra
wires served as the source for the bias field (see also
Fig.~\ref{fig:Guides}).  Guiding along a curved two wire guide has
been achieved in Heidelberg, and experiments to guide atoms along
a spiral are in progress (for the chip design see
Fig.~\ref{f:ChipFabWIS}).

\begin{figure}[tbp]
    \parbox{0.75\columnwidth}{\input epsf \epsfxsize=\hsize \epsfbox{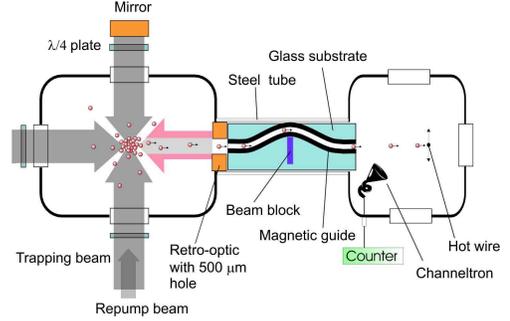} }
    \hfill
    \parbox{0.60\columnwidth}{\input epsf \epsfxsize=\hsize \epsfbox{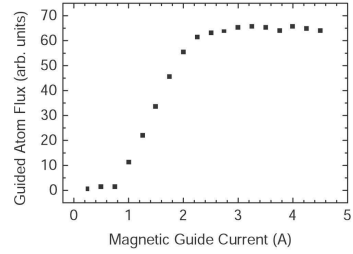} }
    \caption{The JILA setup in which a `low velocity intense source' (LVIS)
    was used to directly load the two wire guide. The data shows the need for
    strong potentials with which the magnetic guide can overcome the kinetic energy in order to deflect
    the atoms thereby bypassing the beam block. Courtesy E. Cornell.}
    \label{jila}
\end{figure}

Several experiments have achieved guiding without any bias field
by trapping the weak field seekers in the minimum existing exactly
in between two parallel wires with co--propagating currents. This
configuration was described in section \ref{II-A-3}. In
Fig.~\ref{jila}, we present such a setup \citep{Mue99-5194}.
Another similar use of this principle (in this case, not surface
mounted), in which a storage ring has been realized is presented
in section \ref{III-B-7}. Although advantageous for the lack of
bias fields, this concept may be hard to implement on miniaturized
atom chips as the atoms would be extremely close to the surface
for $1-2\mu$m thick wires.

\begin{figure}[tbp]
    \infig{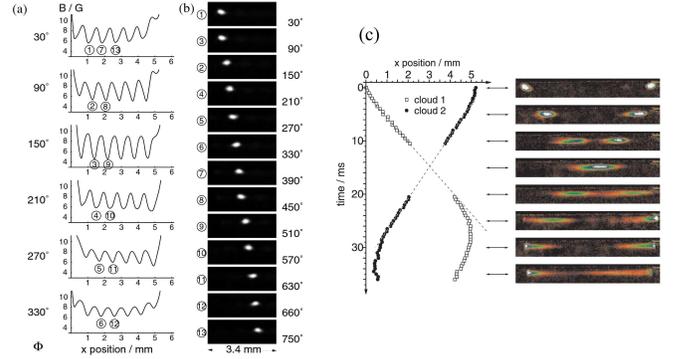}{\columnwidth}
    \caption{Moving atoms using a magnetic conveyor belt. ({\em
    a}) Potential for various phases of the movement.  The
    numbers indicate the position of the atoms as shown in the
    absorption images in column ({\em b})
    ({\em c}) Linear collider experiment.  ({\em left}) Time evolution
    of the centers of mass of the two clouds.  ({\em right})
    Absorption images of the colliding atoms. Courtesy J. Reichel.}
    \label{collider}
\end{figure}

Guiding with semi--permanent magnets has also been achieved
\citep{Roo01,Ven01}. These materials enhance the magnetic fields
coming from current carrying wires (see section \ref{II-A-10} for
a description). Completely permanent magnets are also being
contemplated to avoid current noise. However, to the best of our
knowledge, only atom mirrors have thus far been realized this way
\citep{Hin99-R119}.

A further limitation of the guides described above is that they
rely on thermal velocity.  Much more control can be achieved by
transporting atoms using moving potentials, as described in
section \ref{II-A-7}.  Such a transport device was implemented in
an experiment in Munich.  Using the movable 3-dimensional
potentials of their `motor',  atoms can be extracted from a
reservoir and moved or stopped at will \citep{Hae01-608}
(Fig.~\ref{collider}). This considerably improves the
possibilities of the chip, as demonstrated by the `linear
collider' shown in Fig.~\ref{collider}c, in which the motor was
used to split a cloud in two and then to collide the two halves
\citep{Rei01-81}.

\subsubsection{Beam splitters}
\label{IV-C-3}

\begin{figure}[tbp]
    \infig{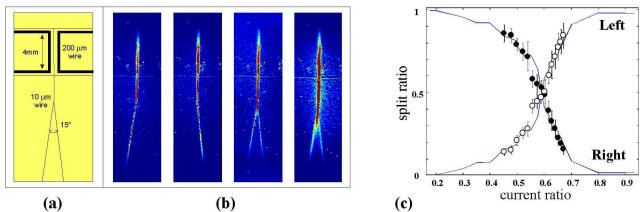}{\columnwidth}
    \caption[]{Beam splitter on a chip. ({\em a}) Chip outline and ({\em b})
     fluorescence images of guided atoms.  Two large U--shaped
     $200\mu$m wires are used to load atoms onto the $10\mu$m
     Y-shaped wire.  In the first two pictures in (b), a
     current ($0.8$A) is driven only through one side of the Y, therefore
     guiding atoms either to the left or to the right; in the next two
     pictures, taken at two different bias fields ($12$G
     and $8$G respectively), the current is divided in equal parts
     and the guided atoms split into both sides.  ({\em c}) Switching
     atoms between left and right is achieved by changing the current
     ratio in the two outputs and keeping the total current constant
     as before.  The points are measured values while the lines are
     obtained from Monte Carlo simulations with a $3$G field along
     the input guide.  The kinks in the lines are due to Monte Carlo
     statistics.}
    \label{chipBS}
\end{figure}

 As discussed in section \ref{II-A-8}, one may
combine the wire guides as described in the previous section to
build more complicated atom optical elements.  One such element is
a beam splitter.  A simple configuration is a Y--shaped wire
(Fig.~\ref{chipBS}a) which creates a beam splitter with one input
guide for the atoms, the central wire of the Y, and two outputs
guides, the right and left arms.  The atoms are split by means of
a symmetric scattering off the potential hill, which they
encounter at the splitting point.  Such a beam splitter on an atom
chip was realized by \cite{Cas00-5483} in Innsbruck. Atoms were
released from a chip microtrap and guided into the beam splitter.
Depending on how the current in the input wire is sent through the
Y, atoms can be switched from output arms of the Y to the other,
or directed to the two outputs with any desired ratio
(Fig.~\ref{chipBS}). Similar beam splitters have been widely used
for the splitting of guided electron waves in solid--state quantum
electronics devices. For example two Y splitters were put back to
back to form an Aharonov-Bohm type interferometer
\citep{Buk98-871}.

A four--port beam splitter has been realized at JILA by the group
of E. Cornell and D. Anderson by making a near x-shape out of two
wires which avoid a full crossing \citep{Mue00-1382}. In this
experiment, two input guides formed by two current carrying wires,
merge at the point of closest approach of the wires so that the
two minima merge into one, and then again split into two
independent minima.

\subsubsection{BEC on a chip}
\label{s:BEC}

A degenerate quantum gas in a microtrap is an ideal reservoir from
where to extract atoms for the experiments on the chip.  For
example a BEC will take a similar role as source of bosonic matter
waves as the Fermi sea has in quantum electronics. A clear
advantage of a BEC has to do with the transfer to the smallest
compressed surface traps, which involves high compression, leading
to large losses for thermal atoms if the trap depth is not
appropriate. The condensate occupies the trap ground state and
should follow any adiabatic compression of the trap. Second, a BEC
in a microtrap also provides the initial atomic state needed to
initiate delicate quantum processes such as interference or even a
well defined entanglement between atoms in two nearby traps. The
latter stands at the base of a two--qubit gate needed for quantum
information processing.

\begin{figure}[tbp]
    \infig{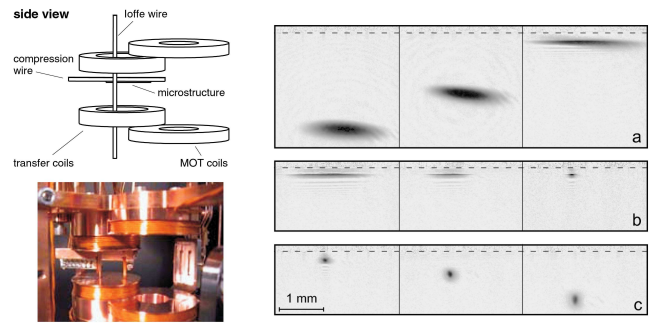}{\columnwidth}
    \caption{
    ({\em right}) The T\"ubingen setup.
    The first pair of coils (right) produced the MOT
    and then the atoms were conveyed to the trap formed by the second pair
    of coils. The chip mounting is visible within the second pair of coils.
    ({\em left})  Absorption images of the compression and final
    cooling stage. (a) compression into the microtrap. (b) RF
    cooling in the microtrap. (c) release of the condensate after
    5,10, and 15 ms time of flight.
    Courtesy C. Zimmermann.}
    \label{f:zimmermann}
\end{figure}

\begin{figure}[tbp]
    \infig{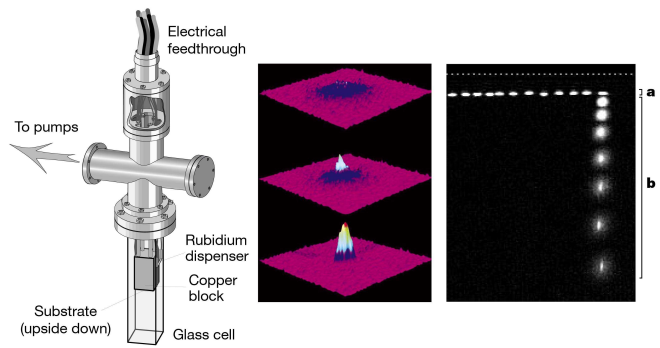}{\columnwidth}
    \caption{Munich atom Chip BEC experiment:
    {\em (left)} Schematics of the simple vapor cell apparatus.
    {\em (center)} Time of flight images showing the formation of
    a BEC.
    {\em (right)} (a) The BEC is transported in a
    movable 3-dimensional potential minimum. (b) At the end it is released and
    observed falling and expanding.
    Courtesy J. Reichel.}
     \label{f:BECmotor}
\end{figure}

In the last year three groups in T\"ubingen, Munich and Heidelberg
succeeded in making and holding a Bose--Einstein condensate in a
surface trap \citep{Ott01-230401,Hae01-498},
and the MIT group managed to transfer a BEC to a surface trap, and
load it into it \citep{Lea02}. These experiments showed that
making the BEC in a surface trap can be much simpler. For example
in very tight micro traps the BEC is formed in much shorter time
as the tightness of the traps allows for fast thermalization and
consequently fast evaporative cooling which relaxes the vacuum
requirements, permitting the use of a very simple one MOT setup to
collect the atoms \citep{Hae01-498}.

In the T\"ubingen experiment \citep{Ott01-230401}, a relatively
large condensate of $4\times 10^5$ $^{87}$Rb atoms has been formed
at a height of some $200\mu$m above the surface. The experiment
made use of a pulsed dispenser as an atom source, allowing ultra
high vacuum ($2\times 10^{-11}$mbar) while the dispenser was off.
This enabled the use of a simple single MOT setup. The experiment
was designed to transfer the atoms magnetically from a distant six
beam MOT to the chip using two adjacent pairs of coils
(Fig.~\ref{f:zimmermann}).  In the chip trap, condensation was
reached after $10$ to $30$s of forced RF evaporative cooling.
Aside from being the first surface BEC, the chip used in
T\"ubingen with its $25$mm long wires provides a highly
anisotropic BEC (aspect ratio $10^5$), approaching a quasi
one--dimensional regime. In recent work, the BEC was taken to a
height of only $20\mu$m without observing substantial heating
\citep{For02}. The smallest structure holding the BEC was a $3
\times 2.5\mu{\rm m}^2$ cross section copper wire with a current
of $0.4$A. The BEC had a lifetime of $100$ms in the compressed
trap (limited by 3--body collisions) and a $1$s lifetime once it
was expanded into a larger trap.

\begin{figure}[tbp]
    \infig{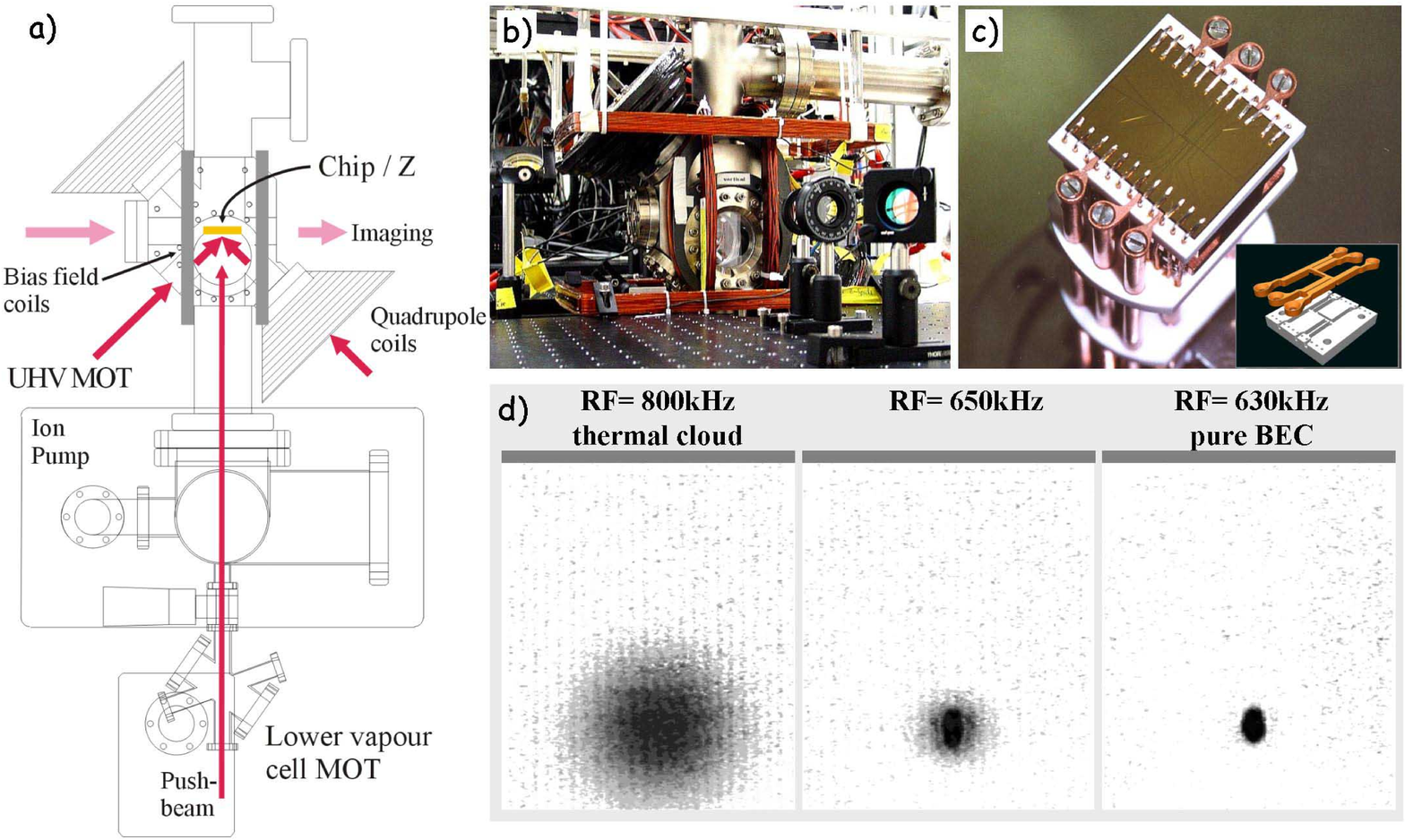}{\columnwidth}
    \caption{Heidelberg atom chip BEC experiment:
    (a) Schematics of the double MOT setup.  Atoms from a lower
    vapor cell MOT are transferred to a UHV mirror MOT using a continuous push beam.
    (b) Photograph of the upper (UHV) chamber. (c) The mounted chip and the U- and Z-shaped wire structure
    underneath the chip (inset).
    (d) Thermal cloud, BEC with thermal background, and pure BEC released and expanded for 15ms.}
     \label{f:BEC-HD}
\end{figure}

The second experiment producing a BEC in a microtrap was performed
in Munich \citep{Hae01-498} (Fig.~\ref{f:BECmotor}). Here an
attractively simple setup with a continuous dispenser discharge
was used. Consequently, the vacuum background pressure was high
($10^{-9}$mbar) and evaporative cooling had to be achieved
quickly. RF cooling times were as short as $700$ms thanks to the
strong compression in the microtrap which results in a high rate
of elastic collisions. The final BEC included some $6000$ atoms at
a height of $70\mu$m. The trapping wire was $1.95$mm long and the
trap aspect ratio was $10^3$.  The wire cross section was $50
\times 7 \,\mu{\rm m}^2$, and the current density approached
$10^6\,{\rm A} / {\rm cm}^2$. Strong heating of the cloud was
observed in this experiment but the source remains elusive
(possibly, current noise). A beautiful feature of this experiment
is the use of the magnetic `conveyer belt' described before
(section \ref{II-A-7} and \ref{GuideTransport},
Fig.~\ref{collider}), in order to transport the BEC during a time
of $100$ms over a distance of $1.6$mm without destroying it (see
Fig.~\ref{f:BECmotor}). Furthermore, the ability of the `motor' to
split clouds was used to show that a BEC survives such a
splitting. Two such halves were then released into free fall
exhibiting interference fringes as they overlapped.

In the third experiment, performed in Heidelberg, the condensate
of typically $3\times 10^5$ $^{87}$Rb atoms was formed either in a
Z-wire joffe pritchard trap, created by a wire structure
underneath the chip, or in a Z trap on the chip
(Fig.~\ref{f:BEC-HD}).
First $>3\times 10^8$ atoms are loaded into a mirror MOT
($<10^{-11}$ torr) created by external quadruople coils using a
double MOT configuration with a continuous push beam. The atoms
are then transferred into a U-MOT, where they are compressed and
after molasses cooling loaded into a Z wire trap. The BEC is
formed by forced RF evaporation in typically 20 seconds. Creating
th BEC using a wire structure underneath the chip allows to place
other surfaces close to the BEC, and still maintaining the high
precision of a micro trap for manipulating the cold atoms.  This
will open up the possibility to either study surfaces with the
cold atoms, and also to transfer the BEC to surface traps based on
dipole forces in light fields created by micro-optic elements and
evanescent fields.

\begin{figure}[tbp]
    \infig{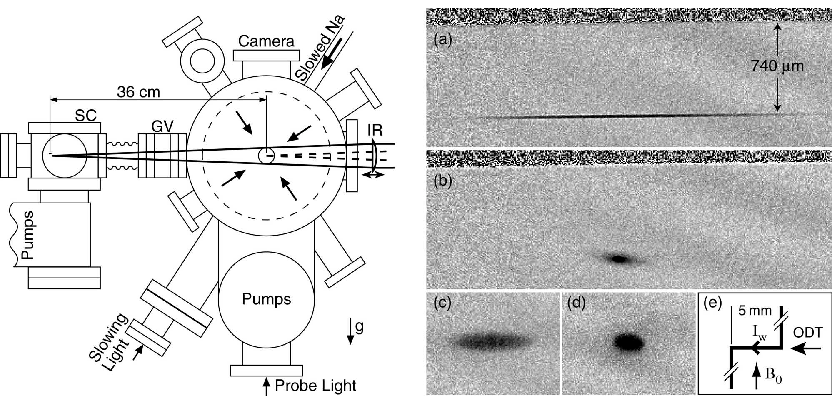}{\columnwidth}
    \caption{Transfer of a BEC to a microtrap:
    {\em (left)} Schematics of the setup with the science chamber housing
    the Z-trap on the far left and the BEC production chamber on the right.
    {\em (right)} Condensates in the science chamber
    (a) optial trab (b) Z-trap.
    The condensate was released from (c) an optical trap after 10 ms
    and (d) wire trap after 23 ms time of flight.
    (e) Schematic of the Z- trap. Courtesy W.~Ketterle}
     \label{f:MIT-BEC}
\end{figure}

The MIT group has transported a BEC of the order of $10^6$ Na
atoms into an auxiliary chamber and loaded it into a magnetic trap
formed by a Z-shaped wire \citep{Lea02}(Fig.~\ref{f:MIT-BEC}).
This was accomplished by trapping the condensate in the focus of
an infrared laser and translating the location of the laser focus
with controlled acceleration. This transport technique avoids the
optical and mechanical access constraints, and the extreme UHV
requirements of conventional condensate experiments.  The BEC was
consequently loaded into a micro structure.

Finally, we would like to note that currently other groups are
also working towards BEC in surface traps, and in a short time we
will see many different successful experiments

\section{Loss, heating and decoherence}
\label{s:loss}

For atom chips to work, three main destructive elements have to be
put under control.

\begin{itemize}
  \item {\em (i) Trap loss}: It is crucial that we are able to keep the atoms
  inside the trap as long as needed.
  \item {\em (ii) Heating}:
  Transfer of energy to our quantum system may result in excitations
  of motional degrees of freedom (e.g. trap vibrational levels), and
  consequently in multi mode propagation which would render the
  evolution of our system ill-defined.
  \item {\em (iii) Decoherence} or dephasing as it is sometimes referred to
  also originates from coupling to the environment. While heating requires
  the transfer of energy, decoherence is more delicate
  in nature \citep{Ste90-3436}. Nevertheless, the effect is
  just as harmful because superpositions with a definite phase
  relation between different quantum states are destroyed. This has
  to be avoided, e.g. for interferometers or quantum information
  processing on the atom chip.
\end{itemize}

In discussing these three points, we focus on the particularities
of atoms in strongly confined traps close to the surface of an
atom chip. The small separation between the cold atom cloud and
the `hot' macroscopic environment raises the intriguing question
of how strong the energy exchange will be, and which limit of atom
confinement and height above the surface can ultimately be
reached. We review theoretical results showing that fluctuations
in the magnetic trapping potential give a fairly large
contribution to both atom loss and heating. In addition, thermally
excited near fields are also responsible for loss and may impose
limits on coherent atom manipulation in very small ($\mu$m sized)
traps on the atom chip. Estimates for the relevant rates are
given, and we outline strategies to reduce them as much as
possible. Experimental data are not yet reliable enough to allow
for a detailed test of the theory, but there are indications that
field fluctuations indeed influence the lifetime of chip traps.

\subsection{Loss mechanisms}
\label{s:loss2}

\subsubsection{Spilling over a finite potential barrier}

Compression of a thermal atom cloud can lead to losses when the
cloud temperature rises above the trap depth. The smallest losses
occur if the compression is adiabatic. Atoms then stay in their
respective energy levels as the level energy increases. They can
nevertheless be lost during trap compression because of the finite
trap depth. It should be noted that this loss occurs for the
highest energies in the trap and can also be used to evaporatively
cool the cloud (see \cite{Lui96-381} for a review).

\subsubsection{Majorana flips}
If the atomic magnetic moment is not able to follow the change in
the direction of the magnetic field, the spin flips, and a weak
field seeking atom can be turned into a strong field seeker which
is not trapped \citep{Maj32-43,Gov00-3989}. This occurs when the
adiabatic limit (Larmor frequency $\omega_{L}$ much larger than
trap frequency $\omega$) does not hold. Majorana flips thus happen
at or near zeros of the magnetic field. For this reason,
additional bias fields are employed to `plug the hole' in the
center of a quadrupole field.

For a magnetic field configuration with a zero, loss can be
reduced if the atoms circle around it. The loss rate is then
inversely proportional to the angular momentum because the latter
determines the overlap with the minimum region
\citep{Ber89-2249,Hin00-033614}.

In Ioffe-Pritchard traps with nonzero field minimum $B_{ip}$,
there is a finite residual loss rate that has been calculated by
\cite{Suk97-2451}. For a model atom with spin 1/2 in the
vibrational ground state, one gets
\begin{eqnarray}
\gamma &=& \frac{ \pi \omega }{ 2 \sqrt{\rm e} } \exp( -
\mu_{\Vert} B_{ip} / \hbar \omega ) \nonumber
\\
&=& 6\times 10^{5}{\rm s}^{-1} \frac{ \omega / 2 \pi }{ 100{\rm
kHz} } \exp\left( - 14 \frac{ ( \mu_{\Vert} / \mu_{B} ) (B_{ip} /
{\rm G}) } { \omega / 2 \pi \,100{\rm kHz} } \right) ,
\label{eq:Majorana-rate}
\end{eqnarray}
where $\mu_{\Vert}$ is the component of the magnetic moment
parallel to the trapping field. Note the exponential suppression
for a sufficiently large plugging field $B_{ip}$, typical of
nonadiabatic (Landau-Zener) transitions. Choosing a Larmor
frequency $\omega_{L} = 2 \mu_\Vert B_{ip} / \hbar > 10\,\omega$,
one gets a lifetime larger than $\simeq 10^4$ trap oscillation
periods. A ratio $\omega_{L} / \omega > 20$ pushes this limit
already to $\simeq 10^8$.

\subsubsection{Noise-induced flips}

Fluctuations in the magnetic trap fields can also induce spin
flips into untrapped states, and lead to losses.  These
fluctuations are produced by thermally excited currents in the
metallic substrate or simply by technical noise in the wire
currents. Fluctuations of electric fields and of the Van der Waals
atom-surface interaction have been shown to be less relevant for
typical atom traps \citep{Hen99-414,Hen99-379}.

The trapped spin is perturbed via the magnetic dipole interaction
and flips at a rate given by second-order perturbation theory:
\begin{equation}
   \gamma = \frac{ 1 }{ 2 \hbar^2 } \sum_{k,l =
   x,y,z} \left\langle i | \mu_k | f \right\rangle \left\langle f |
   \mu_l | i \right\rangle S_B^{kl}( \omega_{L} ) ,
   \label{eq:flip-rate}
\end{equation}
where $S_B^{kl}( \omega_{L} )$ is the noise spectrum of the
magnetic fields, taken at the Larmor frequency $\omega_L$. We use
the following convention for the noise spectrum
\begin{equation}
   S_B^{ij}( \omega ) = 2 \int_{-\infty}^{+\infty}
   \!d\tau \, e^{i \omega \tau }
   \left\langle B_i(t + \tau) B_j( t ) \right\rangle
   ,
   \label{eq:def-noise-spectrum}
\end{equation}
where $\langle \ldots \rangle$ is a time average (experiment) or
an ensemble average (theory). The rms noise is thus given by an
integral over positive frequencies
\begin{equation}
   \left\langle B_{i}(t) B_{j}(t) \right\rangle =
   \int_0^\infty\!\frac{ d\omega }{ 2 \pi } S_B^{ij}( \omega )
   .
\end{equation}
For example, the rms magnetic field in a given bandwidth $\Delta
f$ for a white noise spectrum $S_B$ is given by $B_{\rm rms} =
\sqrt{ \Delta f S_B }$. The spectrum $S_B$ thus has units ${\rm
G}^2 / {\rm Hz}$.

\begin{table}
   \caption[]{Trace of the geometric tensor $Y_{ij}$ that determines
   the loss due to the thermally fluctuating magnetic near field,
   according to the rate~(\ref{eq:estimate-gamma}). The metallic
   layer has a thickness $d$, assumed much smaller than the distance
   $\height$ to the trap center. The wire has a radius $a \ll \height$,
   and $\height \ll \delta$ is assumed where $\delta$ is the skin
   depth of the metal. Taken from \cite{Hen01-73}. A more accurate
   calculation of Tr $Y_{ij}$ corrects the results of
   table~\ref{table:henkel} by a factor of $1/2$ for the half-space
   and the layer \citep{Hen02}.
   \label{table:henkel}}
   \begin{tabular}{ll}
   Geometry & Tr $Y_{ij}$\\
   \hline
   Half-space & $\pi / \height $ \\
   Layer      & $\pi d / \height^2 $ \\
   Wire       & $\pi^2 a^2 / (2 \height^3 )$\\
   \end{tabular}
\end{table}

\vspace{5mm}
\paragraph{Thermally excited currents}
An explicit calculation of the magnetic noise due to substrate
currents (`near field noise') yields the following estimate for
the loss rate \citep{Hen99-379}
\begin{equation}
   \gamma \simeq 75{\rm s}^{-1} \frac{ ( \mu / \mu_{\rm B} )^2 (T_s /
   300{\rm K}) } { ( \varrho / \varrho_{\rm Cu} ) } ( {\rm Tr} Y_{ij}\times
   1\mu{\rm m} )
   \label{eq:estimate-gamma}
   ,
\end{equation}
where $1/\varrho$ is the substrate conductivity (for copper,
$\varrho_{\rm Cu} = 1.7 \times 10^{-6}{\rm\Omega\,cm}$) and $T_s$
the substrate temperature. Note that the Larmor frequency
$\omega_{L}$ actually does not enter the loss rate. The `geometric
tensor' $Y_{ij}$ has dimension (1/length) and is inversely
proportional to the height $\height$ of the trap center above the
surface (table~\ref{table:henkel}). The loss
rate~(\ref{eq:estimate-gamma}) is quite large for a trap microns
above a bulk metal surface. One can reduce the loss by two orders
of magnitude when bulk metal in the vicinity of the trap is
replaced by microstructures. For a thin metallic layer of
thickness $d$, the loss rate (\ref{eq:estimate-gamma}) is
proportional to $d/\height^2$, and for a thin wire (diameter $a
\ll \height$), a faster decrease $\propto a^2 / \height^3$ takes
over (table~\ref{table:henkel}).

\begin{figure}[t]
   \infig{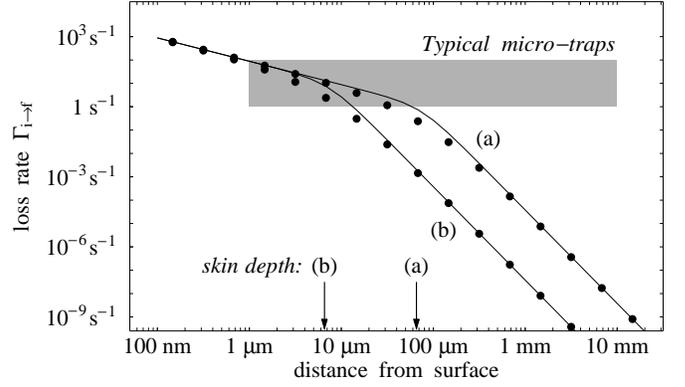}{\columnwidth }
   \caption{Loss rates in a magnetic trap above a copper surface.
   Results for two different Larmor frequencies
   $\omega_L/2\pi = 1$MHz (curve a) and $100$MHz (curve b) are shown.
   The arrows mark
   the corresponding skin depths $\delta( \omega_L )$.
   Eq.(\ref{eq:estimate-gamma}) applies to the region $\height \ll
   \delta( \omega_L )$. See \cite{Hen99-379} for details. Parameters:
   spin $S=1/2$,
   magnetic bias field aligned parallel to the surface. The loss
   rate due to the blackbody field is about $10^{-13} {\rm s}^{-1}$
   at $100$MHz (not shown). Taken from \cite{Hen99-379}.}
   \label{fig:lifetime2}
\end{figure}

The estimates of table~\ref{table:henkel} apply only in an
intermediate distance regime, $a \ll h \ll \delta( \omega_L )$: on
the one hand, when the trap distance $\height$ is smaller than the
size of the metallic structures, one recovers a $1/\height$
behaviour characteristic for a metallic half-space; on the other
hand, steeper power laws take over at large distances, when
$\height$ gets comparable to or larger than the skin depth
\begin{eqnarray}
    \delta( \omega_L ) &=& \sqrt{ 2 \varrho / \mu_0 \omega_L } \nonumber\\
        &=&
    160\mu{\rm m} \, (\varrho / \varrho_{\rm Cu})^{1/2}
    (\omega_L / 2\pi \, 1{\rm MHz})^{-1/2}
    .
\label{eq:estimate-skin}
\end{eqnarray}
Recall that the skin depth characterizes the penetration of
high-frequency radiation into a metal. This crossover can be seen
in Fig.~\ref{fig:lifetime2} where the flip
rate~(\ref{eq:estimate-gamma}) is plotted vs.\ the trap height
$\height$ for a metallic half-space. For details, we refer to
\cite{Hen99-379} and \cite{Hen01-73}. Note that an increase of the
Larmor frequency only helps to reduce the substrate-induced flips
in the regime where $\height \gg \delta( \omega_L )$.
Eq.(\ref{eq:estimate-skin}) shows that this requires, for $\height
\simeq 1\mu{\rm m}$, quite large Larmor frequencies $\omega_{L} /
2\pi \gg 1$GHz, meaning large magnetic (bias) fields.

\vspace{5mm}
\paragraph{Technical noise}
Additional loss processes may be related to fluctuations in the
currents used in the experiment, for example in the chip wires and
in the coils producing the bias and compensation fields. Let us
focus on the wire current and denote $S_{I}( \omega )$ its noise
spectrum. Neglecting the finite wire size, the magnetic field
$B_{w} = \mu_{0} I_{w} / 2\pi \height$ is given by
(\ref{eq:trap_height_in_chap.2}), and we find the following upper
limit for the noise-induced flip rate
\begin{eqnarray}
    \gamma &\simeq& \frac{\mu^2}{ 2 \hbar^2 }
    \left( \frac{\mu_{0}}{2\pi \height} \right)^2 S_{I}( \omega_{L} )
    \nonumber\\
    &\simeq& 2.6{\rm s}^{-1}
    \frac{(\mu / \mu_{B})^2 }{ ( \height / 1\mu{\rm m} )^2 }
    \frac{S_{I}( \omega_{L} ) }{
     S_{\rm SN} }
, \label{eq:technical-flips}
\end{eqnarray}
where the reference value $S_{\rm SN} = 6.4 \times 10^{-19}{\rm
A}^2/{\rm Hz}$ corresponds to shot noise at a wire current of 1A
($S_{\rm SN} = 4 e I_{w}$). This estimate is pessimistic and
assumes equal noise in both field components parallel and
perpendicular to the static trapping field. Nevertheless, it
highlights the need to use `quiet' current drivers for atom chip
traps.
In future chip traps with strong confinement, it may turn out
necessary to reduce current noise below the shot noise level. This
can be achieved with permanent magnets, as discussed in
section~\ref{II-A-10} and reviewed by \cite{Hin99-R119}.
Superconducting wires might provide an alternative solution, but
we are unaware of investigations in this area, an exception being
the relatively old paper by \cite{Var84-4015}.

\subsubsection{Collisional losses}

\paragraph{Background collisions}
Here, collisions between background gas atoms and trapped atoms
endow the latter with sufficient energy to escape the trap. In
order to estimate the loss rate per atom $\gamma$, let us assume
that the background gas is dominated by hydrogen molecules and at
room temperature. We then get:
\begin{eqnarray}
    \gamma &=& n_{\rm bg} \bar v_{\rm bg} \sigma
    \nonumber\\
    & = & 3.8 \times 10^{-3}{\rm s}^{-1}
    \frac{ p_{\rm bg} }{ 10^{-10}{\rm mbar} }
    \frac{ \sigma }{ 1{\rm nm}^2 }
    ,
    \label{eq:collision-loss}
\end{eqnarray}
where $p_{\rm bg}$ is the background pressure. Typical collision
cross sections $\sigma$ are in the $1{\rm nm}^2 = 100\mbox{\AA}^2$
range \citep{Bal99-29}. As a general rule, one gets a trap
lifetime of a few seconds in a vacuum of $10^{-9}$mbar. It is
clear that vacuum requirements will become more stringent as
longer interaction times are required.

\vspace{5mm}
\paragraph{Collisions of trapped atoms}
For traps in UHV conditions, and especially for highly compressed
traps and high density trapped samples, the dominant collisional
loss mechanisms involve collisions between trapped atoms. The
scattering of two atoms leads to a loss rate per atom scaling with
the density, while 3-body collision rates scale with the density
squared.

\vspace{3mm} {\em Spin exchange}  This process corresponds to
inelastic two-body collisions where the hyperfine spin projections
$m_F$ are conserved, but not the spins $F$ themselves. In the
alkali atoms $^7$Li, $^{23}$Na, and $^{87}$Rb, for example, a
collision between two weak field seeking states $| F = 1, m_F =
-1\rangle$ can lead to the emergence of two strong field seekers
$| 2, -1 \rangle$ that are not trapped. This transition requires
an excess energy of the order of the hyperfine splitting to occur,
which is typically not available in cold atom collisions.
Exothermic collisions between the weak field seekers $| 1, -1
\rangle$, $| 2, +1 \rangle$, and $| 2, +2 \rangle$ are not
suppressed, however. The corresponding rate constant is
proportional to $n ( a_S - a_T )^2$ where $a_S$ ($a_T$) are the
scattering lengths in the singlet (triplet) diatomic potential
\citep{Cot94-399}. For $^{87}$Rb, these scattering lengths are
accidentally very close, leading to a very small spin flip rate
\citep{Moe96-R19,Bur97-R2511}. As a consequence, $^{87}$Rb is
practically immune to spin exchange and can form stable
condensates, even of two hyperfine species
\citep{Mya97-586,Jul97-1880,Kok97-R1589}. Spin polarized samples
consisting only of $| 2, +2 \rangle$ cannot undergo spin exchange
because of $m_F$ conservation, the other available states having
smaller $F$. For more details, we refer to the review by
\cite{Wei99-1} and references therein.

\vspace{3mm} {\em Spin relaxation} This process also results from
inelastic two-body collisions, but does not conserve $m_F$. Spin
relaxation is caused by a flip of the nuclear spin and occurs at a
lower rate because of the smaller nuclear magnetic moment. In
$^{87}$Rb for example, the trapped weak field seeker $| 1, -1
\rangle$ may be changed into the untrapped strong field seeker $|
1, +1 \rangle$. More details can be found in the theoretical work
by \cite{Bur97-R2511,Jul97-1880,Tim98-3419}, the experimental work
of \cite{Ger99-1514} and \cite{Sod98-1869} and the review paper by
\cite{Wei99-1}.

\vspace{3mm} {\em Three-body recombination} In this process, two
atoms combine to form a molecule. Although the molecules may have
a definite magnetic moment and be still trapped, the reaction
releases the molecular binding energy that is shared as excess
kinetic energy between the molecule and the third atom. The
binding energy being typically quite large (larger than
100$\mu$eV), both partners escape the trap
\citep{Fed96-2921,Moe96-R19,Moe96-916,Esr99-1751}. For references
to experimental work, see \cite{Bur97-337,Sod99-257} and
\cite{Wei99-1}. We expect three-body processes to be the dominant
collisional decay channel in strongly compressed traps because the
collision rate per atom increases with the square of the atomic
density.

\subsubsection{Tunneling}

Traps very close to the surface might also show loss due to
tunneling of atoms out of the local minimum of the trap towards
the surface.  The rate can be estimated from
 \begin{equation}
 \gamma \sim \omega
 \int_{barrier \; width} \frac{1}{\hbar} \exp (-\sqrt{2m(U
 (z)-E) }) \; dz
  \end{equation}
where $U(z) - E$ is the height of the barrier above the energy of
the trapped particle. Tunneling will therefore only be important
for states close to the top of the potential barrier. Low lying
states in traps where the magnetic field magnitude rises for long
distances will have very little tunneling. Even for atom waveguide
potentials as close as 1$\mu$m from the surface, tunneling
lifetimes of more than 1000s have been estimated
\citep{Sch98-57,Pfau96b}.

\subsubsection{Stray light scattering}
\label{s:optical-flips}

Residual light can flip the atomic spin via optical pumping. For
resonant light, this happens at a rate of the order of $\Gamma
(I_{\rm stray} / I_{\rm sat} )$ where $\Gamma$ is the linewidth of
the first strong electric dipole transition (typically,
$\Gamma/2\pi \simeq 5$MHz) and $I_{\rm sat}$ the saturation
intensity (typically a few mW/cm$^2$). It is highly desirable to
perform atom chip experiments `in the dark': a shielding from any
stray light at the level $10^{-6}\,I_{\rm sat}$ is required for
manipulations on a scale of seconds. For more detailed estimates,
we refer to the review by \cite{Gri00-95} on optical traps. An
overview of the previous loss mechanisms is given in
table~\ref{table:loss}.
\begin{table}[tbf]
   \caption{Loss mechanisms for the atom chip (overview). The columns
   `Scaling' and `Magnitude' refer to loss rates per atom at typical
   atom chip traps: density $n = 10^{10}{\rm
   cm}^{-3}$, height $\height = 10\mu$m,
   trap frequency $\omega/2\pi = 100$kHz,
   Larmor frequency $\omega_L/ 2\pi = 5$MHz.%
   \label{table:loss}}
   \begin{tabular}{lccl}
   Mechanism & Scaling & Magnitude & Workaround
   \\
   \hline
   Spilling over & & & deep trap
   \\
   \hline
  Background collisions\tablenote{Eq.(\ref{eq:collision-loss}).}   & $p_{\rm bg}$ & $0.01{\rm s}^{-1}$ & vacuum \\
   Majorana flips%
   \tablenote{Flip rate~(\ref{eq:Majorana-rate}) from trap
   ground state}   & $\omega \,{\rm e}^{- \omega_L / 2 \omega }$ & $ \simeq 1{\rm s}^{-1}$
   & avoid B=0\\
   %
   Near field & $T_s / \varrho \height^\alpha$ & $10{\rm s}^{-1}$ & little metal \\
   noise%
   \tablenote{Eq.(\ref{eq:estimate-gamma}). The exponent
   $\alpha = 1,2,3$ for metal half-space, layer, and wire
   (see table~\ref{table:henkel}).
   The estimate $10{\rm s}^{-1}$ is for a half-space.}
   \\
   Current noise%
   \tablenote{Eq.(\ref{eq:technical-flips}).}
   & $S_{I}( \omega_{L} ) / \height^2$ & $\simeq 3{\rm s}^{-1}$
   & quiet drivers
   \\
   2-body spin exchange%
   \tablenote{Experimental result for $^{87}$Rb \citep{Mya97-586}}
   & $n$
   & $10^{-4}{\rm s}^{-1}$ & spin polarize
   \\
   2-body spin relaxation%
   \tablenote{Experimental result for Cs and $^7$Li, respectively \citep{Sod98-1869,Ger99-1514}}
   & $n$ &$10^{-2}-10^{-4}$ &
   \\
   3-body collisions%
   \tablenote{Experimental result for $^{87}$Rb and $^7$Li, respectively \citep{Bur97-337,Sod99-257,Ger99-1514}}
   & $n^2$ & $10^{-9}-10^{-7}$ &
   \\
   Tunneling & & $10^{-3}{\rm s}^{-1}$ & thick barrier
   \\
   Stray light & $I_{\rm stray}$ & & keep in dark
   \end{tabular}
\end{table}
We expect that on the route towards $\mu$m sized traps with high
compression, inelastic collisions and magnetic field noise will
dominate the trap loss.

\subsection{Heating}
\label{s:heating}

In the previous section, regarding loss mechanisms, heating was
mentioned in relation to adiabatic compression where some atoms
gain energies larger than the trap depth. Here, we discuss a
different form of heating in which the atom exchanges energy with
the environment. Such heating does not necessarily cause the atom
to be lost, but it is still very harmful as excitations of
vibrational degrees of freedom lead to an ill defined quantum
state of the system. In the case of the atom system and the chip
environment, the environment is always hot compared to the system.
Energy exchange thus increases both the system's mean energy and
its energy spread. In the following, we first describe the
influence of position and frequency noise using the harmonic
oscillator model, then turn to substrate and technical noise, and
finally touch upon the issue of heating due to light fields.

\subsubsection{Harmonic oscillator model} Let us consider the trap
potential to be a one dimensional harmonic potential with angular
frequency $\omega$ and with a ground state size of $a_0 = (\hbar
/(2 \mass \omega))^{1/2}$, where $\mass$ is the mass of the
vibrating atom. Assume for simplicity that the atom is initially
prepared in the oscillator ground state $|0\rangle$. Heating can
occur as a result of a fluctuating trap either in frequency or
position (see for example \cite{Geh98-3914,Tur00-053807}). These
processes may be described by transition rates to higher excited
states of the oscillator. For example, fluctuations in the trap
position (amplitude noise) are equivalent to a force acting on the
atom. They drive the transition $0\to1$ between the ground and
first excited vibrational states, with an excitation rate given by
\citep{Geh98-3914,Hen99-379}
\begin{equation}
\Gamma_{0\to1} = \frac{ a_0^2 }{ 2\hbar^2 } S_F( \omega ) = \frac{
S_F( \omega ) }{ 4 \hbar\omega \mass } \label{eq:heating-rate}
\end{equation}
that is determined by the noise spectrum of the force at the
oscillator frequency $S_F( \omega )$. The rate of energy transfer
to the atom (`heating rate') is simply $\Gamma_{0\to1}\hbar\omega$
or $S_F( \omega ) /4\mass$. Note that this estimate remains valid
for an arbitrary initial state.

We may make contact with the work of \cite{Geh98-3914} by noting
that fluctuations $\Delta x$ of the trap center are equivalent to
a force
\begin{equation}
{F} = \mass \omega^2 \Delta x . \label{eq:force+displacement}
\end{equation}
In terms of the fluctuation spectrum of the trap center $S_{x}(
\omega )$, the excitation rate~(\ref{eq:heating-rate}) is thus
given by
\begin{equation}
\Gamma_{0\to1} = \frac{ \mass \omega^3 }{ 4 \hbar } S_{x}( \omega
) = \frac{ \omega^2 }{ 8 } S_{x / a_{0}}( \omega )
\label{eq:force-displacement-spectra} ,
\end{equation}
which is equivalent to the heating rate (12) of \cite{Geh98-3914},
given our definition~(\ref{eq:def-noise-spectrum}) of the noise
spectrum.

Fluctuations of the trap frequency are described by the
Hamiltonian $\mass x^2 \omega \Delta \omega$ and heat the atom by
exciting the $0 \to 2$ transition. The corresponding transition
rate is \citep{Geh98-3914}
\begin{equation}
\Gamma_{0\to2} = \frac14 S_{\omega}( 2 \omega )
\label{eq:frequency-noise-heating}
\end{equation}
and involves the frequency noise spectrum at twice the trap
frequency. Using the rates given by \cite{Geh98-3914}, one can
show that the heating rate due to frequency fluctuations is equal
to $\Gamma_{0\to2}( 4 \langle E \rangle + \hbar\omega )$, where
the mean energy $\langle E \rangle = \frac12 \hbar \omega$ in the
ground state.

In the following, we differentiate between thermal fluctuations
and technical ones. To get the total heating rate, one simply adds
the force fluctuation spectra $S_F( \omega )$ of all the relevant
sources (e.g. electromagnetic noise from radio stations).

\subsubsection{Thermal fluctuations} Magnetic fields generated by
thermally excited currents in the metallic substrate correspond to
a force given by the gradient of the Zeeman interaction $-
\mbox{\boldmath$\mu$}\cdot{\bf B}$. An explicit calculation of the
magnetic gradient noise gives the following force spectrum
\citep{Hen99-414,Hen99-379}
\begin{equation}
S_{F}( \omega ) = \frac{ \mu_0^2 k_B T_s }{ 32 \pi \varrho }
\frac{ \langle \mbox{\boldmath$\mu$}^2 \rangle + \langle
\mu_{\Vert}^2 \rangle }{ \height^3 } ,
 \label{eq:Zeeman-force-correlations}
\end{equation}
where $\mbox{\boldmath$\mu$}$ is the magnetic moment and
$\mu_{\Vert}$ its component parallel to the static trapping field.
The expression~(\ref{eq:Zeeman-force-correlations}) applies to a
planar metallic substrate (half-space) and an oscillation
perpendicular to its surface. Again, the noise spectrum is
actually frequency independent as long as $\height \ll \delta(
\omega )$ where $\delta( \omega )$ is the skin depth
\ref{eq:estimate-skin}. The average magnetic moment is taken in
the trapped spin state (see \cite{Hen99-379} for details). We thus
obtain the following estimate for the excitation rate
(\ref{eq:heating-rate}):
\begin{eqnarray}
&&\Gamma_{0\to1} \simeq 0.7{\rm s}^{-1} \times
\nonumber\\
&& {} \times \frac{ ( \mu / \mu_B )^2 (T_s / 300{\rm K}) }{ (
\mass / {\rm amu} ) ( \omega / 2\pi \, 100{\rm kHz} ) ( \varrho /
\varrho_{\rm Cu} ) ( \height / 1\mu{\rm m} )^3 } .
\label{eq:estimate-substrate-heating-rate}
\end{eqnarray}
For lithium atoms, a typical trap frequency of $100$kHz and a
height of $\height = 10\mu$m, we estimate a heating rate of
$10^{-4}{\rm s}^{-1}$. For time scales typical of atom chip
experiments (1--100ms), thermal fluctuations thus lead to
tolerable heating only for traps with $h>100$nm.
\begin{figure}
    \infig{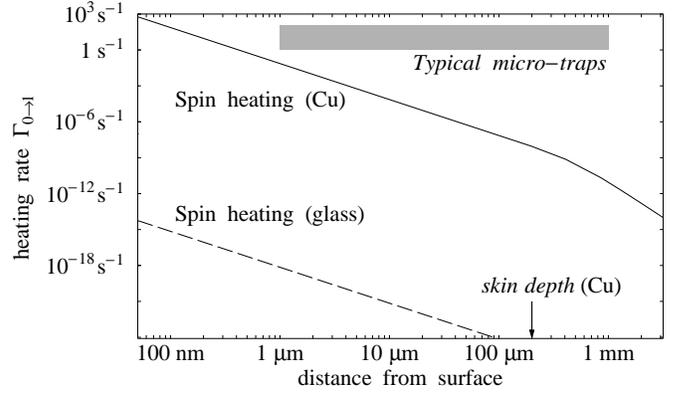}{\columnwidth }
    \caption[]{Heating rate for a trapped spin above copper and glass
    substrates. Parameters: trap frequency $\omega/2\pi = 100{\rm
    kHz}$, $\mass = 40{\rm amu}$, magnetic moment $\mu = \mu_B = $ 1
    Bohr magneton, spin $S = 1/2$. The heating rate due to the
    magnetic blackbody field (not shown) is about $10^{-39} {\rm
    s}^{-1}$. For the glass substrate, a dielectric constant with
    $\mathop{\rm Re}\,\varepsilon = 5$ and a resistivity
    $\varrho = 10^{11}{\rm \Omega}\,{\rm cm}$ are taken.
    Taken from \cite{Hen99-379}.}
    \label{fig:heating}
\end{figure}

\subsubsection{Technical noise}

Heating due to technical noise may arise due to fluctuations in
the currents which are used in the experiment. Noise in the chip
wire currents and in the bias and compensation fields, for
example, randomly shifts the location of the trap center. Let us
focus on fluctuations in the chip wire current $I_w$. Neglecting
finite size effects, the current and the bias field $B_b$ produce
the magnetic trap at a height of $\height = \mu_0 I_w / {2\pi}B_b$
(eq.~\ref{eq:trap_height_in_chap.2}). The conversion from the
current noise spectrum $S_{I}(\omega)$ to the force spectrum
required for the heating rate~(\ref{eq:heating-rate}) is simply
\begin{equation}
S_F( \omega ) = \left( \frac{ \mu_0 \mass \omega^2 }{ 2 \pi B_b }
\right)^2 S_I( \omega ) ,
\end{equation}
and we end up with an excitation rate
\begin{eqnarray}
&&\Gamma_{0\to1} = 2.7 {\rm s}^{-1} \times {} \nonumber
\\
&& {}\times (\mass / {\rm amu}) (\omega / 2 \pi \, 100 {\rm
kHz})^3 \frac{ S_{I}( \omega ) / S_{\rm SN} } { ( B_{b} / {\rm G}
)^2 } . \label{eq:C1-coefficient}
\end{eqnarray}
The reference $S_{\rm SN}$ for the current noise is again the shot
noise level at $I_{w} = 1$A. Note that this rate increases with
the trap frequency: while a strong confinement suppresses heating
from thermal fields
(eq.~\ref{eq:estimate-substrate-heating-rate}), the inverse is
true for trap position fluctuations. This is because in a
potential with a large spring constant, position fluctuations
translate into large forces (eq.~\ref{eq:force+displacement}).
Typical trap parameters ($\omega/2\pi = 100$kHz, $B_b=50$G) lead
for $^7$Li atoms to an excitation rate of $\approx 7.5\times
10^{-3}{\rm s}^{-1} \times S_{I}(\omega) / S_{\rm SN}$. This
estimate shows that even for very quiet currents technical noise
is probably the dominant source of heating on the atom chip.

The fluctuations of the trap center (location proportional to $I_w
/ B_b$) can be reduced by correlating the currents of the bias
field coils and the chip wire so that they have the same
fluctuations, up to shot noise. Heating due to fluctuations in the
trap frequency may then be relevant, as $\omega$ is proportional
to $B_b^2/I_w$ (eq.~\ref{trap_omega_chap.2}). Let us again
calculate an example. For a fixed ratio $I_w / B_b$ (due to
correlated currents), we find for the relative frequency
fluctuations
\begin{equation}
\frac{ \Delta \omega }{ \omega } = \frac{ \Delta I }{ I_w }
\end{equation}
and hence an excitation rate (\ref{eq:frequency-noise-heating})
\begin{equation}
\Gamma_{0\to2} \simeq 10^{-7}{\rm s}^{-1} \frac{ (\omega / 2\pi\,
100 {\rm kHz})^2 }{ ( I_w / {\rm A} )^2 } \frac{ S_I( 2 \omega )
}{ S_{\rm SN} } . \label{eq:estimate-frequency-heating}
\end{equation}
Typical atom chip parameters ($\omega/2\pi = 100$kHz, $I_w=1{\rm
A}$) lead to $\Gamma_{0\to2} \simeq 10^{-7}{\rm s}^{-1}\times S_I(
2\omega )/S_{\rm SN}$ which is negligible when compared to the
rate obtained in (\ref{eq:C1-coefficient}).

\subsubsection{Light heating}

Another source of heating are the external light fields with which
the atoms are manipulated and detected. Here the Lamb Dicke
parameter $\eta$ is a convenient tool, where
\begin{equation}
\eta = \frac{2\pi a_0}{\lambda}
\end{equation}
is the ratio between the ground state size of the trap $a_0$ and
the wave length of the impinging wave. This becomes clear if we
remember that the probability not to be excited $P_{0\rightarrow
0}$ is simply the well known Debye-Waller factor
\begin{equation}
\exp( - \Delta k^2 a_{0}^2 ) \simeq \exp(-\eta^2) ,
\end{equation}
where $\Delta k \approx k$ is the momentum loss of the impinging
photon. Hence, if the atoms are confined below the photon wave
length (the so-called Lamb-Dicke limit $\eta < 1$), they will not
be heated by light scattering. Loss via optically induced spin
flips is still relevant, however, as discussed in
section~\ref{s:optical-flips} and reviewed by \cite{Gri00-95}.

\vspace{7mm}

In table~\ref{table:heating} we give an overview of the heating
mechanisms discussed above. For microscopic traps, we expect noise
from current fluctuations and (to a lesser extent) from the
thermal substrate to be the dominant origins of heating. Note the
scaling with the trap frequency: trap fluctuations due to
technical noise become more important for guides with strong
confinement.

In this subsection, we have restricted ourselves to heating due to
single-atom effects. Collisions with background gas atoms also
lead to heating and rate estimates have been given by
\cite{Bal99-29}. Finally, in a Bose condensate on chip,
fluctuating forces may be expected to drive collective and
quasiparticle excitations, leading to a depletion of the
condensate ground state \citep{Hen02b}. This area deserves further
study in the near future.

\begin{table}[tbf]
\caption{Heating mechanisms for the atom chip (overview). The
columns `Scaling' and `Magnitude' refer to transition rates from
the ground state of a typical atom chip trap: lithium atoms,
height $\height = 10\mu$m, trap frequency $\omega/2\pi = 100$kHz.
Harmonic confinement is assumed throughout.%
\label{table:heating}}
\begin{tabular}{lccl}
Noise & Scaling & \hspace{5mm} Rate \hspace{5mm} & Workaround \\
      & & ${\rm s}^{-1}$ &
\\
\hline
\rule{0pt}{2.8ex}%
Near field%
\tablenote{Eq.(\ref{eq:estimate-substrate-heating-rate}), for a
metal half-space.} & $T_s / \omega \varrho \height^{3}$ &
$10^{-4}$ &
\\
Current%
\tablenote{Eq.(\ref{eq:C1-coefficient}). Note the scalings $\omega
\sim B_b^2/I_w$ and $\height \sim I_w/B_b$ for trap frequency and
height.} & $\omega^3 S_I / B_b^2 $ &
$1$ & correlate \\
 & $ \sim \omega S_I / \height^2$& & currents
\\
Trap frequency%
\tablenote{Eq.(\ref{eq:estimate-frequency-heating}).} & $\omega^2
S_I / I_w^2 $ & $10^{-5}$ &
\\
Light scattering & $1/ \omega \lambda^2$ & & reduce \\
 & $\sim S_I / \height^4$ & &stray light
\end{tabular}
\end{table}

\subsection{Decoherence}
\label{s:decoherence}

We now turn to the destruction of quantum superpositions or
interferences due to the coupling of the atom cloud to the noisy
chip environment. This is an important issue when coherent
manipulations like interferometry or qubit processing are to
succeed on the atom chip. With chip traps being ever closer to the
chip substrate, thermal and technical magnetic noise is expected
to contribute seriously to decoherence, as it does to loss and
heating processes.

The theoretical framework for describing decoherence makes use of
the density matrix for the trapped atoms.  Its diagonal elements
give the occupation probabilities, or populations, in some
preferred basis, usually the stationary trap states.  Their
evolution has been discussed in the previous subsections in terms
of simple rate equations. Decoherence deals with the decay of
off-diagonal elements, or coherences, of the density matrix. Their
magnitude can be related to the fringe contrast one obtains in an
interference experiment. Magnetic fluctuations typically affect
both populations and coherences: field components perpendicular to
the trapping fields redistribute the populations and parallel
components suppress the coherences. The latter case illustrates
that decoherence can occur even without the exchange of energy,
because it suffices that some fluctuations randomize the relative
phase in quantum superposition states \citep{Ste90-3436}. Such
fluctuations are sometimes called `phase noise'.

In this subsection we consider first the decoherence of internal
atomic states and then describe the impact of fluctuations on the
center-of-mass. In the same way as for the heating mechanisms, we
leave aside the influence of collisions on decoherence, nor do we
consider decoherence in Bose-Einstein condensates.

\subsubsection{Internal states}

The spin states of the trapped atom are promising candidates for
the implementation of qubits. Their coherence is reduced by
transitions between spin states, induced by collisions or noise.
The corresponding rates are the same as for the loss processes
discussed in subsec.~\ref{s:loss2}.

In addition, pure phase noise occurs in the form of fluctuations
in the longitudinal magnetic fields (along the direction of the
trapping field). These shift the Larmor frequency in a random
fashion and hence the relative phase between spin states. The
corresponding off-diagonal density matrix element (or fringe
contrast) is proportional to $\langle \exp( {\rm i} \Delta\varphi
) \rangle$ where $\Delta\varphi$ is the phase shift accumulated
during the interaction time $t$. A `decoherence rate' $\gamma_{\rm
dec}$ can be defined by
\begin{equation}
    \gamma_{\rm dec} =
    \frac{ \langle \Delta \varphi^2 \rangle }{ 2 t }
    = \frac{ S_{\dot\varphi}( \omega \to 0 ) }{ 4 }
, \label{eq:def-dephasing-rate}
\end{equation}
where $S_{\dot\varphi}( \omega )$ is the spectrum of the frequency
fluctuations. Two spin states $| m_{F} \rangle, | m_{F}' \rangle$,
for example, `see' a frequency shift $\dot\varphi(t) = g \mu_{B} (
m_{F} - m_{F}' ) \Delta B_{\Vert}(t) / \hbar$, that involves the
differential magnetic moment and the component $\Delta B_\Vert(t)$
of the magnetic field noise parallel to the trap field. The
spectrum $S_{\dot\varphi}( \omega )$ is then proportional to the
spectrum of the magnetic field fluctuations.

Eq.~(\ref{eq:def-dephasing-rate}) is derived in a rotating frame
where the phase shift has zero mean and making the assumption that
the spectral density $S_{\dot\varphi}( \omega )$ is essentially
constant in the frequency range $\omega \le 1/t$. The noise then
has a correlation time much shorter than the interaction time $t$.
We consider, as usual in theory, that $\Delta\varphi$ is a random
variable with Gaussian statistics, and get a fringe contrast
\begin{equation}
\langle {\rm e}^{ {\rm i} \Delta \varphi } \rangle = {\rm e}^{ -
\gamma_{\rm dec} t } \label{eq:fringe-contrast}
\end{equation}
that decays exponentially at the
rate~(\ref{eq:def-dephasing-rate}).

Let us give an estimate for the decoherence rate due to magnetic
noise. If $\Delta{\bf B}( {\bf r}, t )$ are the magnetic
fluctuations at the trap center, the shift of the Larmor frequency
is given by
\begin{equation}
    \Delta\omega_{L}( t ) = -
    \frac{ \langle i | \mu_{\Vert} | i \rangle }{ \hbar }
    \Delta{B}_{\Vert}( {\bf r}, t )
    \label{eq:Larmor-shift}
.
\end{equation}
Here, the average magnetic moment is taken in the spin state $| i
\rangle$ trapped in the static trap field, thus picking the
component $\Delta B_\Vert$ parallel to the trap field. The noise
spectrum of this field component, for thermal near field noise, is
of the same order of magnitude as for the perpendicular component
\citep{Hen99-379} and depends only weakly on frequency. We thus
get a decoherence rate comparable to the loss
rate~(\ref{eq:estimate-gamma}), typically a few $1{\rm s}^{-1}$.
The same argument can be put forward for fluctuations in the wire
current and the bias field. Assuming a flat current noise spectrum
at low frequencies, we recover the
estimate~(\ref{eq:technical-flips}) for spin flip loss (a few
$1{\rm s}^{-1}$). Therefore, keeping the atoms in the trap, and
maintaining the coherence of the spin states requires the same
effort.

We finally note that near field magnetic noise also perturbs the
coherence between different hyperfine states that have been
suggested as qubit carriers. Although these states may have the
same magnetic moment (up to a tiny correction due to the nuclear
spin), excluding pure phase noise, their coherence is destroyed by
transitions between hyperfine states. The corresponding loss rate
(relevant, e.g., for optical traps) has been computed by
\cite{Hen99-379} and is usually smaller than the spin flip rate.

\subsubsection{Motional decoherence}
The decoherence of the center-of-mass motion of a quantum particle
has been put forward as an explanation for the classical
appearance of macroscopic objects since the work of
\cite{Zeh70-69} and \cite{Zur91-36} (see also the book by
\cite{Giu96}). It has be shown that the density matrix of a free
particle subject to a random force field evolves into a diagonal
matrix in the position basis \citep{Zur91-36}
\begin{equation}
   \rho( z, z', t) \simeq \rho( z, z', 0)
   \exp\left[-\frac{(z - z')^2 D t}{
   \hbar ^2}\right]
   .
   \label{eq:free-particle-decoherence}
\end{equation}
Here, the distance $z - z'$ denotes how off-diagonal the element
is, and $D$ is the momentum diffusion coefficient. The `coherence
length' thus decreases like
\begin{equation}
   L_c = \frac{\hbar}{\sqrt{D t}}
   .
   \label{eq:free-particle-coherence-length}
\end{equation}
At the same time, the momentum spread $\Delta p \simeq (2 D t
)^{1/2}$ increases, so that the relation $\Delta p L_{c} \simeq
\hbar$ is maintained at all times. For a particle trapped in a
potential, the density matrix tends to a diagonal matrix in the
potential eigenstate basis if the timescale for decoherence is
large compared to the oscillation time $2\pi/\omega$. Typically,
this applies be for the oscillatory motion in atom chip
waveguides. The opposite case of `fast decoherence' is discussed
by \cite{Zur93-1187} and \cite{Paz93-488} and leads to the
`environment-induced selection' of minimum uncertainty (or
coherent) states.

In the following we discuss different decoherence mechanisms for a
typical separated path atom interferometer on the atom chip.

\subsubsection{Longitudinal decoherence}
\label{s:long-dec}

We focus first on the quasi-free motion along the waveguide axis
(the $z$-axis), using the free particle model mentioned above.
Decoherence arises again from magnetic field fluctuations due to
thermal or technical noise. The corresponding random potential is
given by~(\ref{eq:Larmor-shift}):
\begin{equation}
   V( {\bf r}, t ) = - \langle i | \mu_{\Vert} | i \rangle \Delta
   {B}_{\Vert}( {\bf r}, t )
   ,
   \label{eq:noise-potential}
\end{equation}
where we retain explicitly the position dependence.
\cite{Hen01-73} have shown that for white noise, the density
matrix in the position representation behaves as
\begin{equation}
   \rho( z, z', t ) = \rho( z, z', 0) \exp{\left( -
   \gamma_{\rm dec}(s) t \right)}
   ,
   \label{eq:analytic-solution-2}
\end{equation}
where the decoherence rate $\gamma_{\rm dec}(s)$ depends on the
spatial separation $s = z - z'$ between the two parts of the
atomic wave function being observed:
\begin{equation}
   \gamma_{\rm dec}(s) =
   \frac{ 1 - C(s) }{ 2\hbar^2 } S_V( \height; \omega \to 0 )
   .
   \label{eq:decoherence-rate}
\end{equation}
Here, $C(s)$ is the normalized spatial correlation function of the
potential (equal to unity for $s = 0$), and the noise spectrum
$S_V( \height; \omega \to 0 )$ characterizes the strength of the
magnetic noise at the waveguide center.

For an atom chip waveguide perturbed by magnetic near field noise,
the decoherence rate is of the order of
\begin{equation}
\gamma = \frac{ \langle \mu_{\Vert} \rangle^2 S_B^\Vert( \height;
\omega \to 0 ) }{ 2 \hbar^2 } \label{eq:decoh-rate-2}
\end{equation}
and hence comparable to the spin flip rate~(\ref{eq:flip-rate},
\ref{eq:estimate-gamma}). Decoherence should thus typically occur
on a timescale of seconds. The correlation function $C(s)$ is well
approximated by a Lorentzian, as shown by \cite{Hen00-57}, and the
decoherence rate~(\ref{eq:decoherence-rate}) can be written
\begin{equation}
\gamma_{\rm dec}( s ) = \frac{ \gamma s^2 }{ s^2 + l_c^2 }
\label{eq:decoh-rate-s}
\end{equation}
where $l_c$ is the correlation length of the magnetic noise. This
length can be taken equal to the height $\height$ of the waveguide
above the substrate \citep{Hen00-57}. This is because each volume
element in the metallic substrate generates a magnetic noise field
whose distance-dependence is that of a quasi-static field (a
$1/r^2$ power law). Points at the same height $\height$ above the
surface therefore see the same field if their distance $s$ is
comparable to $\height$. At distances $s \gg h$, the magnetic
noise originates from currents in uncorrelated substrate volume
elements, and therefore $C(s) \to 0$. The corresponding saturation
of the decoherence rate~(\ref{eq:decoh-rate-s}), $\gamma_{\rm
dec}( s \gg l_c ) \to \gamma$, has also been noted, for example,
by \cite{Che99-4807}.

Decoherence due to magnetic noise from technical sources will also
happen at a rate comparable to the corresponding spin flip rate,
as estimated in~(\ref{eq:technical-flips}). The noise correlation
length may be comparable to the trap height because the relevant
distances are below the photon wavelength at typical
electromagnetic noise frequencies, so that the fields produced by
wire current fluctuations are quasi-static, and the same argument
applies. The noise correlation length of sources like the external
magnetic coils will, of course, be much larger because these are
far away from the waveguide. These rough estimates for the spatial
noise properties of currents merit further investigation, in
particular at the shot noise level.

\begin{figure}
   \infig{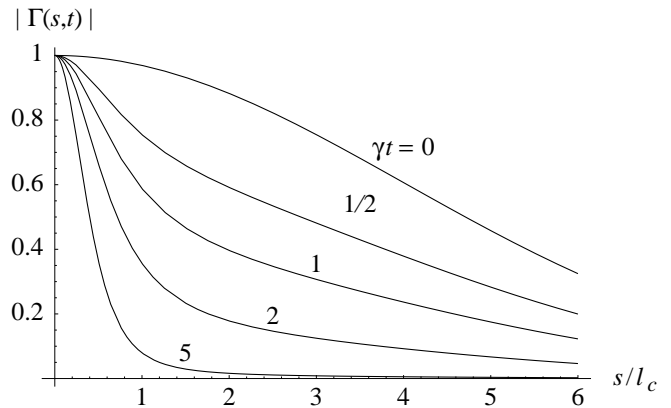}{\columnwidth }
   \caption[]{Illustration of spatial decoherence in an atomic wave guide.
   The spatially averaged coherence function $\Gamma(s, t) =
   \int \!dz \, \rho( z + s, z, t )$ is
   plotted vs. the separation $s$ for a few times $t$. Space is
   scaled to the field correlation length $l_c$ and time to the
   `scattering time' $1/\gamma \equiv 1/\gamma_{\rm dec}( \infty )$.
   A Lorentzian correlation function for the perturbation is assumed.
   Taken from \cite{Hen01-73}.}
   \label{fig:henkel}
\end{figure}

Spatial decoherence as a function of time is illustrated in
Fig.~\ref{fig:henkel} where the density matrix $\rho( z + s, z,
t)$ averaged over $z$ is plotted. Note that this quantity will be
directly proportional to the visibility of interference fringes
when two wavepackets with a path difference $s$ are made to
overlap and interfere. One sees that for large splittings $s \gg
l_c$, the coherence decays rapidly on the timescale $1/\gamma$
given in~(\ref{eq:decoh-rate-2}). This is because the parts of the
split wavepacket are subject to essentially uncorrelated noise. In
a typical waveguide at height $\height = 10\mu$m, fringe contrast
is thus lost after about a second (the spin lifetime) for path
differences $s \gg 10\mu$m. Increasing the height to $\height =
100\mu$m decreases $\gamma$ by at least one order of magnitude as
shown by~(\ref{eq:estimate-gamma}). In addition, the correlation
length grows to $100\mu$m, and larger splittings remain coherent.
Alternatively, one can choose smaller splittings $s \ll l_c$ which
decohere more slowly because the interferometer arms see
essentially the same noise potential. Note, however, that the spin
lifetime will always be the upper limit to the coherence time of
the cloud.

The previous theory allows to recover the decoherence model of
eq.~(\ref{eq:free-particle-decoherence}) at long times $t >
1/\gamma$. In this limit, only separations $s < l_c$ have not yet
decohered, and we can make the expansion
\begin{equation}
\gamma_{\rm dec}( s ) \approx  \gamma \frac{ s^2 }{ l_c^2 }
\label{eq:decoh-rate-s2}
\end{equation}
for the decoherence rate~(\ref{eq:decoh-rate-s}). From the density
matrix~(\ref{eq:analytic-solution-2}), we can then read off the
momentum diffusion constant $D = \hbar^2 \gamma / l_c^2$.

\subsubsection{Transverse decoherence}
We finally discuss the decoherence of a spatially split wavepacket
in an atom chip interferometer, as described in
section~\ref{II-A-9}. The excitation of transverse motional states
in each arm suppresses this decoherence at the same rate as the
heating processes discussed in subsec.~\ref{s:heating} (about
$1{\rm s}^{-1}$). Note that due to the transverse confinement, the
relevant noise frequencies are shifted to higher values compared
to the longitudinal decoherence discussed before.

The coherence between the spatially separated interferometer arms
is suppressed in the same way as the longitudinal coherence
discussed in section~\ref{s:long-dec}. To show this, we use an
argument based on phase noise and focus again on magnetic field
fluctuations, either of thermal or technical origin. Magnetic
fluctuations affect both the bottom of the trap well and the
transverse trap frequency, but are only relevant when they differ
in the spatially separated arms. The well bottoms get
differentially shifted from an inhomogeneous bias field, e.g.,
while the trap frequency shifts due to changes in the field
curvature.

We generalize the formula~(\ref{eq:def-dephasing-rate}) to a phase
shift $\Delta \varphi$ that is the accumulation of energy level
differences $\Delta E(t)$ along the paths in the two arms. The
decoherence (or dephasing) rate is thus given by
\begin{equation}
   \gamma_{\rm dec} =
   \frac{ S_{\Delta E}( \omega \to 0 ) }{ 4 \hbar^2 }
   ,
\end{equation}
where $S_{\Delta E}( \omega \to 0 )$ is the spectral density of
the energy difference, extrapolated to zero frequency.

To make contact with the density matrix formulation of
eq.~(\ref{eq:analytic-solution-2}), we write $\Delta E(t) = E_R(t)
- E_L(t)$ where $E_{R,L}(t)$ are the energy shifts in the right
and left interferometer arms that are `seen' by an atom travelling
through the waveguide. We find
\begin{eqnarray}
   && \left\langle \Delta E(t) \Delta E(t') \right\rangle =
   \left\langle E_R(t) E_R(t') \right\rangle + \left\langle E_L(t)
   E_L(t') \right\rangle
   \nonumber\\
   && \qquad - \left\langle E_R(t) E_L(t') \right\rangle - \left\langle
   E_L(t) E_R(t') \right\rangle
   ,
\end{eqnarray}
where the last two terms contain the correlation between the noise
in both arms. They may therefore be expressed through the
normalized correlation function $C_{RL} \equiv C(s)$ with $s$ the
separation between the left and right arms. The reasonable
assumption that both arms `see' the same white noise spectrum, say
$S_{E}(\omega)$, yields
\begin{eqnarray}
   \left\langle \Delta E(t) \Delta E(t') \right\rangle & = &
   ( 1 - C( s ) )
   S_{E}( \omega \to 0) \, \delta( t - t' )
   ,
   \\
   \gamma_{\rm dec}
   = \gamma_{\rm dec}( s ) &=&
   \frac{ 1 - C( s ) }{ 2 \hbar^2 }
   S_{E}(\omega \to 0)
   ,
   \label{eq:transverse-dec-rate}
\end{eqnarray}
where we recover the decoherence rate~(\ref{eq:decoherence-rate})
obtained for the quasi-free longitudinal motion. We also recover
the trivial result that the contrast stays constant if both
interferometer arms are subject to the same noise amplitude
(perfect correlation $C(s) = 1$).

The previous argument shows that transverse and longitudinal
coherence are affected in a similar way by magnetic noise. Again,
near field noise is a serious threat due to its short correlation
length. Since the decoherence rate is so small that $\gamma_{\rm
dec}( \infty ) t \ll 1$ for interaction times not longer than a
few hundred ms, the phase noise remains small even for widely
separated arms subject to decorrelated noise (separation larger
than the guide height). This is a worst-case estimate: a more
careful approach would take into account the form of the
interferometer, where the arm separation is not constant. Current
noise should neither be underestimated. It is certainly possible
to reduce dephasing by feeding the same current through both left
and right wire guides, as shown by~(\ref{eq:transverse-dec-rate}).
But this does not seem to help at the shot noise level because
each electron randomly follows one or the other wire. The wire
current fluctuations are thus uncorrelated, leading to a
transverse decoherence rate comparable to the longitudinal
decoherence rate. Both rates are thus of the order of the flip
rate~(\ref{eq:technical-flips}), typically a few ${\rm s}^{-1}$.

Let us estimate as another example the dephasing due to technical
noise in a magnetic field gradient. This may be introduced by an
imperfect Helmholtz configuration or coil misalignment. For small
gradients ${\bf b}$, we have
\begin{equation}
   \Delta E(t) = \langle \mu_{\Vert} \rangle \, {\bf s} \cdot
   {\bf b}(t)
\end{equation}
where ${\bf s}$ is the spatial separation between the
interferometer arms. To be precise, ${\bf b}(t)$ gives the
gradient of the bias field component along the direction of the
(static) trapping field. Ignoring a possible anisotropy in the
gradient noise, we find the estimate
\begin{equation}
\gamma_{\rm dec}( s ) \simeq \frac{ \langle \mu_b \rangle^2 s^2 }{
4 \hbar^2 } S_{b}( \omega \to 0 ) ,
\end{equation}
where $S_b( \omega )$ is related to the power spectrum of the
current difference in the Helmholtz coils. We may take as the
worst case completely uncorrelated Helmholtz currents, and a
magnetic gradient $b \simeq B_{b}/R$ where $R$ is the size of the
Helmholtz coils. The dephasing rate is then of the order of
\begin{equation}
   \gamma_{\rm dec}( s ) \simeq 10^{-6} {\rm s}^{-1} \,
   \frac{\langle \mu_b \rangle^2 }{ \mu_B^2 }
   \frac{ s^2 }{ R^2 }
   \frac{ ( B_{b} / G )^2 }{ ( I_b / A )^2 }
   \frac{ S_I( \omega \to 0 ) }{ S_{\rm SN} }
   ,
   \label{eq:dephasing-field-gradients}
\end{equation}
where $I_{b}, \, B_{b}$ are the Helmholtz current and the bias
field. The experimentally reasonable parameters $I_{b}=1{\rm
A},\,s=100\mu{\rm m},\,R=10{\rm cm},\, B_{b}=10{\rm G}$ yield the
small value $\gamma_{\rm dec}( s ) \simeq 10^{-10}{\rm
s}^{-1}\times S_I( \omega \to 0 ) / S_{\rm SN}$. We note that the
residual gradient of imperfect Helmholtz coils is usually less
than $0.1$G/cm which is an order of magnitude below the estimate
$B_{b}/R=1$G/cm taken here.

Finally, let us estimate the phase noise due to fluctuations in
the spring constant of the guide potential. Even in the adiabatic
limit where the transitions between transverse quantum states are
suppressed (no heating), these fluctuations shift the energy of
the guided state. In the harmonic approximation, we have for the
ground state of the guide \( \Delta E = \frac12 \hbar \Delta
\omega \) where $\Delta\omega$ is the relative shift of the
vibration frequency. This gives a dephasing rate
\begin{equation}
\gamma_{\rm dec} = \frac{ 1 }{ 16 } S_{\omega}( \omega \to 0 )
\label{eq:frequency-phase-noise} .
\end{equation}
We have neglected noise correlations between the interferometer
arms that would reduce decoherence because of correlated phase
shifts in both arms. The rate~(\ref{eq:frequency-phase-noise}) is
of the same order as the heating
rate~(\ref{eq:frequency-noise-heating},
\ref{eq:estimate-frequency-heating}) due to frequency noise
($\simeq 10^{-5}{\rm s}^{-1}$). It thus appears that fluctuations
of the trap frequency have a larger impact than bias field
gradients, but still they lead to negligible dephasing.

In table~\ref{table:dec} we give an overview of the different
decoherence mechanisms discussed in this subsection. For
interferometers with large path differences (compared to the
waveguide height), we expect current shot noise and thermal near
field fluctuations to be the dominant sources of decoherence. They
appear quite `rough' (small correlation length) and perturb both
the quasi-free motion along the waveguide axis and the relative
phase between spatially separated wavepackets in an
interferometer. An increase in the trap frequency does not help,
rather the amount of metallic material in the vicinity of the
guide should be kept to a minimum.


\begin{table}[btf]
    \caption{Decoherence mechanisms for atom chip interferometers
(overview). The column `Magnitude' refers to the decoherence rate
$\gamma_{\rm dec}( s )$ for a typical guided interferometer:
lithium atoms, height $\height = 10\mu$m, separation $s =
10\mu{\rm m}$, transverse guide frequency $\omega/2\pi = 100$kHz.
Along the waveguide axis, the
atomic motion is free.%
\label{table:dec}}
\begin{tabular}{lccl}
Noise mchanism& Scaling & \hspace{3mm} Rate \hspace{3mm} & Workaround \\
  &   & ${\rm s}^{-1}$ &
\\
\hline
\rule{0pt}{2.8ex}%
Substrate fields%
\tablenote{Exponent $\alpha = 1, 2, 3$ for metal half-space,
layer, and wire (eq.~\ref{eq:estimate-gamma} and
table~\ref{table:henkel}).} & & &
\\
$\quad s \ll \height$ & $T_s s^2 / \varrho \height^{\alpha+4}$ &
$\ll 10$ & little metal,
\\
$\quad s \gg \height$ &$T_s / \varrho \height^{\alpha+2}$ & $10$ &
small splitting
\\
Current %
\tablenote{Eq.(\ref{eq:C1-coefficient}).} & $\omega^3 S_I / B_b^2$
& $1{\rm s}^{-1}$ & correlate currents
\\
Bias fluctuations%
\tablenote{Eq.(\ref{eq:dephasing-field-gradients}). The bias field
scales as $B_b \sim I_b / R$ where $R$ is the size of the bias
coils.} & $s^2 B_b^2 S_I / R^2 I_b^2 $ & $10^{-8}$ &
\\
 & $\sim s^2 S_I / R^4$\\
Trap frequency %
\tablenote{Eq.(\ref{eq:estimate-frequency-heating}).} & $\omega^2
S_I / I_w^2 $ & $10^{-5}$ & \\
 & $\sim S_I / \height^4$
\end{tabular}
\end{table}

\section{Vision and outlook}
\label{s:outlook}

Much has been achieved in the field of micro-optics with matter
waves in the last 10 years. We have seen a steady development from
free standing wires to micron size traps and guides, from trapping
thermal atoms to the creation of BEC on an atom chip.  Where to go
from here?  What can we expect from future integrated matter wave
devices? There are still many open questions before we can assess
the full promise of integrated microscopic atom optics.

In the following paragraphs we try to pinpoint the relevant future
developments and directions. Some of them like the study of the
influence of the warm thermal surface and the fundamental noise
limits on lifetime, heating and coherence of atoms, are already
under way. Hopefully in a few years we will know how far micro
manipulation of atoms on chips can be pushed.

\subsection{Integrating the atom chip}

\subsubsection{Chip fabrication technology}

We will see continued development of atom chip fabrication
techniques.  Depending on how close to the surface one is able to
place atoms before significant decoherence occurs, the commonly
used technology will be either state of the art nano fabrication
with scale limits below 100nm or thicker and larger wires built by
a combination of less demanding techniques. Another limitation
would be smoothness of fabrication: as fluctuations in wire widths
would cause changing current densities and consequently changing
trap frequencies,  potential ´hills' may appear which may be large
enough to hinder the transport of a BEC or control its phase
evolution (for the same reason bias field stability will have to
be improved in the future).

In the near future many advances are expected. One of the first
steps will be to build multi layer structures that will enable for
example crossing wires in order to realize more elaborate
potentials and give more freedom for atom manipulation.

Thin film magnetic materials should allow to build permanent
magnetic microscopic devices, which can be switched on and off for
loading and manipulation of atoms.  Such structures would have the
advantage that the magnetic fields are much more stable, and
consequently one can expect much longer coherence times, when
compared with current generated fields.

\subsubsection{Integration with other techniques}

With cold atoms trapped close to a surface, integration with many
other techniques of atom manipulation onto the atom chip is
possible.

One of the first tasks will be to integrate present day atom chips
with existing micro optics (see for example \cite{Bir01-67}) and
solid state optics (photonics), for atom manipulation and
detection. We envision for example micro fabricated wave guides
and/or micro fabricated lenses on the atom chip for bringing to
and collecting light from atoms in the atom optical circuits.

Light can also be used for trapping \citep{Gri00-95}. Having cold
atoms close to a surface will allow efficient transfer and precise
loading of atoms into light surface traps, which would be
otherwise difficult because of their small volume and inaccessible
location. For example an atom chip with integrated micro optics,
will allow to load atoms into evanescent wave guides and traps, as
proposed by \cite{Bar00-023608}.  Such traps and guides would be a
way to circumvent the decoherence caused by Johnson noise in a
warm conducting surface (section~\ref{s:loss}).

With the standing wave created by reflecting light off the chip
surface one will be able to generate 2-dimensional traps with
strong confinement in one direction, resembling quantum wells, as
demonstrated by \cite{Gau98-5298}. Adding additional laser beams
or additional electrodes on the surface restrict the atomic motion
further, yielding 2-dimensional devices as in quantum electronics
\citep{Imr87}. Similarly one can build and load optical lattices
close to the surface where each site can be individually addressed
by placing electrodes on the chip next to each site.

In principle, many other quantum optical components can be
integrated on the atom chip. For example high Q cavities combined
with micro traps will allow atoms to be held inside the cavity to
much better than the wavelength of light providing a strong
coupling between light and atoms. For recent experimental work
concerning the manipulation and detection of atoms in cavities, we
refer the reader to
\cite{Ber94,Pin00-365,Hoo00-1447,Osn01-037902,Gut01-49}.

\begin{figure}[tb]
   \infig{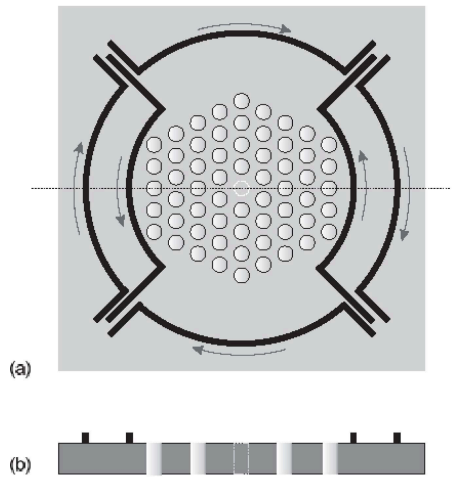}{0.8 \columnwidth }
   \caption{
   A proposed implementation of an integrated nanofabricated high-Q cavity from Cal-Tech.
   The cavity is made of a 2D photonic crystal utilizing holes with diameters of order 100nm.
   A Weinstein-Libbrecht type Ioffe magnetic trap will hold the atom in the cavity.
   Courtesy H. Mabuchi.}
   \label{fig:mabuchi}
\end{figure}

Regarding cavities one can think of examining a wide variety of
technologies ranging from standard high Q cavities consisting of
macroscopic mirrors to optical fiber cavities (with Bragg
reflectors or with mirrors on the ends); from photonic band gap
structures to micro cavities like micro spheres and micro discs
fabricated from a suitable transparent material. One proposed
implementation is presented in Fig. \ref{fig:mabuchi}
\citep{Mab01-7}.

\subsubsection{\label{VI-det}Atom detection}
For future applications, it would be advantageous to have a state
selective single atom detector integrated on the atom chip. Such
detectors could be based on different methods. The most direct
method would be to detect the fluorescent light of the atom using
surface mounted micro optics. More accurate non-destructive
methods could be based on measuring an optical phase shift induced
by an atom in a high Q cavity.

\paragraph{Single atom detection using near field radiation:}
To detect light scattered from single atoms near a chip surface,
the main challenge will be to minimize the stray light scattered
from the surface. One possible solution may be to collect a large
fraction of the light scattered by the atom using near field
apertures and/or confocal microscope techniques. An atom could
also be used to couple light between two wave guides, as used in
some micro optic detectors for molecules and directional couplers
in telecommunication.

\paragraph{Detecting single atoms by selective ionization:} This may be
achieved using a multi step process up to a Rydberg state. The
electron and the location from where it came can then be detected
with a simple electron microscope.  Using a dipole blockade
mechanism as discussed by \cite{Luk01-037901} one should be able
to implement an amplification mechanism, which will allow 100\%
detection efficiency \citep{Sch02-gdansk}

\paragraph{Transmission of resonant light through a small cavity:}
Such a scheme may be used to detect single atoms even for moderate
Q values of the cavity. The cavity could be created by two fibers
with high reflectivity coatings at the exit facets, or even by a
DBR fiber cavity with a small gap for the cold atoms. Fiber ends
molded in a lens shape could considerably reduce the light losses
due to the gap. Having atoms localized in steep traps should allow
a small gap that would reduce the losses even further.

\paragraph{Transmission of light through a high Q cavity:} Here, the transmission is
modified by the presence of single atoms. In the high Q case, the
light may be quite far from atomic resonance and the atoms are
still detected with high probability. The basic mechanism of this
detector is that atoms inside the cavity change the dispersion for
the light. The high Q value makes it possible to detect very small
modifications of the dispersion. In addition the cavities can be
incorporated into integrated optics interferometers to measure the
phase shift introduced by the presence of the atoms. Off resonant
detection would allow for nondestructive atom detection (see for
example \cite{Dom00-115}).

\subsection{Mesoscopic physics}

The potentials created on an atom chip are very similar in scale
and confinement to the potentials confining electrons in
mesoscopic quantum electronics \citep{Imr87}.  There electrons
move {\em inside} semiconductor structures, in our case atoms move
{\em above} surfaces in atom optical circuits.  In both cases they
can be manipulated using potentials in which at least one
dimension is comparable to the de Broglie wavelength of the
guided, trapped particle. To find similarities and differences
between mesoscopic quantum electronics and mesoscopic atom optics
will probably become a very rich and fascinating research field.

Electrons in semiconductors interact strongly with the surrounding
lattice.  It is therefore hard to maintain their phase coherence
over long times and distances.  An atomic system on the contrary
is well isolated.  Furthermore, atoms (especially in a BEC) can be
prepared such that the temperature is extremely low with respect
to the energy level spacing. The consequence is that phase
coherence is maintained over much longer times and distances. This
might enable us to explore new domains in mesoscopic physics,
which are hard to reach with electrons.

\subsubsection{Matter wave optics in versatile potentials}

A degenerate quantum gas in the atom chip will allow us to study
matter wave optics in confined systems with non trivial
geometries, as splitters, loops, interferometers, etc. One can
think of building rings, quantum dots connected by tunnel
junctions or quantum point contacts \citep{Thy99-3762}, or even
nearly arbitrary combinations thereof in matter wave quantum
networks. For many atomic situations the electronic counterparts
can easily be identified. Atom chips will allow to probe a wide
parameter range of transverse ground state widths, confinement and
very large aspect ratios of $10^5$ and more. Atomic flow can be
monitored by observing the expansion from an on-board reservoir
along the conduit. Further perturbations and corrugations can be
added at any stage to the potential by applying additional
electric, magnetic or light fields to modify the quantum wire or
quantum well. In this manner we can also explore how disorder in
the guides may change the atomic behavior.

In the following, we give details regarding three exemplary matter
wave potentials on the atom chip.

\subsubsection{Interferometers}

In the near future it will be essential to develop and implement
interferometers, and to study through them the decoherence of
internal states and external motional states. Atom chip
interferometers have been discussed in detail in section
\ref{II-A-9}. They can be built either in the spatial
\citep{And01-100401} or the temporal domain
\citep{Hin01-1462,Hae01-063607}. Integrated on an atom chip, they
are very sensitive devices that may be used to measure inertial
forces or even to perform computation \citep{And00-052311}.
Coherence properties in more complicated networks can be studied
by observing interference and speckle patterns.

Interferometers can also serve as probes for the understanding of
surface--atom interactions allowing for a quantitative test of the
limits imposed on the atom chip by the warm surface for both
internal state and external (motional) state coherence. Since many
of the important parameters scale with the spin flip life time in
a trap (see section \ref{s:loss}), a first important step would be
to measure the (BEC) lifetime in a micro trap as a function of
distance to the surface. Aside from heating and spin flips, the
surface also induces `phase noise'. Interferometers will be able
to measure this subtle effect as a function of surface material
type and temperature as well as atom-surface distance and spatial
spread of the atomic superposition, through a reduction in the
fringe visibility. Finally, by coupling micro traps (atomic
quantum dots) to one of the interferometer arms, similar to the
mesoscopic electron experiments \citep{Buk98-871}, subtle
interaction terms may be investigated, e.g., 1/r second order
dipole interactions discussed by \cite{Ode00-5687}.

Internal state superpositions of atoms close to surfaces can be
studied using internal state interferometers. Using Raman
transitions or microwave transitions we can create superpositions,
observe their lifetime and put theoretical estimates to the test.

\subsubsection{Low dimensional systems}

Much is known about the behavior of fermions in low dimensional
strongly confining systems (one and two-dimensional systems) from
mesoscopic quantum electronic experiments. By designing low
dimensional experiments using atoms (weakly interacting bosons or
fermions) we expect to obtain further insight also about
electronic phenomena.

The role of interactions inside an atomic matter wave can range
from minimal in a very dilute system to dominating in a very dense
system. Low dimensional systems are especially interesting in this
context, since it is expected that the interactions between the
atoms will change for different potentials. The study of the
dependence of the interactions (scattering length) on the
dimensionality and the degree of confinement of the system
\citep{Ols98-938,Goe01-130402,Pet00-3745}, will benefit due to the
variety of potentials available on the atom chip.

\subsubsection{Non-linear phenomena}

Another example of an interesting regime for the study of
atom-atom interaction or non-linearity are multi-well potentials.
Again, as mentioned in the context of interferometers, the
splitting of a cloud of atoms into these multi sites can be either
temporal or spatial. Here calculations beyond mean field theory
are relevant, and new insight may be acquired. For example, one
expects a crossover from coherent splitting to number splitting in
different potential configurations, depending on the height of the
potential barrier, the density, and the scattering length
\citep{Men01-023601,Var01-568,Orz01-2386,Gre02-39}.

\subsubsection{Boundary between macroscopic and microscopic description}

Let us end this sub-section concerning mesoscopics by noting that
the ability to change the number of atoms in a system, or
alternatively to address specific atoms in an interacting
ensemble, will allow us to probe the boundary between the
macroscopic and microscopic description. Starting from a large
system, we will try to gain more and more control over the system
parameters, imprinting quantum behavior onto the system. On the
other hand we can try and build larger and larger systems from
single quantum objects (in modern lingo called qubits), and keep
individual control over the parameters. Success in such an
undertaking would bring us much closer to implementing quantum
information transfer and quantum information processing as
discussed below.

\subsection{Quantum information}

The implementation of quantum information processing requires
\citep{DiV00-771}: (i) storage of the quantum information in a set
of two-level systems (qubits), (ii) the processing of this
information using quantum gates, and (iii) reading out the
results. For a review of quantum computation we refer the reader
to \cite{Bou00}.

We believe that quantum optical schemes where the qubit is encoded
in neutral atoms can be implemented using atom optics on
integrated atom chips \citep{Sch02-gdansk}. These promise to
combine the outstanding features of quantum optical proposals, in
particular quantum control and long decoherence times, with the
technological capabilities of engineering micro-structures
implying scalability, a feature usually associated with solid
state proposals. Let us review some of the requirements:

\paragraph{The Qubit}
Using neutral atoms, the qubit can be encoded in two internal,
long lived states (e.g. two different hyperfine electronic ground
states). Single-qubit operations are induced as transitions
between the hyperfine states of the atoms.  These are introduced
by external fields, using RF pulses like in NMR or in
Ramsey-Bord\'e interferometers, Raman transitions or adiabatic
passage.

One method to realize a qubit is to write the qubits into single
atoms, which requires selective cooling and filling of atoms into
the qubit sites. However, recently it was proposed that single
qubits can also be written into an ensemble of atoms using `dipole
blockade' \citep{Luk01-037901}. This may be simpler as it avoids
the need for single atom loading of traps. As it will be pointed
out, the dipole blockade mechanism can also be used to manipulate
the qubit.

\begin{figure}[tb]
   \infig{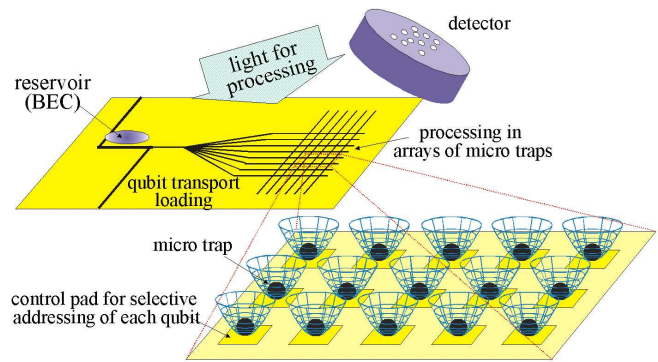}{\columnwidth }
   \caption{
   A possible implementation of a neutral atom qubit processor on an
   \emph{atom chip} which includes a reservoir of cold atoms in a
   well defined state (for example a BEC or a degenerate Fermi gas).
   From there the atoms can be transported using guides or moving
   potentials to the processing sites.  Either single atoms, or small
   ensembles of atoms are then loaded into the qubit traps.
   Each qubit can be addressed either by bringing light to each individual site
   separately, or by illuminating the whole processor and addressing the
   single qubits by shifting them in and out of resonance using local
   electric or magnetic fields created by the nano-structures on the
   atom chip. We think that electric fields are preferable, since
   magnetic fields might produce qubit state dependent phase shifts,
   which have to be corrected.  A different method would also be to
   address the single sites using field gradients like in NMR.}
   \label{fig:DipBlockQIProcessor}
\end{figure}

\paragraph{Entangling Qubits}
The fundamental two-qubit quantum gate requires state selective
interaction between two qubits, which is more delicate to
implement. A two-qubit quantum gate is a state dependent operation
such as a control not gate:
 \bea
  |0\rangle|0\rangle&\rightarrow&|0\rangle|0\rangle,\nonumber\\
  |0\rangle|1\rangle&\rightarrow&|0\rangle|1\rangle,\nonumber\\
  |1\rangle|0\rangle&\rightarrow&|1\rangle|1\rangle,\nonumber\\
  |1\rangle|1\rangle&\rightarrow&|1\rangle|0\rangle.\label{controlnot}
 \eea
A good way to implement such a quantum gate is by state selective
interactions, which can be switched on and off at will. This
interaction can be between the qubits themselves, or mediated by a
`bus'.  Neutral atoms naturally interact with each other.  To
achieve different phase shifts for different qubit states, either
the interaction between the qubits has to be state selective, or
it has to be turned on conditioned on the qubit state. There are
different ways to implement quantum gates in atom optics:
depending on the interaction, we distinguish between (a) the
generic interactions between the atoms, like the Van der Waals
interaction \citep{Jak99-1975,Cal00-022304,Bri00-415},
dipole-dipole interactions \citep{Bre99-1060,Bre00-062309} and (b)
interactions which can be switched on and off, for example
dipole-dipole interactions \citep{Jak00-2208,Luk01-037901} between
Rydberg states.

\paragraph{Dipole-blockade quantum gates between mesoscopic atom
ensembles} \cite{Luk01-037901} devised a technique for the
coherent manipulation of quantum information stored in collective
excitations of many-atom mesoscopic ensembles by optically
exciting the ensemble into states with a strong atom-atom
interaction. Under certain conditions the level shifts associated
with these interactions can be used to block the transitions into
states with more than a single excitation. The resulting
dipole-blockade phenomenon closely resembles similar mesoscopic
effects in nanoscale solid-state devices. It can take place in an
ensemble with a size that can exceed many optical wavelengths and
can be used to perform quantum gate operations between distant
ensembles, each acting as a single qubit.

\paragraph{Cavity QED}
The 2-qubit processing operation may be realized through a direct
interaction (entanglement) between two atoms or through an
intermediate `bus'. A light mode of a high Q cavity can serve as
such a `bus' acting on an array of atoms trapped inside the cavity
\citep{Pel95-3788}. Atoms in high Q cavities which in turn are
connected with fibers, can also act as a converting device between
`flying' qubits (photons) which transverse distances, and storage
qubits (atoms). The same principle can be used for entangling
atoms in different cavities for a `distributed' computation
process \citep{Enk98-205,Enk99-2659}. In all of the above, the
atom chip promises to enhance the feasibility of accurate
atom-cavity systems.

\paragraph{Input/Output}
Even without high Q cavities, an integrated atom chip, with atoms
trapped in well controlled microtraps and with individual site
light elements, can probably provide input/output processes by
making use of techniques such as light scattering from trapped
atomic ensembles \citep{Dua01-413}, slow light
\citep{Hau99-594,Vit00-445}, stopped light
\citep{Phi01-783,Liu01-490,Fle02-022314} or macroscopic spin
states \citep{Dua00-5643,Jul01-400}.

\vspace{5mm}

Let us summarize the road map for quantum computation with the
atom chip: one would need to implement, {\em (a)} Versatile traps
to accurately control atoms up to the stage of entanglement, {\em
(b)} Controlled loading of single qubits (atoms or excitations)
into these traps in well defined internal and external states,
{\em (c)} Manipulation and detection of individual qubits, {\em
(d)} Control over decoherence and {\em (e)} Scalability to be able
to achieve controlled quantum manipulation of a large number of
qubits.

\section{Conclusion}

Neutral-atom manipulation using integrated micro-devices is a new
and extremely promising experimental approach. It promises to
combine the best of two worlds:  the ability to use cold atoms - a
well controllable quantum system, and the immense technological
capabilities of nanofabrication, micro optics and micro
electronics to manipulate and detect the atoms.

In the future, a final integrated atom chip will have a reliable
source of cold atoms with an efficient loading mechanism, single
mode guides for coherent transportation of atoms, nano-scale
traps, movable potentials allowing controlled collisions for the
creation of entanglement between atoms, high resolution light
fields for the manipulation of individual atoms, and internal
state sensitive detection of atoms.  All of these, including the
bias fields and possibly even the light sources and the read-out
electronics, could be on-board a self-contained chip.  Such a
robust and easy to use device, would make possible advances in
many different fields of quantum physics: from applications such
as clocks, sensors and implementations of quantum information
processing and communication, to new experimental insight into
fundamental questions relating to decoherence, dissorder,
non-linearity, entanglement and atom scattering in low dimensional
physics.

\section*{Acknowledgement}
\label{s:Acknowledgement}

Foremost we would like to thank all the members of the Innsbruck,
now Heidelberg, atom chip group for their enthusiasm and enormous
effort they put into the experiments. We would like to personally
thank our long time theoretical collaborators Peter Zoller,
Tommaso Calarco and Robin C\^{o}t\'{e}. The atom chips for the
Innsbruck-Heidelberg experiments were fabricated by Thomas Maier
at the Institut f\"{u}r Festk\"{o}rperelektronik, Technische
Universit\"{a}t Wien, Austria, and by Israel Bar-Joseph at the
Sub-micron center, Weizmann Inst. of Science, Israel. We would
also like to extend a warm thanks to the entire atom chip
community for responding so positively to our requests for
information and figures. Our work was supported by many sources,
most notably the Austrian Science Foundation (FWF), project SFB
S065-05, F15-07, the Deutsche Forschungsgemeinschaft
Schwerpunktprogramme: Quanten Informationsverarbeitung and
Wechselwirkungen in ultrakalten Atom- und Molek\"{u}lgasen, the
Institute for Quantum Information, and the European Union,
contract numbers IST-1999-11055 (ACQUIRE), HPRI-CT-1999-00069
(LSF), TMRX-CT96-0002, and HPMF-CT-1999-00235.

\bibliography{references_v700}

\end{document}